\documentstyle[epsf,aasmod]{chithesis}
\newcommand{\spacing}[1]{\renewcommand{\baselinestretch}{#1}\large\normalsize}

\spacing{2}
\title{DATA REDUCTION AND ANALYSIS OF THE PYTHON V COSMIC MICROWAVE BACKGROUND ANISOTROPY EXPERIMENT}
\author{Kimberly Ann Coble}
\university{The University of Chicago}
\faculty{The Faculty of The Division of The Physical Sciences}
\department{Department of Astronomy and Astrophysics}
\degreename{Doctor of Philosophy}
\placename{Chicago, Illinois}
\degreedate{December 1999}  
\begin{document}                   
%
%
%
%
\newcommand{\be}{\begin{equation}}
\newcommand{\ee}{\end{equation}}
\newcommand{\etal}{et\ al.\ }
\newcommand{\mdot}{\dot{\rm M}}
%
\maketitle			   
\makecopyright			   
%
%
\newpage
\thispagestyle{empty}
\setcounter{page}{3}
\null\vfil
\begin{center}
{\it `Helped are those who love the entire cosmos... for to them will be shown the unbroken web of life and the meaning of infinity.'}
\end{center}
\rightline{Alice Walker, \it{The Temple of My Familiar}}
\vspace{1.5in}
%
%
%
%
\topmatter{Acknowledgments}
{\small `The spoken word is a jacket too tight.'

\rightline{Poi Dog Pondering, {\it Collarbone}}}

I cannot adequately express in words my gratitude to the people
who have helped me along my path.

I thank my advisor, Scott Dodelson, for being a good teacher, able to explain
in a simple way difficult concepts, and for advice at crucial moments.
I am indebted to the other memebers of the Python collaboration,
especially Lloyd Knox for explaining sophisticated analysis techniques,
John Kovac for explaining the experiment, and Ken Ganga, who did
much of the work on the Python V foregrounds.
I thank the MSAM collaboration, especially Steve Meyer, Jason Puchalla,
Grant Wilson, Dave Cottingham, and Ed Cheng. Much of my work on
window functions was done in the context of working with their group.
I thank Dave Cottingham and Grant Wilson for many conversations,
cross-checks and useful memos on window functions. I also thank
Bharat Ratra for helpful discussions on window functions.

The many other people working on the CMB at Chicago
have sparked interesting discussions and I have learned much from them.
John Carlstrom taught a fantastic course on experimental techniques.
The graduate students here have been unendingly supportive,
sharing scientific knowledge and always being
there for each other.  Aparna Venkatesan and Brad Holden especially
have comiserated and celebrated with me throughout the years.
I don't know how any of us would make it without Sandy ``Mom'' Heinz.

My family has been a continual source of love and support. Thanks Mom, Dad,
and Kellie.

To Sarah Lord, your cleverness and caring inspire me.
To Chris Newell, thank you for feeding my body and soul. Your
love of life and thoughtfulness are contagious.
To Erik Leitch, thank you for countless conversations on the
CMB, on making the transition from graduate student to graduate,
and on life. You make the best pizza in Chicago.

And finally, I thank Mark Dragovan, without whom this thesis would not be
possible. Thank you for giving me the opportunity to work on Python,
for your encouragement, and for being a Mentor.

%
\tableofcontents
\listoffigures
\listoftables
%
%

\topmatter{Abstract}
             
Observations of the microwave sky using the Python telescope 
in its fifth season of operation at the Amundsen-Scott South Pole Station
in Antarctica are presented.
The system consists of a 0.75 m off-axis telescope 
instrumented with a
HEMT amplifier-based radiometer having continuum sensitivity 
from 37-45 GHz in two frequency bands. With
a $0.91^{\circ} \times 1.02^{\circ} $ beam the instrument fully 
sampled 598 deg$^2$ of sky, including fields measured during the previous
four seasons of Python observations. Interpreting the observed
fluctuations as anisotropy in the cosmic microwave background, we
place constraints on the angular power spectrum of fluctuations
in eight multipole bands up to $l \sim 260$. The observed spectrum
is consistent with both the COBE experiment and previous Python results.
Total-power Wiener-filtered maps of the CMB are also presented.
There is no significant contamination from known foregrounds.
The results show a discernible rise in the angular
power spectrum from large ($l \sim 40$) to small ($l \sim 200$)
angular scales.

%
%
%
\chapter{Introduction}

{\small `We know of an ancient radiation
that haunts dismembered constellations,
a faintly glimmering radio station.'

\rightline{Cake, {\it Frank Sinatra}}}

The Cosmic Microwave Background Radiation (CMB),
which formed when the universe was $\sim$ 300,000
years old (redshift $z \sim 1100$), is the 
signature of a young, hot, dense universe. Along with
big bang nucleosynthesis and the Hubble expansion, the CMB is one of
the key observational pillars that support the big bang model
of the universe.

Since its discovery (Penzias and Wilson 1965) and interpretation
(Dicke et al. 1965), the CMB has been one of the most powerful tools
for discriminating among cosmological models.
The COBE satellite opened up a new era in CMB measurement,
by detecting anisotropy on the largest scales with the
DMR instrument (Smoot et al. 1992) and by establishing
the blackbody nature (T=2.728 $\pm$ 0.004 K today)
of the CMB spectrum to high precision with the
FIRAS instrument (Mather et al. 1990, Fixsen et al. 1996).
Measurement of anisotropy in the CMB
directly probes conditions of the early universe.
The anisotropies in the CMB are the
seeds of the large scale structures (galaxies, clusters of galaxies,
and super-clusters) that we observe today.
Observations of the
angular power spectrum of CMB temperature fluctuations can be used to test
theories of structure formation and constrain cosmological
parameters.

There are three classes of measurements that
could be made of the CMB: spectral,
temperature anisotropy, and polarization anisotropy.
This thesis describes a specific CMB temperature anisotropy experiment,
Python V, with an emphasis on data reduction and analysis techniques.
It is an expansion of Coble et al. (1999).

\section{Theory}

The theory of how CMB anisotropies grew via gravitational
instability from primordial fluctuations is now well understood and has been
discussed at length by other authors (see, for example, Hu et al. 1997).
Before redshift $z \sim 1100$, photons and baryons were tightly
coupled via Compton scattering, in a photon-baryon fluid.
Acoustic oscillations in the fluid were created by the resistance
of the photon pressure against gravitational compression.
As the universe cooled,
neutral hydrogen formed and radiation decoupled from
matter and began free-streaming. The epoch of decoupling
occurred at $z \sim 1100$ for most cosmologies.
Regions of compression and rarefaction in
\linebreak
\\
\noindent the photon-baryon fluid
at decoupling correspond to hot
and cold spots that we observe in the CMB.

\begin{figure}[h!]
\centerline{\epsfxsize=13.9cm \epsfbox{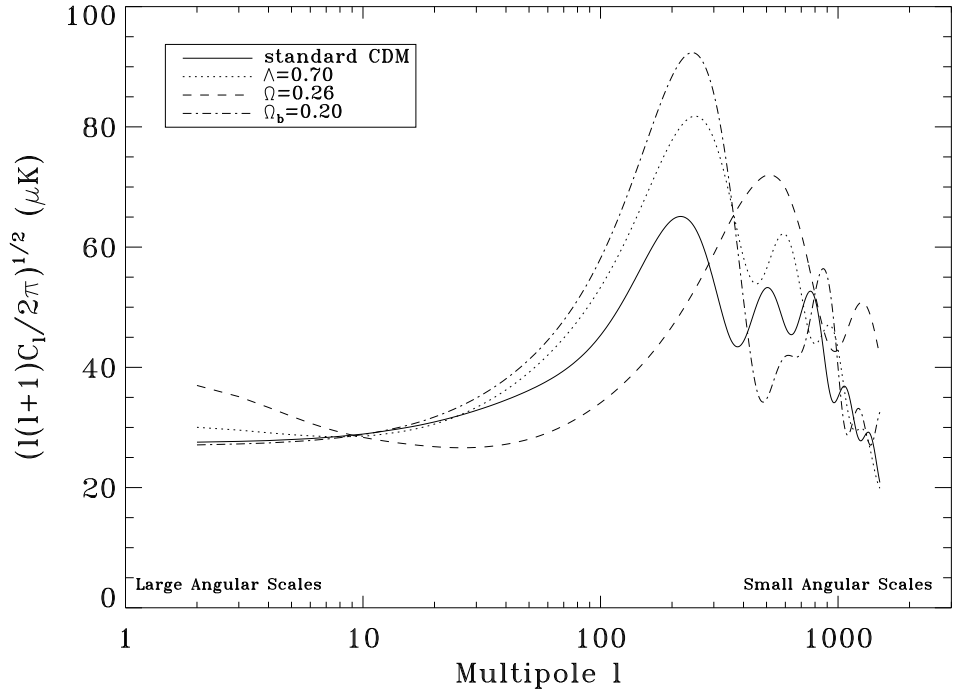}}
\ssp
\caption[Model CMB angular power spectra for various cosmological parameters.]{Model CMB angular power spectra for various cosmological parameters. Large $l$ corresponds to small angular scales. Shown are a standard CDM model (solid line, density $\Omega=1.0$, Hubble constant $H_0=50$ km/s/Mpc, baryon density $\Omega_b=0.05$, and cosmological constant $\Lambda=0.0$), a model with high baryon content (dot-dashed line, $\Omega=1.0$, $H_0=50$ km/s/Mpc, $\Omega_b=0.20$, and $\Lambda=0.0$), a model with a non-zero cosmological constant (dotted line, $\Omega=1.0$, $H_0=50$ km/s/Mpc, $\Omega_b=0.02$, and  $\Lambda=0.70$), and an open model (dashed line, $\Omega=0.26$, $H_0=50$ km/s/Mpc, $\Omega_b=0.06$, and  $\Lambda=0.0$). The models were computed using CMBFAST (Seljak and Zaldarriaga 1996).}
\label{fig:clfig}
\end{figure}

CMB angular power is usually expressed in terms of {\em $C_{l}'s$}. If
the sky temperature is expanded as
$T(\theta,\phi)=\Sigma_{lm}a_{lm}Y_{lm}(\theta,\phi)$,
then the multipole moments, $C_{l}$, are defined as
$C_{l}=<|a_{lm}|^{2}>$. Large $l$ corresponds to small angular scales.
Measuring the angular power spectrum of temperature anisotropies
in the CMB is a
powerful tool for determining many cosmological parameters,
including the expansion rate of the universe ($H_0$),
the total density of the universe ($\Omega$), the amount of
baryonic matter ($\Omega_b$) in the universe, the
amount and nature of dark matter, the spectral index of primordial
fluctuations ($n$), the cosmological constant ($\Lambda$), and
the ionization history of the universe.
Figure \ref{fig:clfig} shows the dependence of
the angular power spectrum on various
cosmological parameters.

Traditional cosmological tests, such as galaxy and cluster
distributions, cannot probe these parameters directly, because
they only determine the distribution of visible mass. They must
assume an underlying bias, the unknown relation between mass and light,
whereas CMB anisotropy corresponds directly to density fluctuations
(White et al. 1994).
CMB measurements are also independent of the extra-galactic distance
ladder and probe parameters on a global scale.

As one example of the dependence of the CMB angular
power spectrum on a cosmological parameter, the angular scale of
the first acoustic peak (Figure \ref{fig:clfig}) is a sensitive
indicator of the total density, $\Omega$.
The horizon scale at the epoch of decoupling, which corresponds
to the scale of the first acoustic peak, is given by
\begin{equation}
\Delta\theta \approx 0.87^{\circ} \Omega^{1/2}.
\label{eq:zrec}
\end{equation}
As $\Omega$ decreases, the peak occurs at smaller angular scales. For
a flat universe ($\Omega$=1), the corresponding angular scale
is $\sim 1^{\circ}$.

\section{Temperature Anisotropy Experiments}

As with most astronomical measurements, CMB observations are
usually conducted from locations inhospitable to humans, namely
places that are high, dry, and cold, in order to reduce the interference
from the atmosphere.
CMB experiments have been and will be conducted from satellites,
such as COBE, MAP (see the MAP website), and Planck (Planck website).
The advantage to satellite missions
is that they can map the whole sky with high precision
and without interference from the atmosphere.
The disadvantages are that they often have a long
development period and are costly.
A number of balloon-borne experiments have been conducted, such
as ARGO (de Bernardis et al. 1994, Masi et al. 1996),
BAM (Tucker et al. 1996), FIRS (Ganga et al. 1994),
MAX (Lim et al. 1996, Tanaka et al. 1996), MSAM (Cheng et al. 1997), and
QMAP (Devlin et al. 1998, Herbig et al. 1998).
These balloon flights typically observe
the sky for one night per year. The next generation of balloon-borne
experiments, long-duration ballooning (LDB) experiments, are currently
under way. The BOOMERANG (BOOMERANG website)
experiment circumnavigated Antarctica, observing for
approximately 10 days in December 1998. The TopHat (TopHat website)
experiment is scheduled
for an LDB flight in December 1999.
A plethora of experiments have detected or constrained CMB
anisotropy from the ground. The South Pole is an especially
attractive site for CMB measurement because of its stable
atmosphere and high altitude. The Python (Dragovan et al. 1994,
Ruhl et al. 1995, Platt et al. 1997, Kovac et al. 1999),
Viper (Peterson et al. 1998), and
ACME (Gaier et al. 1992, Schuster et al. 1993, Gundersen et al. 1995)
telescopes have made observations from the South Pole.
Other ground-based experiments include CAT (Scott et al. 1996),
IAB (Piccirillo and Calisse 1993), IAC (Femenia et al. 1997),
MAT (Torbet et al. 1999),
OVRO NCP (Readhead et al. 1989), OVRO RING5M (Leitch et al. 1998), 
Saskatoon (SK) (Netterfield et al. 1996),
SuZIE (Church et al. 1997) and Tenerife (Hancock et al. 1997).

Since anisotropies are temperature differences,
CMB experiments must difference (chop) in some way.
The chopping, whether internal or external,
must be quick, and come back to the same place on the sky
faster than the 1/f noise in the electronics and atmosphere.
Often this is accomplished with a
chopping secondary mirror (the chopper) and/or an internal Dicke switch.
Some simple chopping strategies include a single difference (2-point chop),
for which $\Delta T_i = T_{i+1}-T_i$, a double difference (3-point chop), 
for which $\Delta T_i = T_i-\frac{1}{2}T_{i-1}-\frac{1}{2}T_{i+1}$,
and a 4-point chop, for which
$\Delta T_i = -\frac{1}{4}T_{i}+\frac{3}{4}T_{i+1}-\frac{3}{4}T_{i+2}+\frac{1}{4}T_{i+3}$.
Other experiments perform a continuous (fast) scan with the chopper
and modulate the data in software, scan slowly
with the telescope and read out the
detectors quickly, or spin rapidly.
Chopping strategies are
sometimes called lockins, harmonics, modulations, or demodulations.

There are 2 basic detector technologies used for CMB experiments:
bolometers and HEMTs. SIS detectors are also sometimes used.
HEMTs have the advantage that they are robust and easy to use.
They are inherently polarization sensitive, coherent detectors
that typically operate at frequencies $\lesssim$ 90 GHz.
They can operate at room temperature, but become more sensitive as
they are cooled down to $\sim$ 10 K.
Bolometers are more sensitive than HEMTs, but they are less robust
and must be cooled to at least 0.25 K, so much
work must go into the cryogenics.
Bolometers typically operate at frequencies $\gtrsim$ 90 GHz and are often
used on balloon-borne experiments, where their sensitivity is
better than the atmospheric noise.
It is advantageous to make measurements at several frequencies
in order to spectrally distinguish possible contamination
by galactic foregrounds (e.g. dust, synchrotron).

CMB anisotropies can also be measured
using interferometers, such as CAT, which has detected anisotropy,
DASI (White et al. 1997), which
will observe at large to medium angular scales
from the South Pole, and CBI (CBI website), which will
observe at small angular scales from Chile.
Interferometers are advantageous because they have low
systematics.

In choosing an observing strategy (how much time to spend
on which region of the sky and in what order), the theoretically
optimal signal-to-noise ratio S/N=1, is too optimistic. In reality,
experiments should have S/N$>$1, in order to perform consistency
checks on the data.
Interlaced observing strategies are desirable because they
reduce striping in the maps obtained, but that is not always
possible given observing sites and conditions.

\begin{figure}[h!]
\centerline{\epsfxsize=13.9cm \epsfbox{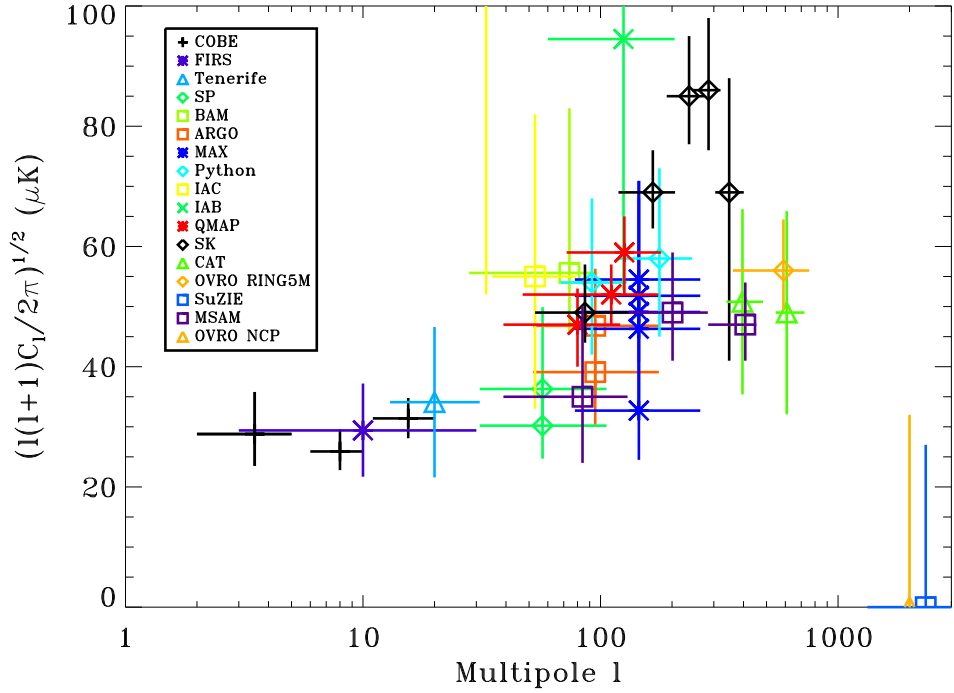}}
\ssp
\caption[CMB anisotropy measurements as of Feb 1999.]{CMB anisotropy measurements as of Feb 1999. The points are COBE (Hinshaw et al. 1996), FIRS (Ganga et al. 1994), Tenerife (Hancock et al. 1997), SP (Gundersen et al. 1995), BAM (Tucker et al. 1996), ARGO (de Bernardis et al. 1994, Masi et al. 1996), MAX (Tanaka et al. 1996), Python I-III (Platt et al. 1997), IAB (Piccirillo and Calisse 1993), IAC (Femenia et al. 1997), QMAP (de Oliveira-Costa et al. 1998), SK (Netterfield et al. 1996), CAT (Scott et al. 1996), OVRO RING5M (Leitch et al. 1998), SuZIE (Church et al. 1997), MSAM (Wilson et al. 1999), and OVRO NCP (Readhead et al. 1989).}
\label{fig:expt_summary}
\end{figure}

Results from the COBE satellite (Smoot et al. 1992) tightly
constrain the angular power spectrum at the
largest angular scales. Several experiments
have measured the angular power spectrum at degree angular
scales (e.g., SP, MAX, Python I-IV, MSAM, SK, QMAP).
Figure \ref{fig:expt_summary}
is a compilation of measurements of the angular power spectrum previous
to the publication of the PyV measurements.
Collectively the data show a rise in
power towards smaller angular scales; individually no experiment covers
a wide range of angular scales from COBE scales to degree scales,
and most cover only small regions of the sky.

The dataset from the fifth Python observing season (hereafter PyV)
has sufficient sky coverage to probe the smallest scales
to which COBE was sensitive, while having a small enough beam to
detect the rise in angular
power at degree angular scales.

\chapter{The Python Experiment}

{\small `Python is a large non-venomous Old World snake that coils around and suffocates its prey.'

\rightline{Oxford American Dictionary}}

\section{Introduction to the Python Experiment}

In its first four seasons the Python experiment detected significant
anisotropy in the CMB (Dragovan et al. 1994 (PyI), Ruhl et al. 1995 (PyII),
Platt et al. 1997 (PyIII), Kovac et al. 1999 (PyIV)). Observations from the
first three seasons were made at 90 GHz with a bolometer system
and a 4-point chop scan strategy, yielding CMB detections at
angular scales of $l \sim 90$ and $l \sim 170$. During the PyIV season
measurements were made using the same scan strategy with
a HEMT amplifier-based radiometer, confirming PyI-III detections 
in a 37-45 GHz frequency band. 

Observations were made from November 1996 through February 1997 in
the fifth Python observing season.
In order to increase the range of observed angular scales, a smoothly scanning
sampling scheme was implemented. As a result,
PyV is sensitive to the CMB angular power spectrum
from $l \sim$ 40 to $l \sim$ 260.

Table \ref{tbl:pyth_obs} summarizes the
observing parameters from all five Python seasons.

\begin{table*}
\begin{center}
\begin{tabular}{|cccccccc|}
\tableline
Season & dates & $\nu_{c}$ & FWHM & chop & throw & $l_{eff}$ & fields \\
\tableline
PyI & 12/92 & 90.3 & 0.82$^{\circ }$ & 4-pt & 8.25$^{\circ }$ & 87 & 15 \\ 
PyII & 12/93 & 90.3 & 0.82$^{\circ }$ & 4-pt & 8.25$^{\circ }$ & 87 & 31\\ 
PyIII-L & 10/94-12/94 & 90.3 & 0.82$^{\circ }$ & 4-pt & 8.25$^{\circ }$ & 87 & 110 \\ 
PyIII-S & 10/94-12/94 & 90.3 & 0.82$^{\circ }$ & 4-pt & 2.75$^{\circ }$ & 170 & 140 \\ 
PyIV & 2/96 & 40.1 & 1.06$^{\circ }\times $1.12$^{\circ }$ & 4-pt & 8.25$^{\circ }$ & 85 & 14 \\ 
PyV & 11/96-2/97 & 40.3 & 0.91$^{\circ }\times $1.02$^{\circ }$ & triangle & 11.0$^{\circ }$ & 40-260 & 598 deg$^{2}$ \\
\tableline
\end{tabular}
\end{center}
\ssp
\caption[Python observing parameters.]{Python observing parameters (adapted from Kovac et al. 1999). $\nu_{c}$ is the central frequency of the observations, FWHM is the FWHM of the beam,  chop is the chopping strategy, throw is the extent of the chopper sweep, $l_{eff}$ is the multipole (angular scale) at which observations were made, and fields is the number of fields observed or the sky coverage.}
\label{tbl:pyth_obs}
\end{table*}

\section{Instrument}

\begin{figure}[t!]
\centerline{\epsfxsize=13.0cm \epsfbox{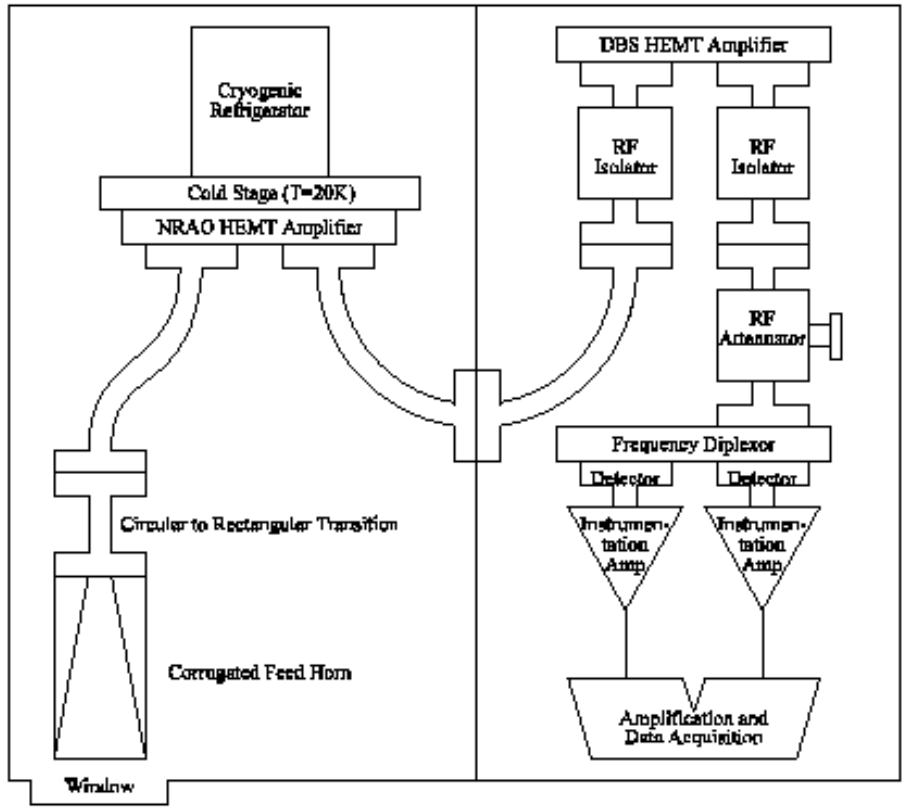}}
\ssp
\caption[Schematic diagram of the receiver.]{Schematic diagram of the receiver (Alvarez 1996). The signal path is shown for only one of the two feeds.}
\label{fig:hemt_scheme}
\end{figure}

\begin{figure}[ht!]
\centerline{\epsfxsize=13.9cm \epsfbox{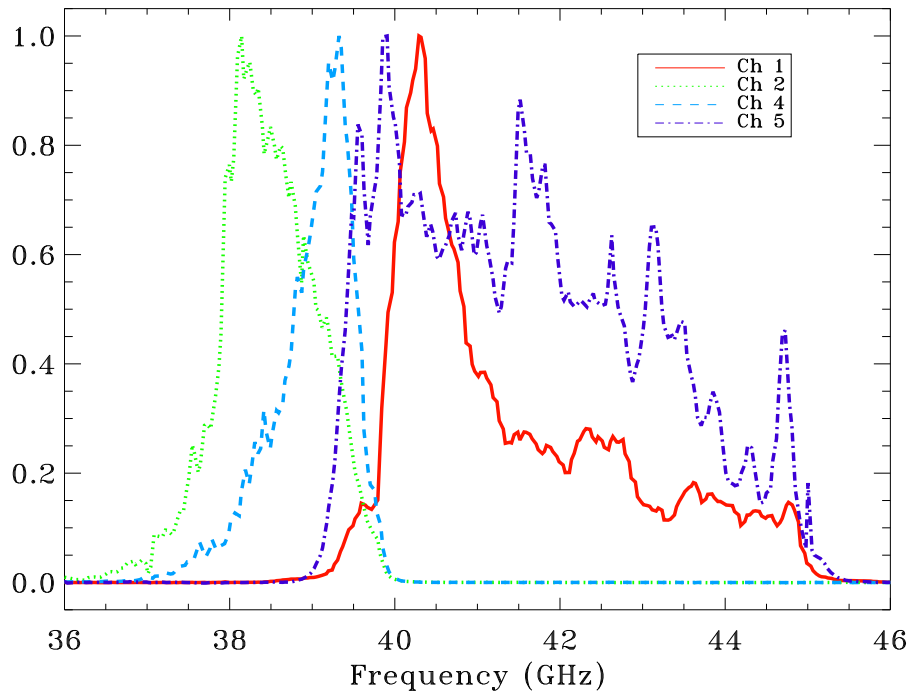}}
\ssp
\caption[Bandpasses for the 4 data channels.]{Bandpasses for the 4 data channels. Channel 3 is a dark channel. Channels 1 and 2 observe the same points on the sky as do channels 4 and 5. The bandpasses are similar in frequency range, so that the reduced data from channels which observe the same points on the sky are eventually co-added (Chapter 3).}
\label{fig:bandpass}
\end{figure}

The PyV measurements were made using the same receiver as the PyIV system
as described in Alvarez (1996) and Kovac et al. (1999).
A schematic of the receiver is shown in Figure \ref{fig:hemt_scheme}.
The receiver consists of  two focal-plane feeds, each with a
single 37-45 GHz HEMT amplifier.
A diplexer splits each signal at $\sim$ 41 GHz before detection,
giving four data channels.  The analysis reported here
eventually combines signals from all four channels,
resulting in a thermal radiation centroid $\nu_{c} = 40.3$ GHz and effective
passband $\Delta \nu = 6.5$ GHz for the PyV dataset. The bandpasses for the four data channels are shown in Figure \ref{fig:bandpass}. 

The receiver is mounted on a 0.75 m diameter
off-axis parabolic telescope (Dragovan et al. 1994),
which is surrounded by a large ground shield to
block stray radiation from the ground and Sun.
The beams corresponding to the two feeds observe the same
elevation and are separated by 2.80$^{\circ}$ on the sky. These beams
are scanned horizontally across the sky by a large rotating vertical flat
mirror, the chopper, at 5.1 Hz. 
The new scan strategy motivated two changes from the
instrument configuration described in Kovac et al. (1999):
the frequency response of the data system was
extended by switching to 100 Hz-rolloff antialiasing Bessel filters,
and the data recording rate was correspondingly
increased, to 652.8 samples/sec for each channel.

\begin{figure}[h!]
\centerline{\epsfxsize=13.9cm \epsfbox{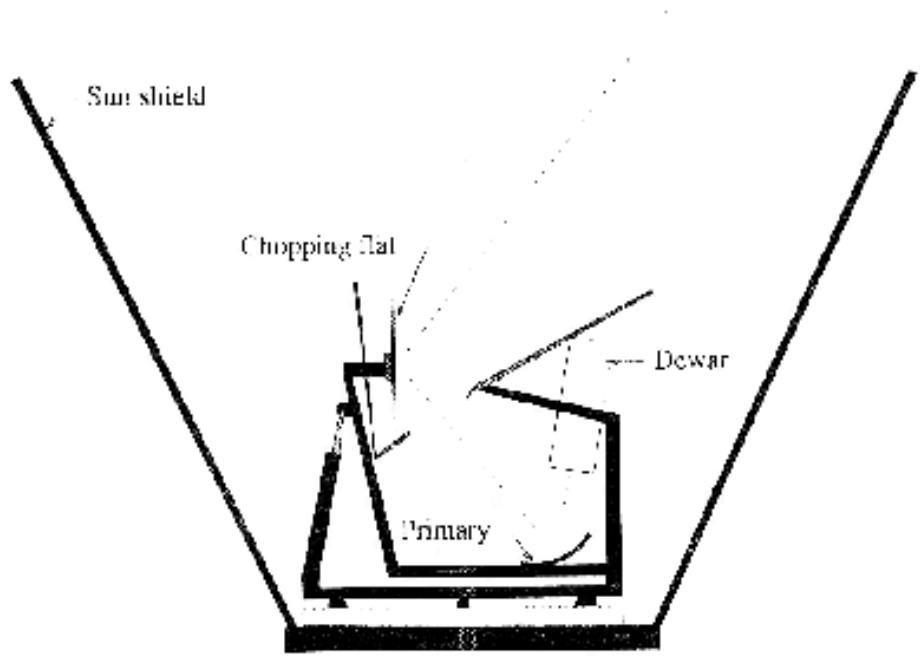}}
\ssp
\caption[Schematic diagram of the Python telescope.]{Schematic diagram of the Python telescope (Dragovan et al. 1994). The telescope is surrounded by a large ground/Sun shield. The chopping flat (chopper) is oriented vertically and scans in azimuth.}
\label{fig:py_schematic}
\end{figure}

\begin{figure}[h!]
\centerline{\epsfxsize=13.9cm \epsfbox{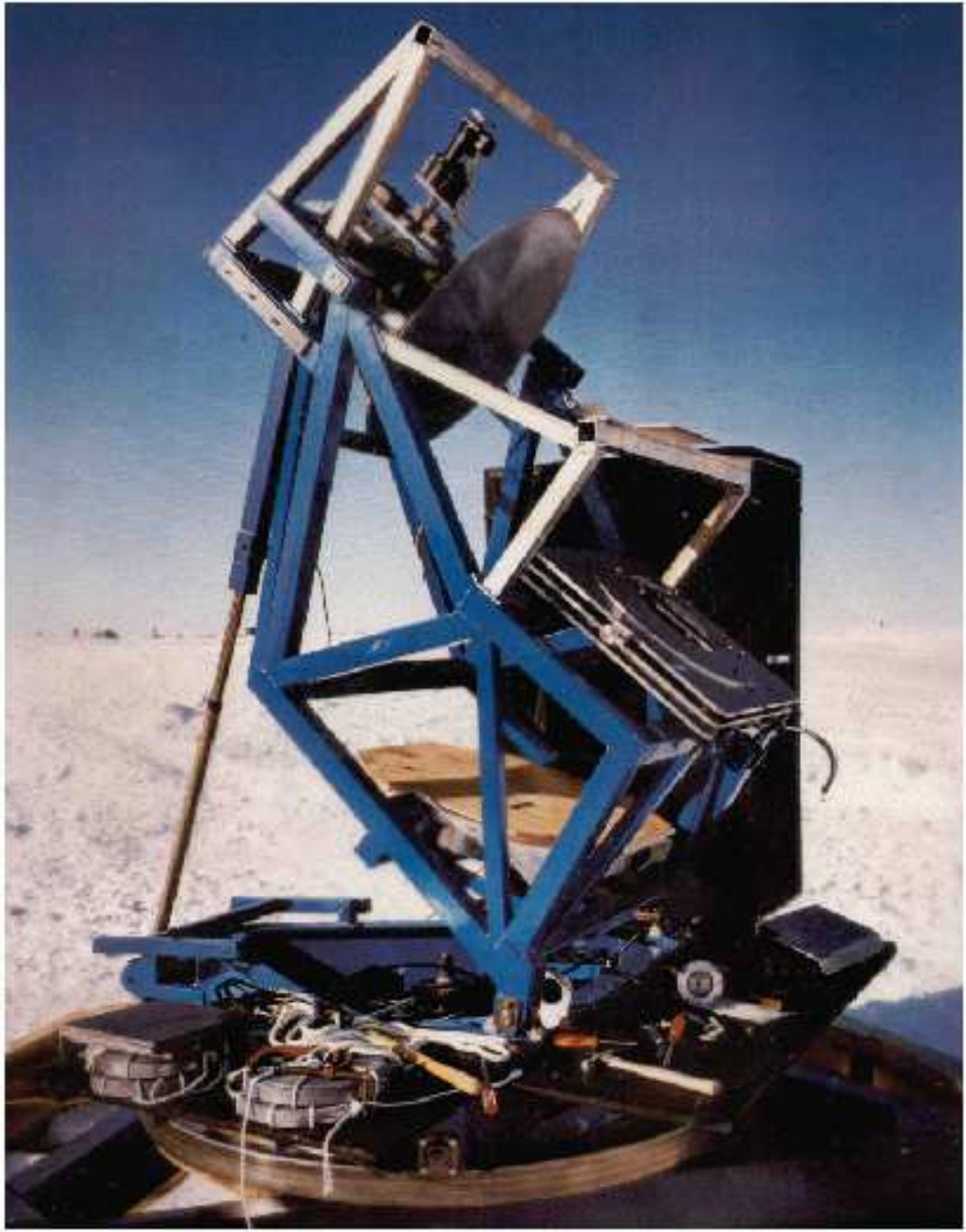}}
\ssp
\caption[Picture of the Python telescope without the ground shield.]{Picture of the Python telescope without the ground shield or dewar. For CMB observations the telescope is oriented such that the chopper is vertical.}
\label{fig:python_tel}
\end{figure}

\begin{figure}[h!]
\centerline{\epsfxsize=13.9cm \epsfbox{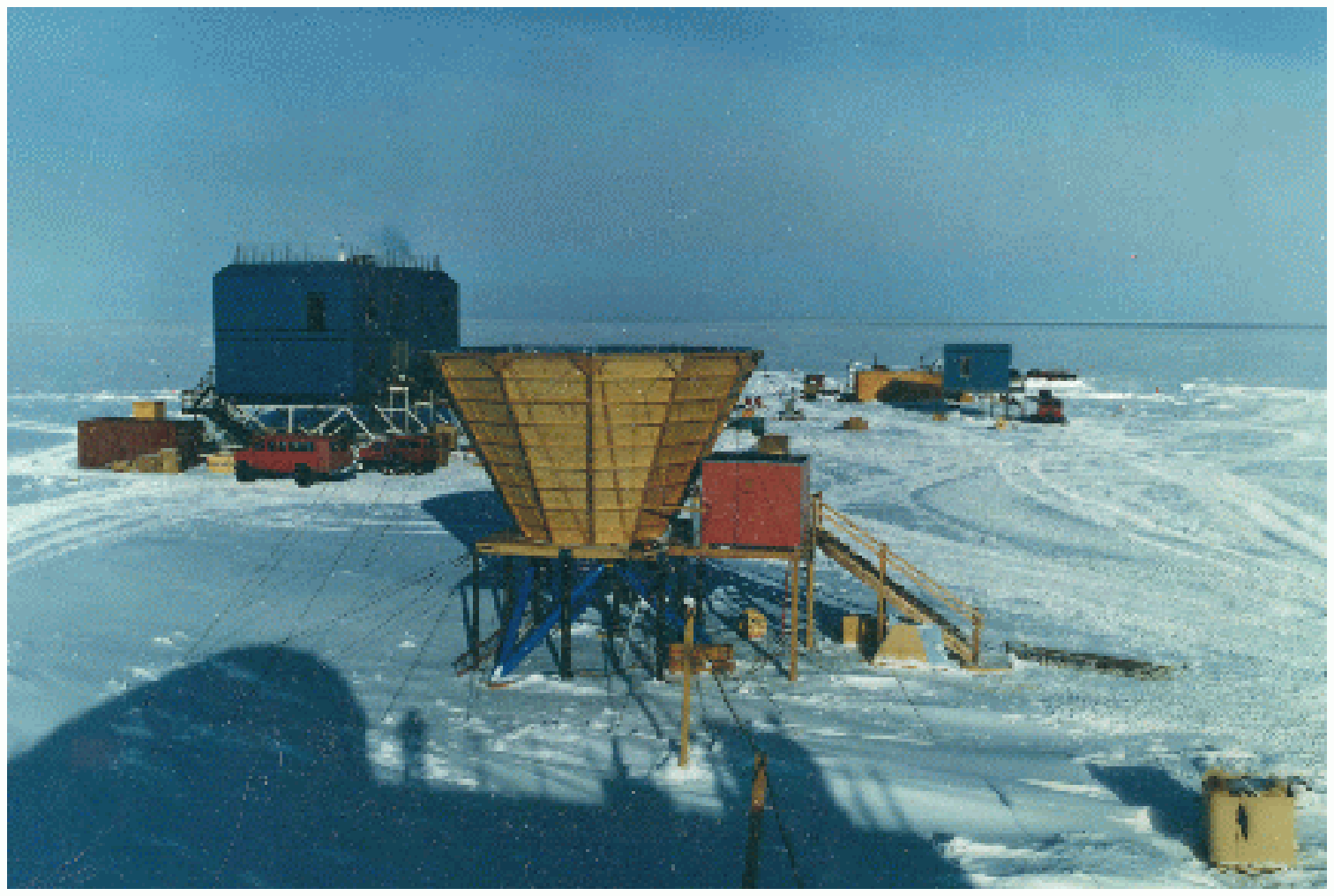}}
\ssp
\caption[A view of the Python telescope and surrounding labs.]{A view of the Python telescope and surrounding labs.}
\label{fig:sp_dslabs}
\end{figure}

A schematic of the telescope is shown in Figure \ref{fig:py_schematic}.
Pictures of the telescope with and without the ground shield are shown
in Figures \ref{fig:python_tel} and \ref{fig:sp_dslabs} respectively.

\spacing{2}

\section{Calibration}

As in previous Python seasons, the primary DC calibration of the detectors
was derived using liquid nitrogen, liquid oxygen, and ambient
temperature thermal loads external to the receiver (Dragovan et al. 1994, Ruhl et al. 1995).
Load calibrations were performed approximately once per day,
and gains were found to be consistent over the
entire season to within $\pm$ 2\%, with no discernible trends.
Gain compression, which was a source of systematic
uncertainty in the calibration of
PyI-PyIII, is measured to be negligible for
the Python HEMT receiver.  Systematic uncertainty
in the DC load calibration is estimated to be $\pm$ 10\%.

Several efficiencies must be estimated to relate
the load calibrations to celestial response
in the main beam, which account for power losses
in the atmosphere, in the sidelobes,
and in the telescope, and they are calculated using data from skydips and 
from various beam measurements.
In order to account for the net effect of atmospheric attenuation and
beam efficiency, the data must be multiplied by a factor of 1.10.
The resulting systematic calibration uncertainty of $^{+10\%}_{-4\%}$ is 
asymmetric, due to the fact that the individual losses
are small positive numbers and hence the errors in their
estimation follow skewed distributions (Kovac et al. 1999).

The dynamic response of the system was calculated from laboratory measurements
of the transfer functions for the AC coupling and
antialiasing filters in the data system,
and confirmed on the telescope by comparison of observations
made of the moon using normal and slow chopper speeds.
The response speed of the detectors is not a concern for this
calibration or for its uncertainty.
An appropriate response correction factor is applied
to each modulation of the data.
The uncertainty on these factors is small, and
is dominated by a $\pm$ 5\% systematic uncertainty on
their common normalization.

The overall uncertainty in the calibration of this
dataset is estimated to be $^{+15\%}_{-12\%}$.
Antenna temperature has been converted to units
of $\delta T_{\rm CMB}$ throughout.
The calibrations, efficiencies and their uncertainties are summarized
in Table \ref{tbl:calib}.

\begin{table*}
\begin{center}
\begin{tabular}{|ll|}
\tableline
Calibration & Uncertainty (\%) \\
\tableline
DC load systematic & $\pm$ 10 \\
DC load statistical & $\pm$ 2 \\
atmospheric attenuation &  \\
plus beam efficiency &  +10, -4 \\
dynamic response & $\pm$ 5 \\
{\bf TOTAL} & {\bf +15, -12} \\
\tableline
\end{tabular}
\end{center}
\caption{Calibrations, efficiencies and corresponding uncertainties.}
\label{tbl:calib}
\end{table*}

\section{Observations}

Observations for the PyV season were taken from November 1996
through February 1997.

\subsection{Observing Regions}

\begin{figure}[t!]
\centerline{\epsfxsize=13.9cm \epsfbox{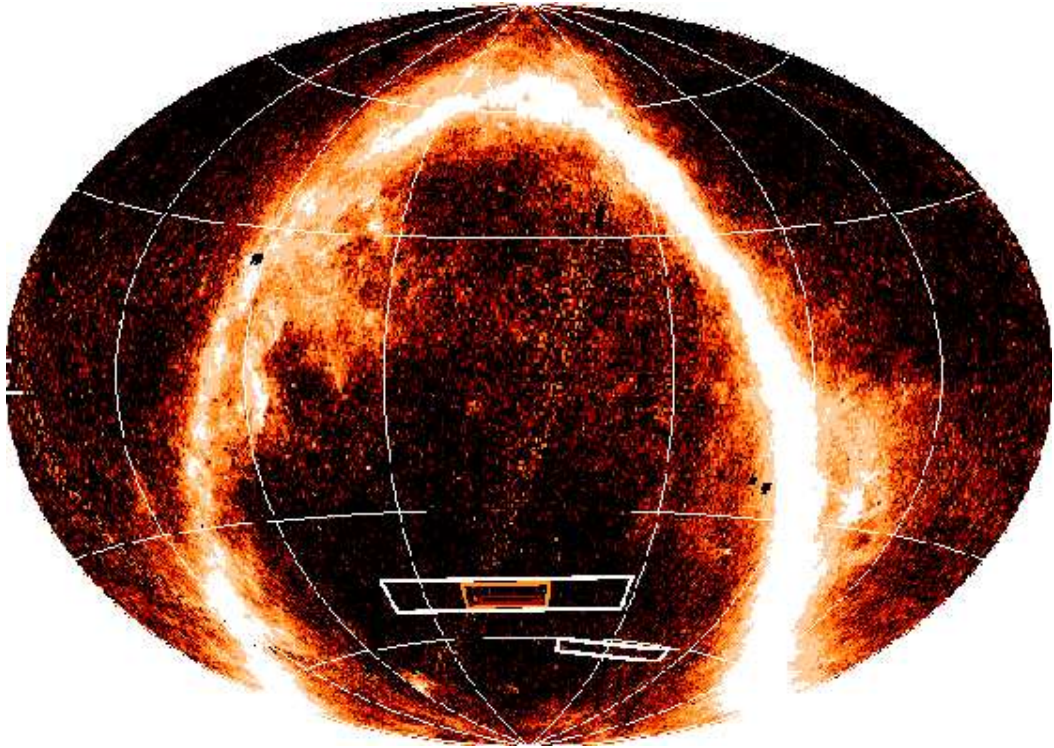}}
\ssp
\caption[IRAS 100 $\mu m$ view of the galaxy with the Python observing regions superimposed.]{IRAS 100 $\mu m$ view of the galaxy with the Python observing regions superimposed. The 2 large white boxes indicate the PyV regions, the medium-sized orange box PyIII, and the small red box PyIV. This demonstrates the large sky coverage achievable from the South Pole in one of the most clean regions of sky.}
\label{fig:iras}
\end{figure}

Two regions of sky were observed:  the PyV main
field, a $7.5^{\circ} \times 67.7^{\circ}$ region of sky centered at
$\alpha=23.18^{h}$, $\delta=-48.58^{\circ}$ (J2000) which includes 
fields measured during the previous four seasons of Python observations and
a $3.0^{\circ} \times 30.0^{\circ}$ region of sky centered
at $\alpha=3.00^{h}$,
$\delta=-62.01^{\circ}$ (J2000), which encompasses the region observed
with the ACME telescope (Gundersen et al. 1995).
The total sky coverage for the PyV regions is 598 deg$^2$,
greater sky coverage than previous degree-scale CMB experiments.
Figure \ref{fig:iras} shows the
IRAS 100 $\mu m$ view of the galaxy with the
Python observing regions superimposed.

\subsection{Observing Strategy}

\begin{figure}[h!]
\centerline{\epsfxsize=13.9cm \epsfbox{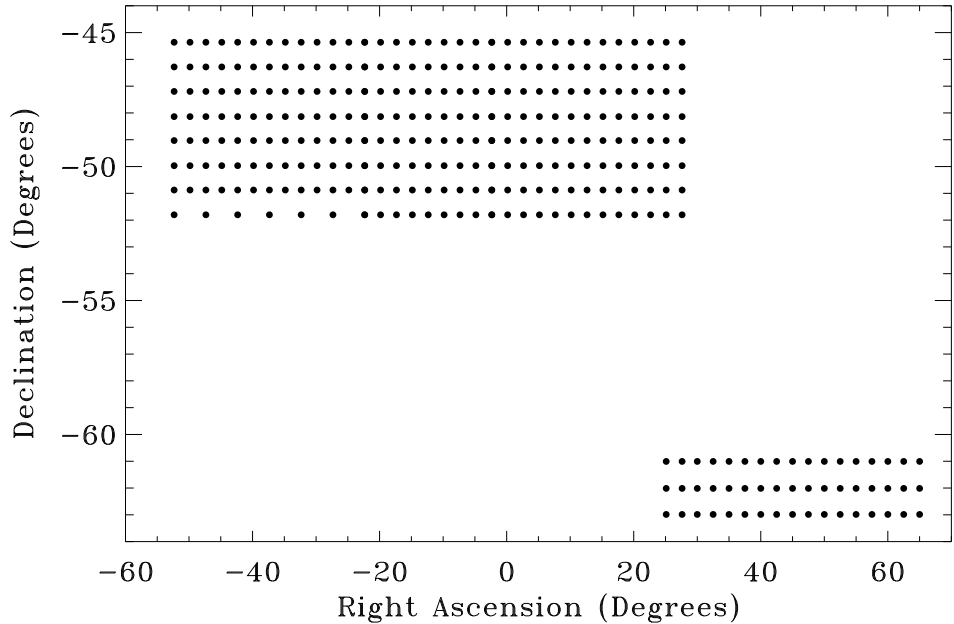}}
\ssp
\caption[Center positions of the Python V fields.]{Center positions of the Python V fields. The fields are 2.5$^{\circ}$ apart in right ascension, which corresponds to 1.6$^{\circ}$ on the sky at a declination of $-$50$^{\circ}$. The fields are
0.92$^{\circ}$ apart in declination.}
\label{fig:py_strategy}
\end{figure}

\begin{figure}[h!]
\centerline{\epsfxsize=11.0cm \epsfbox{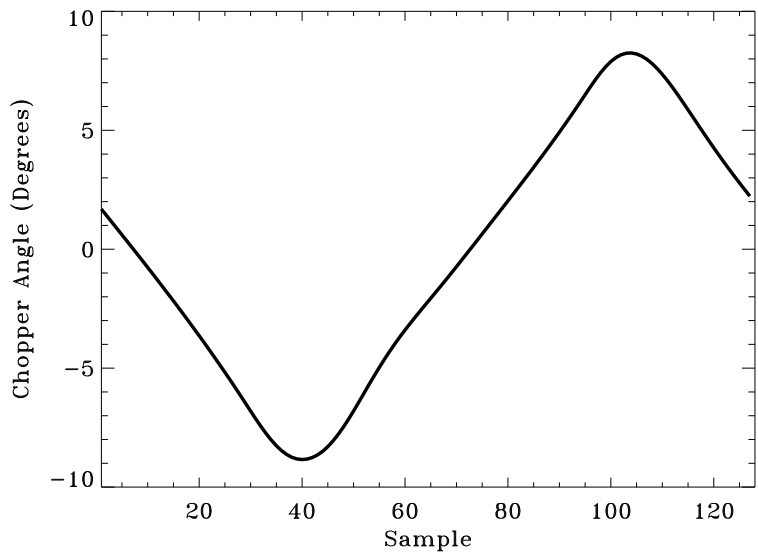}}
\ssp
\caption[Motion of the chopper.]{Motion of the chopper. The chopper
smoothly scans in azimuth in a nearly triangular wave pattern at 5.1 Hz.
The waveform is not perfectly triangular because the chopper slows
down when turning around and because the LVDT is non-linear.}
\label{fig:chop_motion}
\end{figure}

Both PyV regions are fully sampled 
with a grid spacing of 0.92$^{\circ}$ in elevation
and 2.5$^{\circ}$ in right ascension, corresponding to a distance
of 1.6$^{\circ}$ on the sky at a declination of $-50^{\circ}$.
A grid of the center positions of the PyV fields is shown in Figure
\ref{fig:py_strategy}.
The telescope is positioned on one of the fields
and the chopper smoothly scans the beams
in azimuth in a nearly triangular wave
pattern (Figure \ref{fig:chop_motion}). The chopper throw is
17$^{\circ}$ in azimuth, corresponding to 11$^{\circ}$ on the sky
at a declination of $-50^{\circ}$.
The telescope tracks on each field; it does not drift scan.

There are 128 data samples for each detector channel
in a complete chopper cycle, and 164 chopper
cycles ($\sim$ 32 s) of data (one stare)
are taken of a given field before the telescope
is positioned on the next field in the set.
One data file consists of 164 chopper cycles for each field in the set.
For a set which contains 13 fields, one data
file is $\sim$ 32 s $\times$ 13 $\sim$ 6.9 min long.

\begin{figure}[h!]
\centerline{\epsfxsize=13.9cm \epsfbox{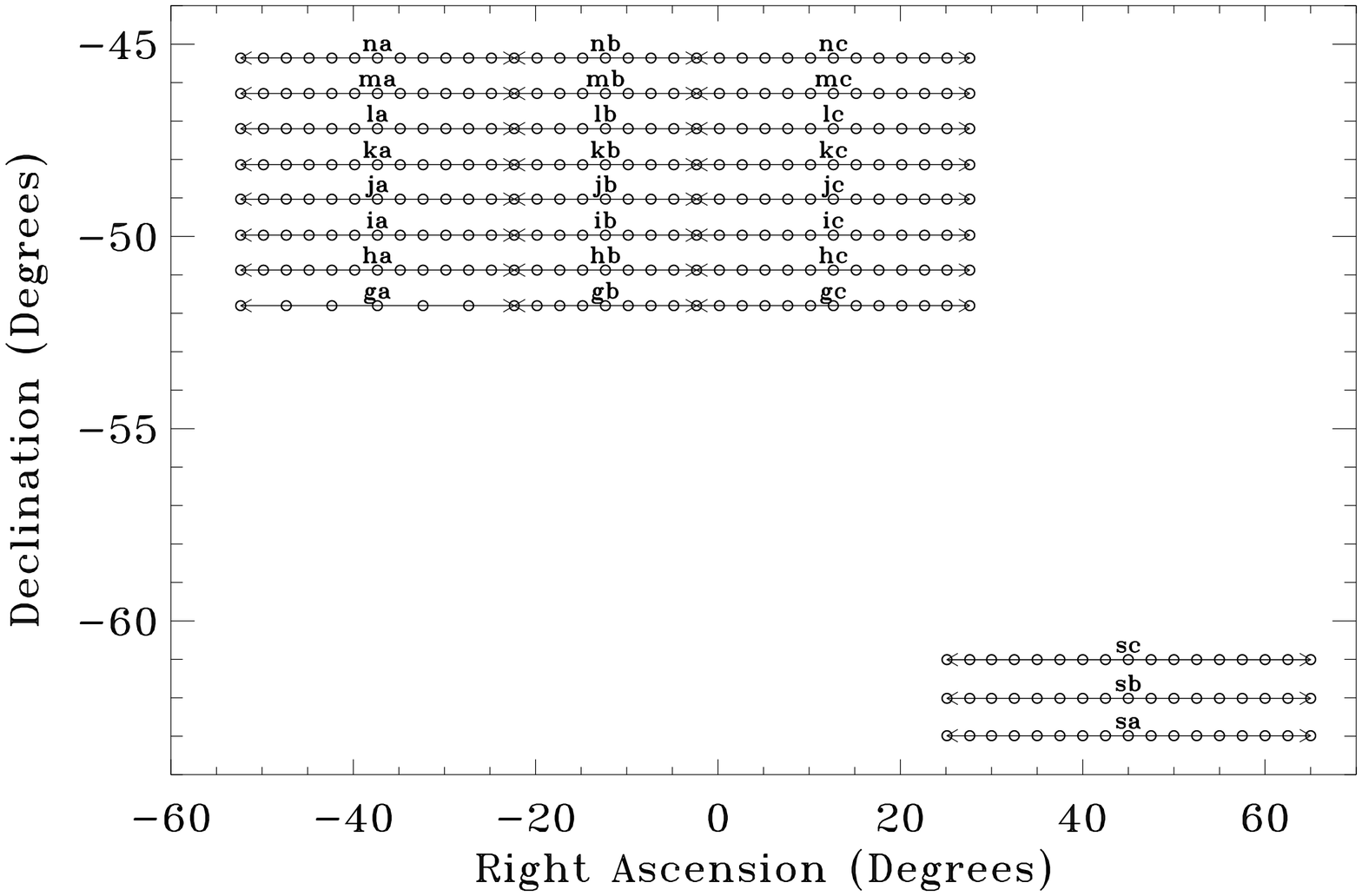}}
\ssp
\caption[Observing sets.]{Observing sets. Each set is observed for
approximately
a day (leaving $\sim$ 13 hours of good data) before moving on to the next
set. Neighboring sets in the main PyV region overlap
by 1 field. Circles represent the fields and arrows point to the
end fields in each set.
In addition to the sets shown, sets ib, jb, kb, and lb were
observed with a scan pattern of 5 fields per file.}
\label{fig:py_strategy_detail}
\end{figure}

Approximately 13 hours of good data (100 files) are taken of a set
of fields before the telescope moves on to the next set of fields.
A total of 309 fields are observed in 31 sets of 5--17 fields.
Some fields are observed in more than one set, yielding 345 effective fields.
Figure \ref{fig:py_strategy_detail} labels the observing sets with
respect to the grid shown in Figure \ref{fig:py_strategy} and
Table \ref{tbl:sets} lists the number of fields in each set,
the number of files of good data taken in each set, and the declination
at which each set was taken.

\begin{table*}
\begin{center}
\begin{tabular}{|cccc|}
\tableline
Set & $N_{\rm fields}$ &  $N_{\rm files}$ & $\delta$\\
\tableline
sa & 17 & 41 & -62.99\\
sb & 17 & 103 & -62.01\\
sc & 17 & 33 & -61.01\\
ga & 7 & 302 & -51.80\\
gb & 9 & 184 & -51.80\\
gc & 13 & 95 & -51.80\\
ha & 13 & 100 & -50.88\\
hb & 9 & 84 & -50.88\\
hc & 13 & 104 & -50.88\\
ia & 13 & 45 & -49.97\\
ib & 9 & 130 & -49.97\\
ib & 5 & 52 & -49.97\\
ic & 13 & 119 & -49.97\\
ja & 13 & 120 & -49.03\\
jb & 9 & 269 & -49.03\\
jb & 5 & 250 & -49.03\\
jc & 13 & 109 & -49.03\\
ka & 13 & 81 & -48.14\\
kb & 9 & 186 & -48.14\\
kb & 5 & 295 & -48.14\\
kc & 13 & 59 & -48.14\\
la & 13 & 191 & -47.20\\
lb & 9 & 298 & -47.20\\
lb & 5 & 343 & -47.20\\
lc & 13 & 122 & -47.20\\
ma & 13 & 113 & -46.28\\
mb & 9 & 198 & -46.28\\
mc & 13 & 124 & -46.28\\
na & 13 & 153 & -45.36\\
nb & 9 & 50 & -45.36\\
nc & 13 & 107 & -45.36\\
\tableline
\end{tabular}
\end{center}
\caption{Observing sets.}
\label{tbl:sets}
\end{table*}

\subsection{Pointing}

\begin{figure}[b!]
\centerline{\epsfxsize=11.0cm \epsfbox{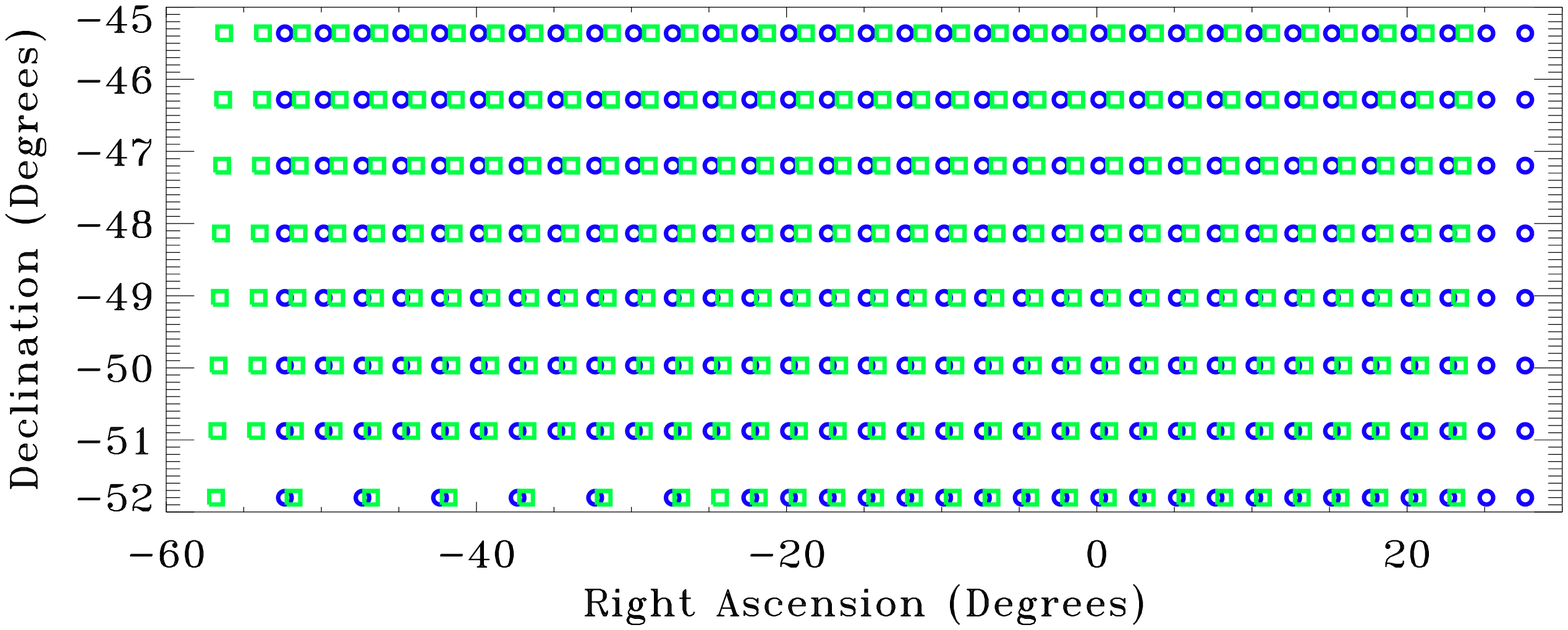}}
\ssp
\caption[Pointing of the two feeds on the sky.]{Pointing of the two
feeds on the sky in the main PyV region. The beams corresponding to the two feeds
observe the same elevation and are separated by 2.80$^{\circ}$ on the sky.
The squares represent the positions of channels 1 and 2 and the circles represent
the positions of channels 4 and 5.}
\label{fig:py_strategy_2ch}
\end{figure}

The combined absolute and
relative pointing uncertainty is estimated to be $0.15^{\circ}$, 
as determined by measurements of the
moon and the Carinae nebula ($\alpha=10.73h, \delta=-59.65 ^{\circ}$).
The beams corresponding to the 2 feeds observe the same elevation and
are separated by 2.80$^{\circ}$ on the sky (Figure \ref{fig:py_strategy_2ch}).

\subsection{Beam}

The beam is determined from scans of the Carinae nebula and the Moon.
Scans of the Carinae nebula were obtained from the Viper telescope
(Griffin and Peterson 1998), which has the same receiver system
as PyV and a 0.25$^{\circ}$ beam. The Viper scans were convolved
with asymmetric Gaussian beams of widths $\sigma_x$ and $\sigma_y$.
The widths were varied as free parameters and best fits to
the observed PyV beams were obtained.
This convolution procedure works better than a deconvolution procedure
because of the noise in the measurements. The results are not dependent on
size of the Carina nebula assumed in the fits.
The PyV beam is well approximated by an asymmetric Gaussian
of FWHM $0.91^{+0.03}_{-0.01} \times 1.02^{+0.03}_{-0.01}$
degrees ($az \times el$).

\chapter{Data Reduction}

%

\section{Data Cuts}

A total 719.52 hours of data were taken during the PyV season.
After cutting 45.6\% of the data for file errors, tracking errors and weather,
391.32 hours of data remained for use in the CMB analysis.
Table \ref{tbl:data_cuts} summarizes the data cuts.

A file is cut:
\begin{enumerate}
\item [$\bullet$] for file errors if any of the data, pointing, or time
matrices are empty, if the time matrices from the two data computers
did not agree, or if the number of stares in the data, pointing and time
matrices did not agree.
\item [$\bullet$] for a relative pointing error
if the stare that is furthest from where it should be relative to
the scan center is more than 3$\arcmin$ from where it should be.
\item [$\bullet$] for an absolute pointing error if
its scan center deviates by more than 6$\arcmin$
from the scan center of all other files in the set.
\item [$\bullet$] if the mean chopper waveform in any stare deviates
by more than 1\% from the benchmark waveform
made from the whole data set.
\item [$\bullet$] for anomalously low noise levels, which could be
caused by the power failing and the dewar heating up or the
gain being set to the wrong level.
\item [$\bullet$] for anomalously high noise levels, which are
caused by weather. There is a clear distinction between good and bad weather.
\end{enumerate}

\begin{table*}
\begin{center}
\begin{tabular}{|lcc|}
\tableline
Cut & hours & percent\\
\tableline
internal file errors & 10.42 & 1.45\\
relative pointing & 49.39 & 6.86 \\
absolute pointing & 73.43 & 10.21 \\
chopper error & 7.47 & 1.04 \\
low levels & 24.48 & 3.40 \\
high levels & 163.01 & 22.65 \\
${\bf TOTAL CUTS}$ & ${\bf 328.20}$ & ${\bf 45.61}$ \\
\tableline
\end{tabular}
\end{center}
\caption{Data cuts.}
\label{tbl:data_cuts}
\end{table*}

\section{Modulations}

\begin{figure}[h!]
\centerline{\epsfxsize=13.9cm \epsfbox{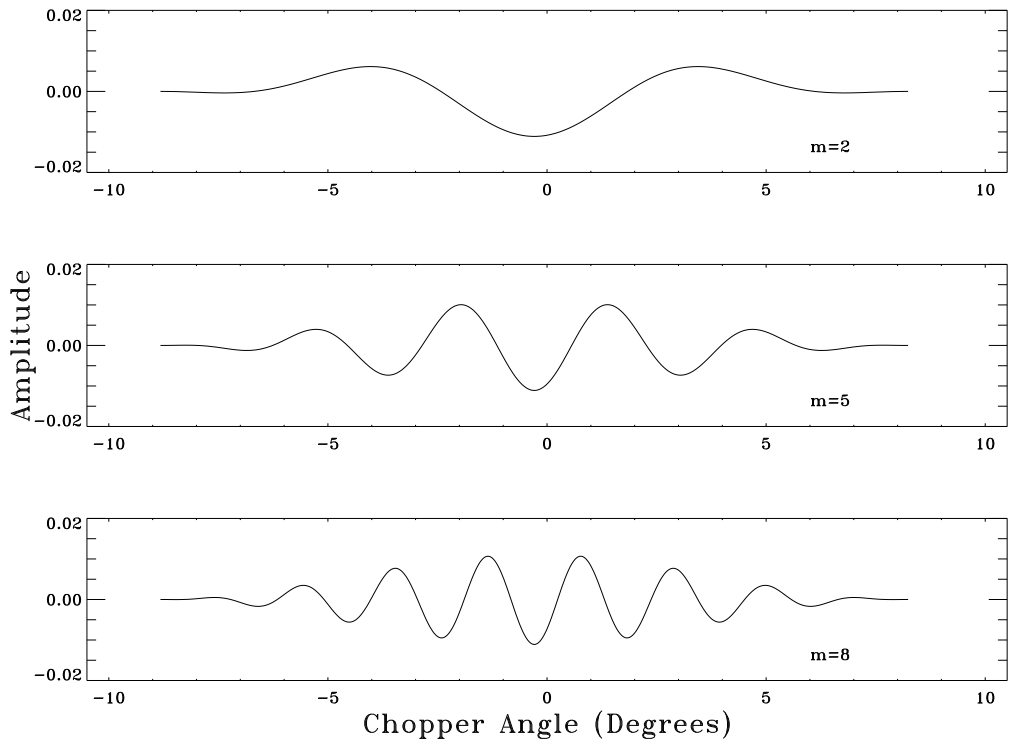}}
\ssp
\caption[The data are modulated with cosines which have been apodized by a Hann window. Three of the modulations are shown.]{The data are modulated with cosines which have been apodized by a Hann window. Three of the modulations are shown.}
\label{fig:mods}
\end{figure}

The data are modulated using
\begin{equation}
M_{m}(\theta)=cos(m \pi \theta/\theta_c) \times
\left\{
\begin{array}{ll}
1 & m=1\\
H(\theta) & m=2 \ldots 8
\end{array}
\right.
\label{eq:modeqn}
\end{equation}
where $m$ is the modulation number, $\theta$ is the
chopper angle, $\theta_c$ is
the extent of the chopper throw and
$H(\theta)~=~0.5(1~-~cos(2 \pi \theta/\theta_c))$ is a Hann window
(Figure \ref{fig:mods}).
The $m=2 \ldots 8$ modulations are apodized with the Hann window 
in order to reduce the ringing of the window functions in $l$-space.
Data taken during the right- and left-going portions of the chopper
cycle are modulated separately, to allow for cross-checks of the data.
Sine modulations are not used in the analysis because they are anti-symmetric
and are thus sensitive to gradients on the sky.

\begin{figure}[h!]
\centerline{\epsfxsize=13.9cm \epsfbox{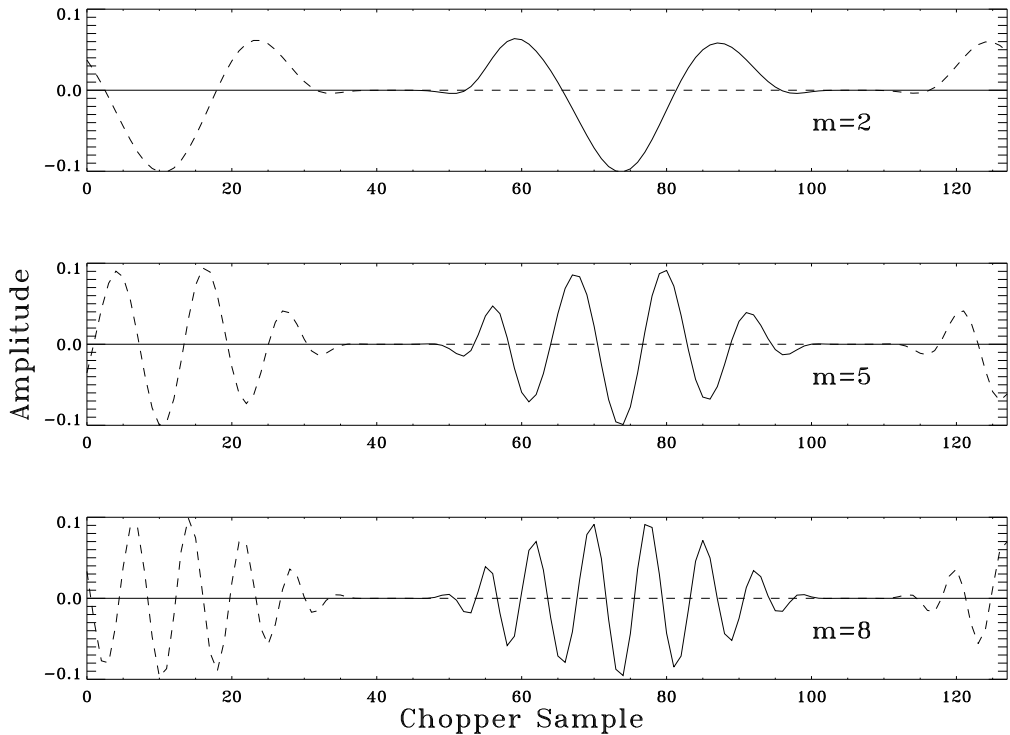}}
\ssp
\caption[Timestream modulations.]{Timestream modulations. Since the chopper waveform is non-linear, the timestream modulations are created so that the spatial response will be cosines or apodized cosines. These 3 timestream modulations correspond to the 3 spatial modulations shown in Figure \ref{fig:mods}.}
\label{fig:masks}
\end{figure}

The modulations in equation \ref{eq:modeqn} and Figure \ref{fig:mods}
are cosines or apodized cosines in space, not in time. Since the chopper
waveform (Figure \ref{fig:chop_motion}) is non-linear, the timestream
modulations are not strictly cosines or apodized cosines in time (or chopper
sample). Rather, they are created to produce the spatial response in
equation \ref{eq:modeqn} and Figure \ref{fig:mods}. The timestream
modulations as a function of chopper sample are shown in Figure
\ref{fig:masks}.
The timestream modulations, $M_m(s)$ are normalized such
that $\sum_s M = 0$ (so that a uniform temperature does not contribute)
and $\sum_s |M| = 2$ where $s$ is chopper sample,
following the single difference convention.

\section{Instrument Transfer Function}

\begin{figure}[h!]
\centerline{\epsfxsize=11.0cm \epsfbox{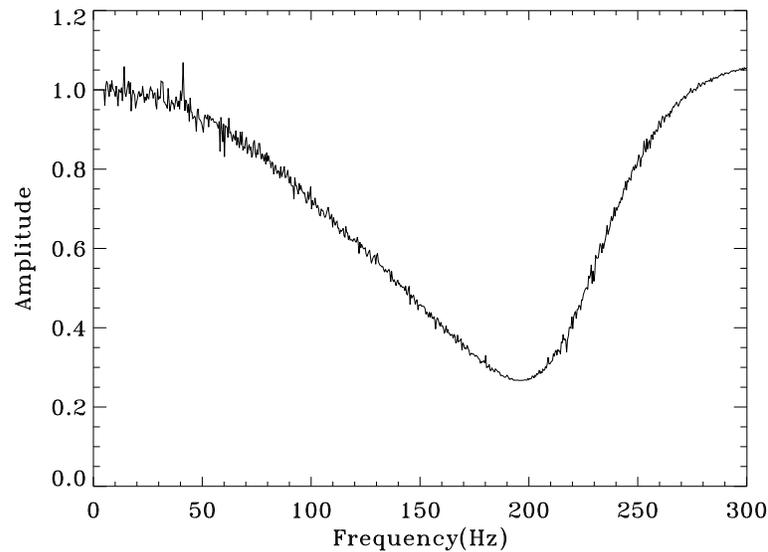}}
\ssp
\caption[Frequency dependence of the instrument transfer function.]{Frequency dependence of the instrument transfer function.}
\label{fig:tf}
\end{figure}

The instrument transfer function produces two main effects on the data:
a roll-off with frequency and a phase shift between data taken
during the right and left-going portions of the chopper cycle
(hereafter right and left data). Figure \ref{fig:tf} shows the
frequency dependence of the transfer function. 
Since each modulation
is well-isolated in frequency space, the effect of the frequency
roll-off of the transfer function can be accounted for with 1 number
for each modulation (Table \ref{tbl:tf}). The modulated data are
divided by $T$ to remove the frequency dependence of the transfer function.
\begin{table*}[h!]
\begin{center}
\begin{tabular}{|ccc|}
\tableline
mode & $f$ (Hz) & $T$\\
\tableline
1	&	13	&	1.0\\
2	&	22	&	.99\\
3	&	33	&	.98\\
4	&	43	&	.95\\
5	&	55	&	.92\\
6	&	66	&	.88\\
7	&	77	&	.83\\
8	&	88	&	.79\\
\tableline
\end{tabular}
\end{center}
\ssp
\caption[Instrument transfer function vs. mode.]{Instrument transfer function vs. mode. Frequency $f$ is the peak frequency of each modulation.}
\label{tbl:tf}
\end{table*}

\begin{figure}[h!]
\centerline{\epsfxsize=11.0cm \epsfbox{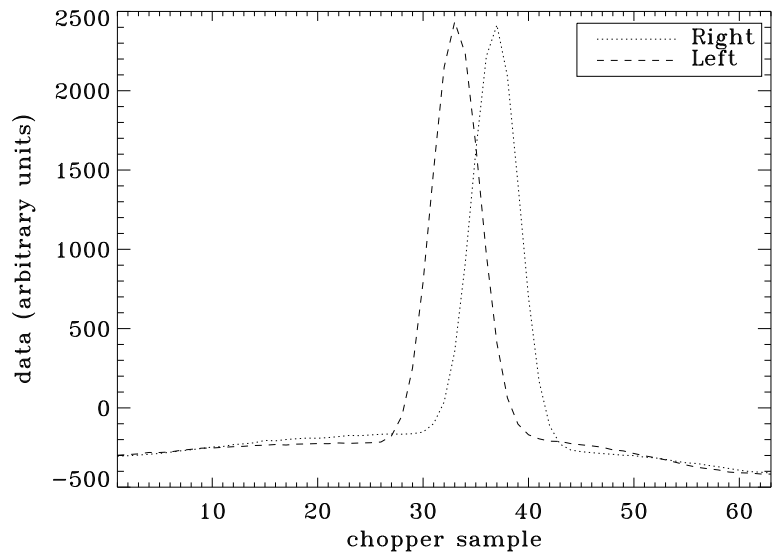}}
\ssp
\caption[Unmodulated moon data showing effect of the transfer function
on the right-left phase.]{Unmodulated moon data showing effect of the
transfer function on the right-left phase. Data from one stare and channel is shown as an example.}
\label{fig:mphase}
\end{figure}

\begin{figure}[h!]
\centerline{\epsfxsize=11.0cm \epsfbox{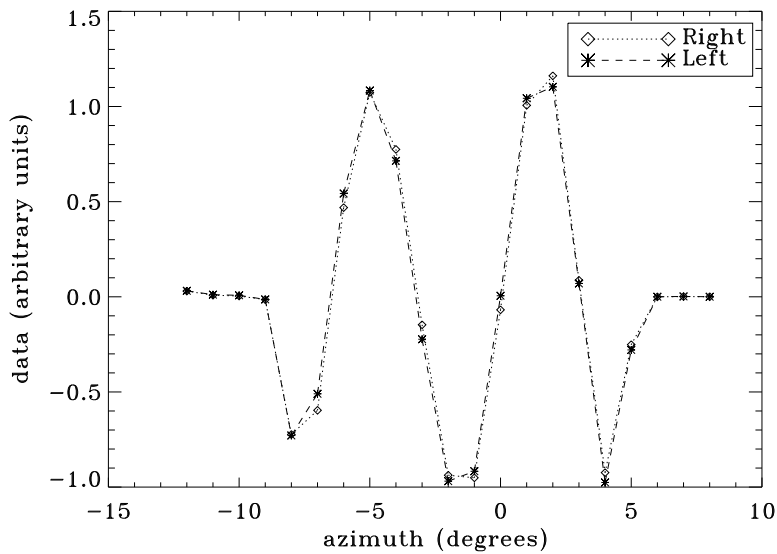}}
\ssp
\caption[Checking the phase of right and left data on the moon.]{Checking the phase of right and left data on the moon. Data from channel 1, modulation 3 is shown as an example. If the right and left data are not properly phased, the signal will be washed out.}
\label{fig:moonmods_RL}
\end{figure}

Figure \ref{fig:mphase} shows the phase effect of the transfer function
on right and left unmodulated moon data and Figure \ref{fig:moonmods_RL}
shows modulated moon data, in which right and left data are properly phased.
If the right and left data are not properly phased, the signal will
be washed out.

\section{Chopper Synchronous Offset}

\begin{figure}[h!]
\centerline{\epsfxsize=10.0cm \epsfbox{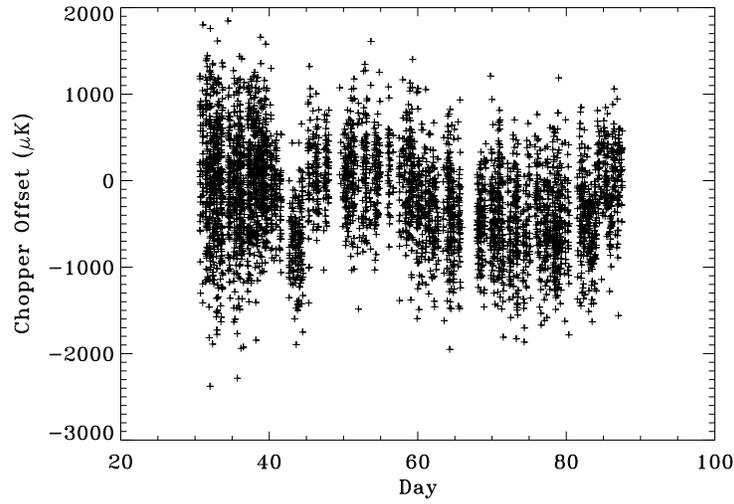}}
\ssp
\caption[Chopper offset vs. time.]{Chopper offset vs. time for channel 1, modulation 2, right-going data. Each point corresponds to the offset removed from one file.}
\label{fig:plot_chop_offset}
\end{figure}

\begin{figure}[b!]
\centerline{\epsfxsize=10.0cm \epsfbox{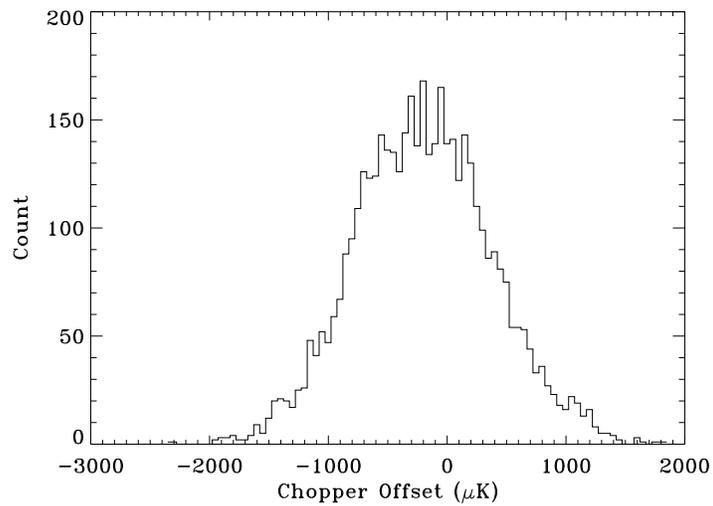}}
\ssp
\caption[Histogram of chopper offset.]{Histogram of chopper offset for channel 1, modulation 2, right-going data.}
\label{fig:hist_chop_offset}
\end{figure}

Data in a given file, field, channel, and modulation are co-added
over all chopper cycles.
A chopper synchronous offset, caused by differential
spillover past the chopper, is removed from each data file
by subtracting the average of all of the fields
in a file. This is not just a DC offset; there is an offset removed
for each modulation (or equivalently, for each sample of the chopper
waveform) and channel. Figure \ref{fig:plot_chop_offset} shows
the chopper offset as a function of time for one of the channels
and modulations as an example. Figure \ref{fig:hist_chop_offset}
shows a histogram of the chopper offset for that same channel
and modulation. The large scatter in both figures is due to noise.


\begin{figure}[t!]
\centerline{\epsfxsize=10.0cm \epsfbox{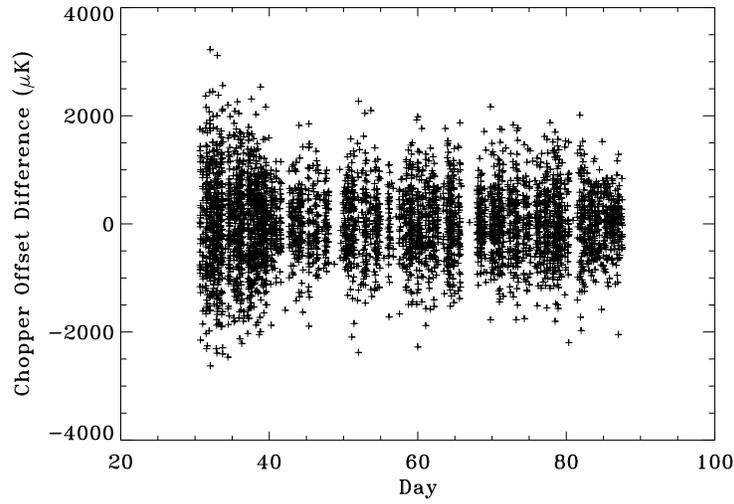}}
\ssp
\caption[Chopper offset difference vs. time.]{Chopper offset difference vs. time for channel 1, modulation 2, right-going data. Each point corresponds to the offset difference between adjacent files in time.}
\label{fig:coffdiff}
\end{figure}

\begin{figure}[b!]
\centerline{\epsfxsize=10.0cm \epsfbox{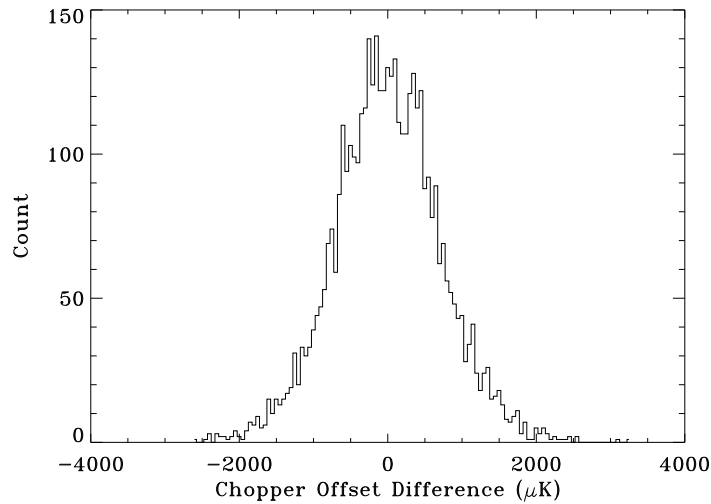}}
\ssp
\caption[Histogram of chopper offset difference.]{Histogram of chopper offset difference for channel 1, modulation 2, right-going data.}
\label{fig:coffdiff_hist}
\end{figure}

To test that the chopper offset is stable over a file, we take the
difference in chopper offset between two adjacent
files in time. A plot of the chopper offset difference is
shown in Figure \ref{fig:coffdiff} and a histogram is
shown in Figure \ref{fig:coffdiff_hist}. Again, there is a large
scatter due to noise.


\begin{figure}[h!]
\centerline{\epsfxsize=11.0cm \epsfbox{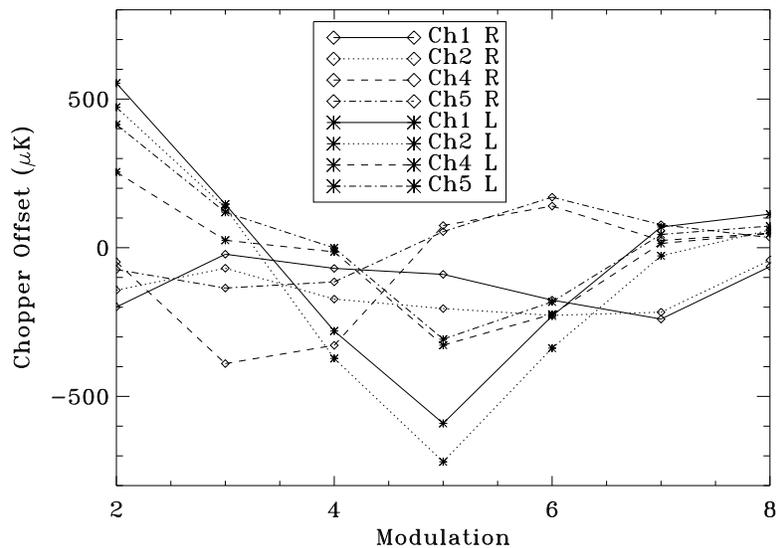}}
\ssp
\caption[Mean chopper offset over the data set for all channels, right and left-going data vs. modulation.]{Mean chopper offset over the data set for all channels, right and left-going data vs. modulation. A different chopper offset is subtracted from each file.}
\label{fig:coff_all}
\end{figure}

\begin{figure}[h!]
\centerline{\epsfxsize=11.0cm \epsfbox{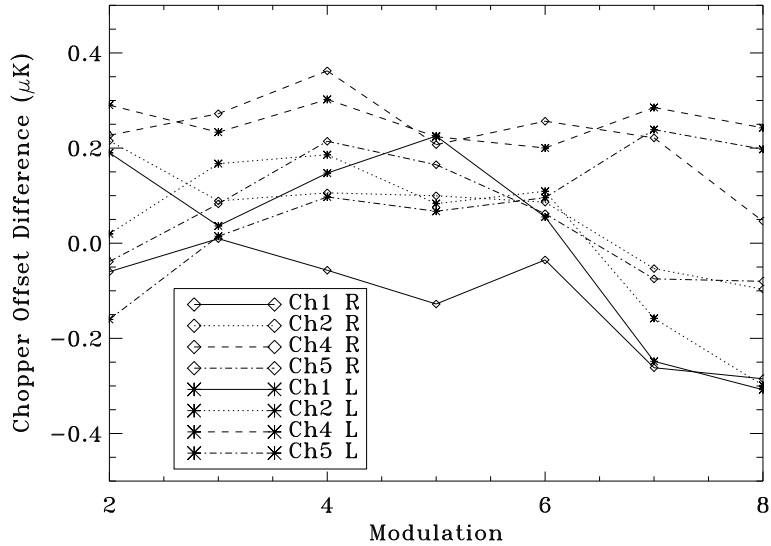}}
\ssp
\caption[Mean chopper offset difference over the data set for all channels, right and left-going data vs. modulation.]{Mean chopper offset difference over the data set for all channels, right and left-going data vs. modulation. The chopper offset difference is significantly smaller that the chopper offset (Figure \ref{fig:coff_all}), indicating that the chopper offset is stable over a file. The vertical scale in this figure covers a smaller range than that of Figure \ref{fig:coff_all}.}
\label{fig:coff_diff_all}
\end{figure}

In order to compare the size of the chopper offset and the chopper offset
difference with the scatter due to noise reduced, we compute
the mean chopper offset and mean chopper offset difference
for each channel and modulation
over the whole data set.
As illustrated in Figures \ref{fig:coff_all} and \ref{fig:coff_diff_all}, the
chopper offset difference is significantly smaller (by at least 3 orders
of magnitude) than the chopper offset.
Further, when the differencing procedure is repeated
as a function of file lag, the results indicate that the chopper
offset is stable over $\sim$ 5 files. Thus our assumption that
the chopper offset is stable over 1 file is a correct and
even a conservative assumption.



\section{Ground Shield Offset}

When the modulated data are binned in azimuth, a periodic signal due
to the 12 panels of the ground shield is evident, especially
on larger angular scales. The signal of period 30$^{\circ}$ is fit for
an amplitude and is subtracted. Removal of the ground shield offset
has less than 4\% effect on the final angular power spectrum
in all modulations, because when the data are binned in RA, the effect
averages out.

\begin{figure}[h!]
\centerline{\epsfxsize=13.9cm \epsfbox{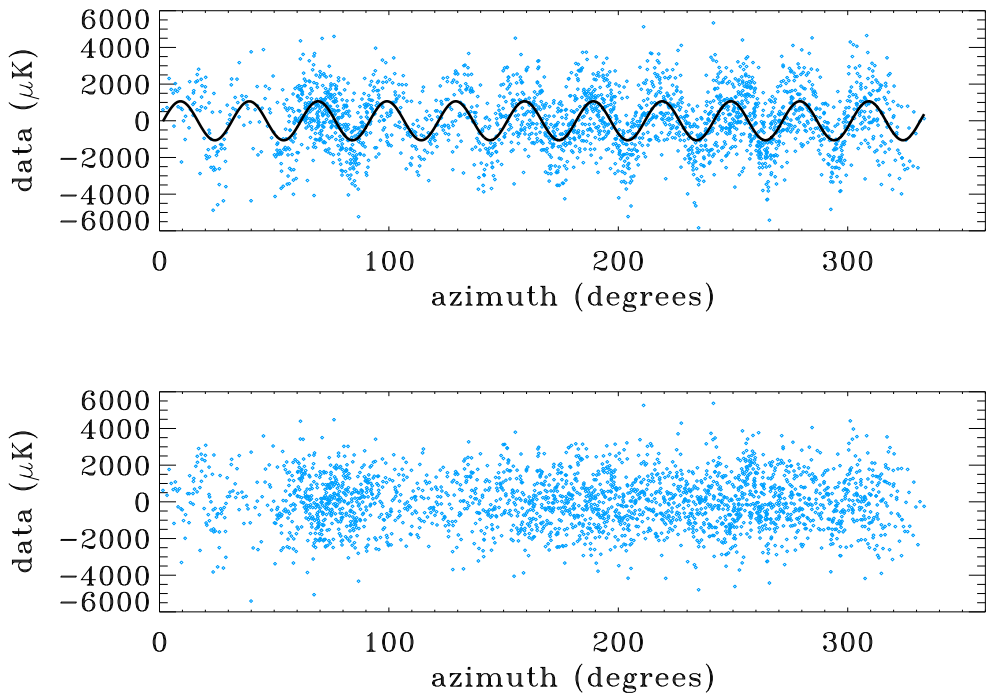}}
\ssp
\caption[Azimuthally binned data. The signal caused by the ground shield is evident.]{Top-- Azimuthally binned data for channel 1, left-going chopper
data, modulation 1, set jb9. There are 2421 data points
(269 files $\times$ 9 fields) plotted.
The periodic signal caused by the 12 panels
of the ground shield is evident and the best fit curve is plotted
as a solid line. Bottom-- the same data with the ground shield
offset subtracted.}
\label{fig:gs_indiv}
\end{figure}

\begin{figure}[h!]
\centerline{\epsfxsize=11.0cm \epsfbox{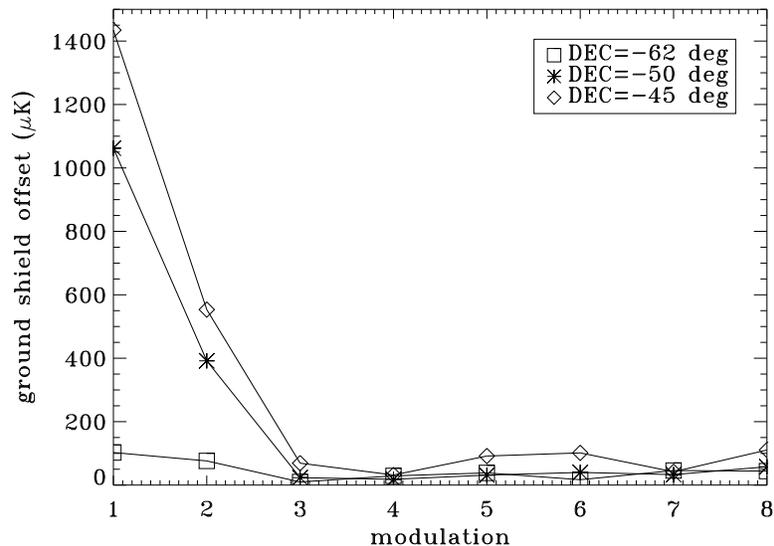}}
\ssp
\caption[Amplitude of the ground shield offset as a function of declination and
modulation.]{Amplitude of the ground shield offset as a function of declination and modulation for channel 1, left-going chopper cycles, file sets sb, jb, and nb. The offset is larger for data taken at a lower elevation, i.e. when the telescope is pointed closer to the ground shield.}
\label{fig:gs_ampl}
\end{figure}

The ground shield offset is computed from all
fields and files in an observing set.
A different ground shield offset is fit and removed for each channel,
modulation, and set.
Figure \ref{fig:gs_indiv} shows the azimuthally binned data for one set,
the fit to the ground shield synchronous signal, and the same data with the
ground shield offset removed. 
The ground shield offset is larger for data taken when the telescope
is pointed closer to the ground shield and for modulations which
probe larger angular scales, as illustrated in Figure \ref{fig:gs_ampl}.

\newpage

\section{Co-adding}

After the data have been modulated and offsets removed, the right and
left-going data, which have been properly phased, are co-added, as are
data from channels which observe the same points on the sky. The bandpasses
are similar enough in frequency range that the data can be co-added
without loss of information.
Hereafter channel 12 will refer to the co-added channel 1 and 2 data
and channel 45 will refer to the co-added channel 4 and 5 data.
Since the two feeds observe different points on the sky they
cannot be
\linebreak
\\
\noindent co-added; the theoretical and noise covariances between them
are included in the likelihood analysis of the angular power spectrum.

\section{60 Hz}

The South Pole Station power nominally operates at a
frequency of 60 Hz, so there is
possible contamination at 60 Hz and its harmonics. However,
the chopper runs at 5.1 Hz, which is incommensurate with 60 Hz.
To verify that a 60 Hz signal would not contaminate the data,
we created a 60 Hz signal for one stare of data, modulated it,
averaged over 164 cycles and found that the resulting signal is negligible
for all modulations.   
\chapter{Data Analysis Techniques}

\section{Bayesian Likelihood Analysis}

In order to estimate parameters from data, a Bayesian
likelihood analysis (see for example Readhead et al. 1989) is used.
What we want to report from any experiment
is the probability of the theory being true given the data,
$P(theory|data)$. What we can actually measure is the likelihood,
${\cal L} = P(data|theory)$. Applying Bayes' theorem,
\begin{equation}
P(theory|data) \propto P(data|theory)P(theory).
\label{eq:likedef}
\end{equation}
The prior probability, $P(theory)$, is conservatively
taken to be uniform (un-informative).
This technique assumes the data are
drawn from a Gaussian distribution on the sky.

The likelihood (${\cal L}$) can be written:
\begin{equation}
{\cal{L}}={(2\pi)^{-N/2}}
{det({\bf C})}^{-1/2}exp{(-\chi^{2}/2)}
\label{eq:like}
\end{equation}
where $\chi^{2} = \vec d^t {\bf C^{-1}} \vec d $,
${\bf C}={\bf C^{T}+C^{N}+C^{C}}$,
${\vec d}$ is the data,  ${\bf C^T}$ is the theory
covariance matrix, ${\bf C^N}$ is the noise covariance
matrix, and ${\bf C^C}$ is
the constraint matrix for offset subtraction.
${\bf C^C}$ for PyV is discussed in section 4.4 and ${\bf C^N}$
for PyV is discussed in sections 5.2 and 6.1.

\begin{figure}[t!]
\centerline{\epsfxsize=11.0cm \epsfbox{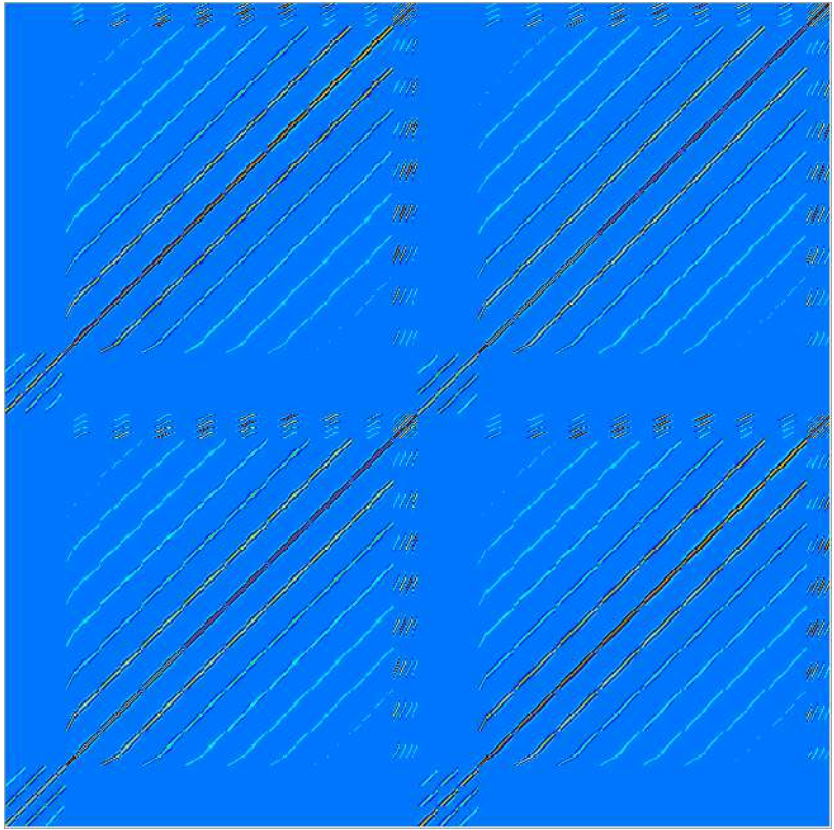}}
\ssp
\caption[Theory covariance matrix, ${\bf C^T}$, for PyV modulation 2.]{Theory covariance matrix, ${\bf C^T}$, for PyV modulation 2. $C^T(1,1)$ is in the bottom left. The matrix is 690 x 690 for the 345 fields and 2 channels in the reduced, co-added data. Points 1:345 correspond to channel 45 and points 346:690 correspond to channel 12. There are significant theory correlations between channels because they are separated by $2.80^{\circ}$ on the sky.}
\label{fig:ct_im}
\end{figure}

The theory covariance matrix, ${\bf C^T}$,
depends on the model CMB angular power spectrum $C_l$
and the experimental window functions $W_{lij}$.
The elements of ${\bf C^T}$ are given by
\begin{equation}
C_{ij}^{T} = \sum_{l} {(2l+1) \over 4\pi} C_{l} W_{lij}.
\label{eq:cth}
\end{equation}
A sample theory covariance matrix is shown in Figure \ref{fig:ct_im}.

\section{Window Functions}
The window functions are a measure of experimental sensitivity as a function
of angular scale $l$. They are generated
from the experimental beam map, modulations, and observing strategy.

\subsection{A Method For Calculating Window Functions}

\begin{figure}[t!]
\centerline{\epsfxsize=13.9cm \epsfbox{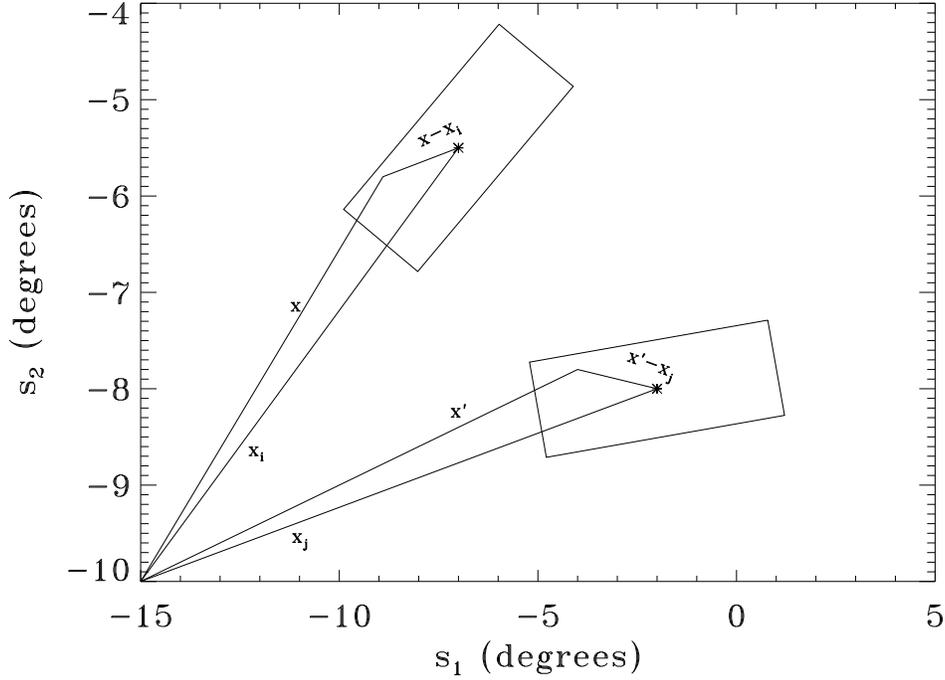}}
\ssp
\caption[Position of beam maps for computing $W_{lij}$.]{Position of beam maps for computing $W_{lij}$. $s_1$ and $s_2$ are flat space coordinates.}
\label{fig:wlij_beam_pos}
\end{figure}

The following is a method for calculating the window
functions for a general beam map and observing strategy in the flat sky
approximation. Section 4.2.3 discusses the importance of
using the exact beam map and observing strategy for
certain experiments. The flat sky approximation is valid
for many current and past CMB experiments.
Another discussion of window functions
can be found in White and Srednicki (1995).
Figure \ref{fig:wlij_beam_pos} illustrates the geometry of
the beam map and observing strategy used in the derivation below.

\begin{enumerate}
\item [$\bullet$]The (modulated) beam map is B, which is 2-dimensional.
$B_{m}(\vec x - \vec x_{i})$ is the beam map
for modulation $m$ corresponding to an observation at pixel $i$.

\item [$\bullet$]$\theta_i$ is the twist of the beam map for pixel $i$ with
respect to the $s_1$-axis in Figure \ref{fig:wlij_beam_pos}.

\item [$\bullet$]The rotation matrix acting
on $(\vec x - \vec x_{i})$ is ${\bf R_i}=\left( 
\begin{array}{ll}
cos(\theta_i) & -sin(\theta_i)\\
sin(\theta_i) & cos(\theta_i)
\end{array}
\right ) $.

\item[$\bullet$]The measured anisotropy at a point is a convolution of the
true anisotropy with the beam map:
$\Delta T_{i}=
\int B_m({\bf R_{i}}(\vec x - \vec x_{i})) \Delta T(\vec x) d\vec x $.

\item[$\bullet$]The underlying temperature field $\Delta T(\vec x)$
can be expanded in spherical harmonics:
$\Delta T(\vec x)=\Sigma_{lm}a_{lm}Y_{lm}(\theta,\phi)$,
and the multipole moments, $C_{l}$, are
$C_{l}=<|a_{lm}|^{2}>$.

\item[$\bullet$] The elements of the theory covariance
matrix are $C^T_{ij} = \langle \Delta T_{i} \Delta T_{j} \rangle $.

\end{enumerate}

Using the above definitions, the elements of the theory
covariance matrix are:
\begin{equation}
C^T_{ij} = \int \int  B_m({\bf R_i}(\vec x - \vec x_{i}))
B_{m^{\prime}}({\bf R_j}(\vec x^{\prime} - \vec x_{j}))
\langle \Delta T(\vec x) \Delta T (\vec x^{\prime}) \rangle
d\vec x d\vec x^{\prime}.
\label{eq:cij1}
\end{equation}
Assuming rotational symmetry and averaging over the whole sky,
\begin{equation}
\langle \Delta T(\vec x) \Delta T (\vec x^{\prime}) \rangle
=\sum { {(2l+1) \over 4\pi} C_{l} P_{l}(cos(q))},
\label{eq:legendre}
\end{equation}
where $P_l$ are the Legendre polynomials associated with the
spherical harmonic expansion, $q = | \vec q |$, and
\begin{equation}
\vec q = \vec x - \vec x^{\prime}.
\label{eq:qdef}
\end{equation}
Defining dummy variables,
\begin{equation}
\begin{array}{ll}
\vec y = & {\bf R_i}(\vec x - \vec x_{i})\\
\vec y^{\prime} = & {\bf R_j}(\vec x^{\prime} - \vec x_{j}),
\end{array}
\label{eq:dummyvars1}
\end{equation}
we have
\begin{equation}
\begin{array}{ll}
\vec x & = {\bf R_i}^{-1}(\vec y) + \vec x_{i}\\
\vec x^{\prime} & = {\bf R_j}^{-1}(\vec y^{\prime}) + \vec x_{j}\\
d\vec x & = {\bf R_i}^{-1}(d\vec y)\\
d\vec x^{\prime} & = {\bf R_j}^{-1}(d\vec y^{\prime})
\end{array}
\label{eq:dummyvars2}
\end{equation}
Substituting these into equation \ref{eq:cij1}, we have
\begin{equation}
C^T_{ij}=\int \int B_m(\vec y) B_{m^{\prime}}(\vec y^{\prime})
\sum { {(2l+1) \over 4\pi} C_{l} P_{l}(cos(q))}
{\bf R_i}^{-1}(d\vec y) {\bf R_j}^{-1}(d\vec y^{\prime})
\label{eq:cij2}
\end{equation}
Defining the window functions $W_{lij}$ by
\begin{equation}
C^T_{ij} = \sum_{l} {(2l+1) \over 4\pi} C_{l} W_{lij}
\label{eq:windef}
\end{equation}
and combining equations \ref{eq:cij2} and \ref{eq:windef},
we get an expression for the window functions:
\begin{equation}
W_{lij} = \int \int
B_m(\vec y) B_{m^{\prime}}(\vec y^{\prime}) P_{l}(cos q)
{\bf R_i}^{-1}(d\vec y) {\bf R_j}^{-1}(d\vec y^{\prime}).
\label{eq:wlij1}
\end{equation}

This is a four-dimensional integral, which is computationally
intensive. In order to reduce the dimensionality of the
integral to be able calculate the window functions in a
reasonable amount of time on a workstation, we make the
flat sky approxmation
that for large l,
\begin{equation}
P_{l}(cos q) = J_{0}(lq).
\label{plapprox}
\end{equation}
The expression for the window functions then becomes
\begin{equation}
W_{lij} = \int \int 
B_m(\vec y) B_{m^{\prime}}(\vec y^{\prime}) J_{0}(lq)
{\bf R_i}^{-1}(d\vec y) {\bf R_j}^{-1}(d\vec y^{\prime}).
\label{eq:wlij2}
\end{equation}
Using the identity
\begin{equation}
J_{0}(lq) = {(1/2\pi)} \int d\phi e^{-ilqcos\phi},
\label{eq:j0iden}
\end{equation}
and substituting into equation \ref{eq:wlij2}, we have
\begin{equation}
W_{lij} = {(1/2\pi)} \int \int \int
B_m(\vec y) B_{m^{\prime}}(\vec y^{\prime}) e^{-ilqcos\phi}
d\phi {\bf R_i}^{-1}(d\vec y) {\bf R_j}^{-1}(d\vec y^{\prime}).
\label{eq:wlij3}
\end{equation}
Defining a vector $\vec k$ such that $| \vec k | = l$ and
$\vec k \cdot \vec q = lqcos\phi$, equation \ref{eq:wlij3} becomes
\begin{equation}
W_{kij} = {(1/2\pi)} \int \int \int
B_m(\vec y) B_{m^{\prime}}(\vec y^{\prime}) e^{-i \vec k \cdot \vec q}
d\phi {\bf R_i}^{-1}(d\vec y) {\bf R_j}^{-1}(d\vec y^{\prime}).
\label{eq:wkij1}
\end{equation}
From equations \ref{eq:qdef} and \ref{eq:dummyvars2}, we have
\begin{equation}
\vec q = {\bf R_i}^{-1}(\vec y) + \vec x_{i}
     - {\bf R_j}^{-1}(\vec y^{\prime}) - \vec x_{j}.
\label{eq:qdummy}
\end{equation}
Substituting equation \ref{eq:qdummy} into equation \ref{eq:wkij1}, the window
functions become
\begin{equation}
W_{kij} = {(1/2\pi)} \int \int \int
B_m(\vec y) B_{m^{\prime}}(\vec y^{\prime}) e^{-i \vec k \cdot
({\bf R_i}^{-1}(\vec y) + \vec x_{i}
- {\bf R_j}^{-1}(\vec y^{\prime}) - \vec x_{j})}
d\phi {\bf R_i}^{-1}(d\vec y) {\bf R_j}^{-1}(d\vec y^{\prime}).
\label{eq:wkij2}
\end{equation}
Since we are integrating over all $\vec y$ and since the determinant
of the rotation matrix is 1, we can write this as
\begin{equation}
W_{kij} = {(1/2\pi)} \int \int \int
B_m(\vec y) B_{m^{\prime}}(\vec y^{\prime}) e^{-i \vec k \cdot
({\bf R_i}^{-1}(\vec y) + \vec x_{i}
- {\bf R_j}^{-1}(\vec y^{\prime}) - \vec x_{j})}
d\phi d\vec y d\vec y^{\prime}.
\label{eq:wkij3}
\end{equation}
Rearranging a few terms,
\begin{equation}
W_{kij} = {(1/2\pi)} \int e^{-i \vec k \cdot (\vec x_{i} - \vec x_{j})} d\phi
\int e^{-i \vec k \cdot ({\bf R_i}^{-1}(\vec y))} B_m(\vec y) d\vec y
\int e^{+i \vec k \cdot {\bf R_j}^{-1}(\vec y^{\prime})}
B_{m^{\prime}}(\vec y^{\prime}) d\vec y^{\prime}.
\label{eq:wkij4}
\end{equation}
Equation \ref{eq:wkij4} can be written
in terms of the Fourier transformed beam maps, yielding
the final expression for the window functions:
\begin{equation}
W_{kij} = {(1/2\pi)} \int
e^{-i \vec k \cdot (\vec x_{i} - \vec x_{j})}
\tilde B_m(\vec k {\bf R_i}^{-1})
\tilde B_{m^{\prime}}^{*}(\vec k {\bf R_j}^{-1}) d\phi .
\label{eq:wkij5}
\end{equation}

The four-dimensional integral of equation \ref{eq:wlij1} has been reduced
to a one-dimensional integral and a Fourier transform, saving computing time.
The window functions can be saved to disk for quick computation of
${\bf C^T}$ for various input $C_l$ spectra using equation \ref{eq:windef}.

We will refer to the array $W_{lij}$ as the window functions and to
$W_l=W_{lii}$ as the diagonal window functions.

\subsection{Normalization of the Beam Map}

The modulated beam map is created by convolving the
(spatial) modulation pattern with the beam and is
usually normalized by $(1/{2 \pi \sigma^2})$, where
$\sigma$ characterizes the width of a symmetric Gaussian beam, or
$\sigma = \sqrt{\frac{1}{2}(\sigma_x \sigma_y)}$ in the case
of an asymmetric Gaussian. The normalization of the modulations
usually follows the convention discussed in 3.2.

Some experiments do not follow this beam normalization.
For example,
for the MSAM-I experiment \cite{msam}, which does not
have a Gaussian beam, the beam map and data are normalized
consistently with each other, but the data are in units of
$\mu K deg^2$ rather than $\mu K$.

\subsection{Approximations}

\begin{figure}[t!]
\centerline{\epsfxsize=13.9cm \epsfbox{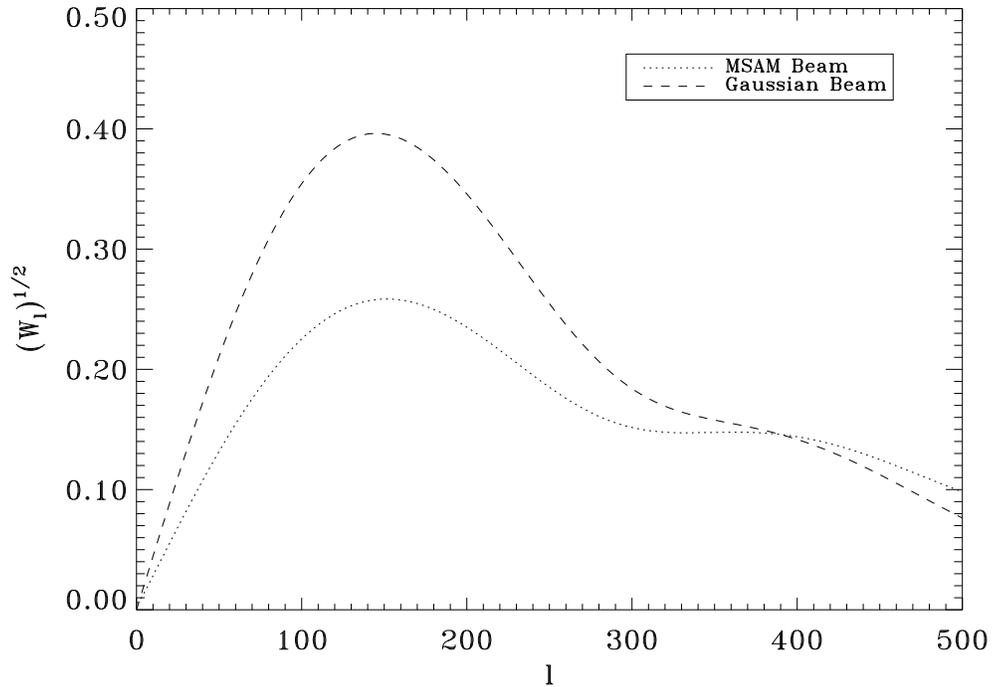}}
\ssp
\caption[Diagonal window functions for MSAM-I with a Gaussian beam and with the true beam.]{Diagonal window functions for MSAM-I with a Gaussian beam and with the true beam.}
\label{fig:msam_diag}
\end{figure}

\begin{figure}[t!]
\centerline{\epsfxsize=13.9cm \epsfbox{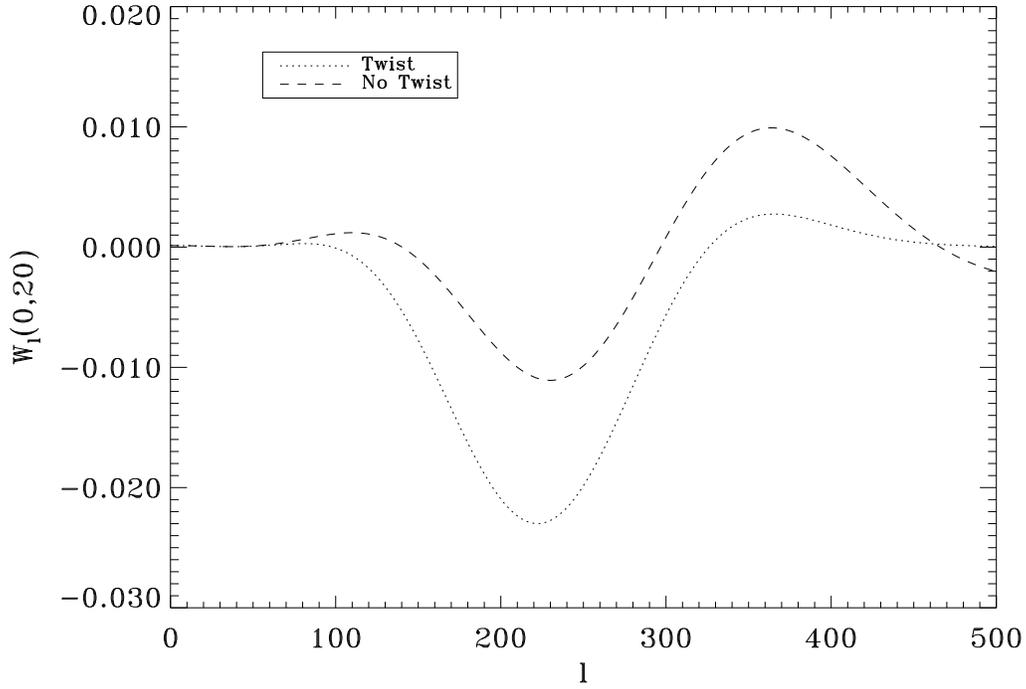}}
\ssp
\caption[Effect of twist on the window functions.]{Effect of twist on the window functions. Including twist is important for the MSAM-I experiment.}
\label{fig:msam_twist}
\end{figure}

Depending on the experiment, some approximations can affect the
window functions. For example, for the MSAM-I experiment,
a Gaussian beam is a poor approximation. Its impact on the window
functions is shown in Figure \ref{fig:msam_diag}.
Ignoring the twist of the MSAM-I experiment is also a poor approximation,
as shown in Figure \ref{fig:msam_twist}.

\subsection{Analytic Expressions}
This section provides some concrete examples of window functions
for observing strategies and beam maps which can be treated analytically.
Examples will be given for 3 cases of analytic expresssions: (1) calculation
of the full window function array in the flat space approximation,
(2) calculation of the diagonal window functions in the flat space
approximation, and (3) calculation of the diagonal window functions
without the flat space approximation.

$\bullet$ Case 1: Two examples of the full window function array given
by equation \ref{eq:wkij5} are given below.

Example 1.1. Gaussian beam.

The beam map for a symmetric Gaussian beam of width $\sigma$ is given by
\begin{equation}
B(\vec a)=\frac{1}{2\pi\sigma^2}e^{-a^2/(2\sigma^2)},
\label{wlij_sd_1beam_1}
\end{equation}
so the Fourier transform of the beam map is
\begin{equation}
\tilde B(k)=e^{-k^2\sigma^2/2}.
\label{wlij_sd_1beam_2}
\end{equation}
Using equation \ref{eq:wkij5} the window functions are:
\begin{equation}
\begin{array}{ll}
W_{kij}= & \frac{1}{2\pi} e^{-k^2\sigma^2} \int_0^{2\pi} d\theta
e^{-ik | \vec x_i - \vec x_j | cos(\theta)}\\
& = e^{-l^2\sigma^2}J_0(l( | \vec x_i - \vec x_j | ).
\end{array}
\label{wlij_sd_1beam_4}
\end{equation}

Example 1.2. Single difference Gaussian beam.

The beam map for a symmetric Gaussian beam modulated with a
single difference chop is given by
\begin{equation}
B(\vec a)=\frac{1}{2\pi\sigma^2}
[e^{-| \vec a - \vec a_0 |^2/(2\sigma^2)}
-e^{-| \vec a + \vec a_0 |^2/(2\sigma^2)}]
\label{wlij_sd_2beam_1}
\end{equation}
and its Fourier transform is
\begin{equation}
\tilde B(k)=e^{-k^2\sigma^2/2}
[e^{-i \vec k \cdot \vec a_0} - e^{+i \vec k \cdot \vec a_0}].
\label{wlij_sd_2beam_2}
\end{equation}
Using equation \ref{eq:wkij5}, the window functions are
\begin{equation}
W_{kij}=\frac{1}{2\pi} e^{-k^2\sigma^2} \int_0^{2\pi} d\theta
e^{-i \vec k \cdot (\vec x_i - \vec x_j)}
[e^{-i \vec k_i \cdot \vec a_0} - e^{+i \vec k_i \cdot \vec a_0}]
[e^{+i \vec k_j \cdot \vec a_0} - e^{-i \vec k_j \cdot \vec a_0}]
\label{wlij_sd_2beam_3}
\end{equation}
where $\vec k_i = \vec k {\bf R_i}^{-1}$. Simplifying using
equation \ref{eq:j0iden}, we have
\begin{equation}
\begin{array}{ll}
W_{lij} = e^{-l^2\sigma^2}
& [J_0(l | \vec x_i - \vec x_j + \vec a_{0i} - \vec a_{0j} | )\\
& +J_0(l | \vec x_i - \vec x_j - \vec a_{0i} + \vec a_{0j} | )\\
& -J_0(l | \vec x_i - \vec x_j - \vec a_{0i} - \vec a_{0j} | )\\
& -J_0(l | \vec x_i - \vec x_j + \vec a_{0i} + \vec a_{0j} | )]
\end{array}
\label{wlij_sd_2beam_4}
\end{equation}
where $\vec a_{0i}={\bf R_i}^{-1} {\vec a_0}$.

$\bullet$ Case2: For just the diagonal window functions,
equation \ref{eq:wkij5} simplifies to 
\begin{equation}
W_{kii} = {(1/2\pi)} \int d\phi | \tilde B_m(\vec k) |^2.
\label{wkij5_diag}
\end{equation}

The following are examples of equation \ref{wkij5_diag}
for simple beams and chopping strategies.

Example 2.1.Single difference Gaussian beam.

The diagonal window functions are:
\begin{equation}
W_l=\tilde B_g^2(l)2[1-J_0(2l\theta_c)]
\label{w2gauss}
\end{equation}
where
\begin{equation}
\tilde B_g(l)=e^{-l^2\sigma^2/2}
\label{eq:bgauss}
\end{equation}
is the Fourier transform of an (unmodulated) symmetric Gaussian beam.

Example 2.2. Double difference Gaussian beam.

The diagonal window functions are:
\begin{equation}
W_l=\tilde B_g^2(l)[\frac{3}{2}-2J_0(l\theta_c)+\frac{1}{2}J_0(2l\theta_c)]
\label{w3gauss}
\end{equation}

Example 2.3. Single difference top hat beam.

The diagonal window functions are:
\begin{equation}
W_l=\tilde B_t^2(l)2[1-J_0(2l\theta_c)]
\label{w2top}
\end{equation}
where
\begin{equation}
\tilde B_t(l)=[2J_1(l \theta_t)]/(l \theta_t)
\label{btop}
\end{equation}
is the Fourier transform of a top hat beam of characteristic size $\theta_t$.

Example 2.4. Double difference top hat beam.

The diagonal window functions are:
\begin{equation}
W_l=\tilde B_t^2(l)[\frac{3}{2}-2J_0(l\theta_c)+\frac{1}{2}J_0(2l\theta_c)]
\label{w3top}
\end{equation}

$\bullet$ Case 3: The diagonal window functions for a
general modulation pattern, without assuming the flat sky appoximation
are given by
\begin{equation}
W_{lmm^{\prime}}=\tilde B^2(l)
\sum_s \sum_{s^{\prime}}
M_{ms} M_{m^{\prime} s^{\prime}}
P_l(cos(\theta_s-\theta_{s^{\prime}}))
\label{eq:wl_gen}
\end{equation}
where $\theta$ is the chopper angle as a function of sample $s$
and $M_m$ is the amplitude of modulation $m$. $\tilde B^2(l)$
is the spherical transform of the (unmodulated) beam map.
The beam must be symmetric.
Some examples of equation \ref{eq:wl_gen} are given below.

Example 3.1. Single difference Gaussian beam.

The modulation is given by:
\begin{equation}
M = 
\left\{
\begin{array}{llll}
+1 & at & \theta=-\theta_c & (s=1)\\
-1 & at & \theta=+\theta_c & (s=2).
\end{array}
\right.
\label{eq:wl2beam_0}
\end{equation}
and the spherical transform of a Gaussian beam is given by
\begin{equation}
\tilde B(l)=e^{-\frac{1}{2}l(l+1)\sigma^2}.
\label{eq:bgauss2}
\end{equation}
Thus, from equation \ref{eq:wl_gen}, the diagonal window functions are
\begin{equation}
\begin{array}{ll}
W_l & = \tilde B_g^2(l) \sum_{s=1}^2 \sum_{s^{\prime}=1}^2
M_{ms} M_{m^{\prime} s^{\prime}}
P_l(cos(\theta_s-\theta_{s^{\prime}}))\\

 &  = \tilde B_g^2(l)[(+1)(+1)P_l(cos(-\theta_c+\theta_c))\\
 & +(+1)(-1)P_l(cos(-\theta_c-\theta_c))\\
 & +(-1)(+1)P_l(cos(+\theta_c+\theta_c))\\
 & +(-1)(-1)P_l(cos(+\theta_c-\theta_c))]\\

 & = \tilde B_g^2(l)[P_l(cos(0))-P_l(cos(-2\theta_c))\\
 & -P_l(cos(2\theta_c))+P_l(cos(0))]
\end{array}
\label{eq:wl2beam_3}
\end{equation}
\begin{equation}
W_l= \tilde B_g^2(l) 2[1-P_l(cos2\theta_c)]
\label{eq:wl2beam_4}
\end{equation}

Example 3.2. Double difference Gaussian beam.

The modulation is given by
\begin{equation}
M = 
\left\{
\begin{array}{llll}
-1/2 & at & \theta=-\theta_c & (s=1)\\
+1 & at & \theta=0 & (s=2)\\
-1/2 & at & \theta=+\theta_c & (s=3)
\end{array}
\right.
\label{eq:wl3beam_0}
\end{equation}
and the diagonal window functions are given by
\begin{equation}
W_l= \tilde B_g^2(l) [\frac{3}{2}-2P_l(cos\theta_c)+\frac{1}{2}P_l(cos2 \theta_c)].
\label{eq:wl3beam_1}
\end{equation}

Example 3.3. Four-point-chop Gaussian beam.

The modulation is given by
\begin{equation}
M = 
\left\{
\begin{array}{llll}
-1/4 & at & \theta=-\frac{3}{2}\theta_c & (s=1)\\
+3/4 & at & \theta=-\frac{1}{2}\theta_c & (s=2)\\
-3/4 & at & \theta=+\frac{1}{2}\theta_c & (s=3)\\
+1/4 & at & \theta=+\frac{3}{2}\theta_c & (s=4)
\end{array}
\right.
\label{eq:wl4beam_0}
\end{equation}
and the diagonal window functions are given by
\begin{equation}
W_l= \tilde B_g^2(l)
[\frac{5}{4}
-\frac{15}{8}P_l(cos \frac{2}{3} \theta_c)
+\frac{3}{4}P_l(cos \frac{4}{3} \theta_c)
-\frac{1}{8}P_l(cos 2 \theta_c)].
\label{eq:wl4beam_1}
\end{equation}

\begin{figure}[h!]
\centerline{\epsfxsize=12.0cm \epsfbox{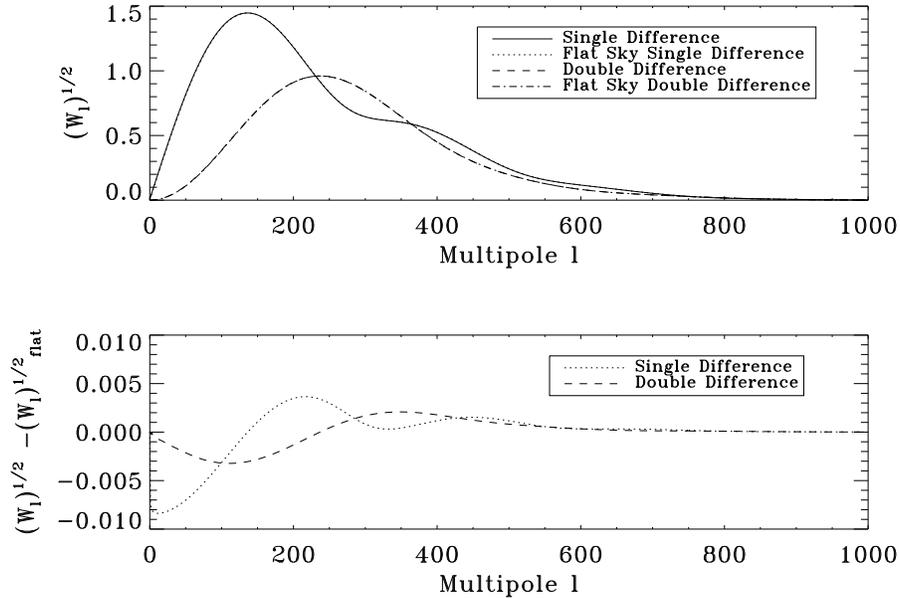}}
\ssp
\caption[Diagonal window functions with and without the flat sky approximation.]{Diagonal window functions with and without the flat sky approximation. Shown in the top panel are $W_l^{1/2}$ for single and double difference chops of $\theta_c=0.7^{\circ}$ and a Gaussian beam of width $\sigma=0.212^{\circ}$. Shown in the bottom panel are the residuals.}
\label{fig:wl_analyt_23}
\end{figure}

A summary of the diagonal window functions for simple chopping strategies
is given in Table \ref{tbl:wl_chops}.
The diagonal window functions in the flat space approximation are similar to
the ones calculated not assuming the flat space approximation
except that the Legendre polynomials $P_l(cos\theta_c)$ are
replaced by Bessel functions $J_0(l\theta_c)$. Figure \ref{fig:wl_analyt_23}
shows the diagonal window functions for single and double difference
Gaussians with and without the flat space approximation.

\begin{table*}[!h]
\begin{center}
\begin{tabular}{|cc|}
\tableline
modulation & $W_l$ \\
\tableline
2-point & $2[1-P_l(cos2\theta_c)]$ \\
3-point & $[\frac{3}{2}-2P_l(cos\theta_c)+\frac{1}{2}P_l(cos2 \theta_c)]$ \\
4-point & $[\frac{5}{4}
-\frac{15}{8}P_l(cos \frac{2}{3} \theta_c)
+\frac{3}{4}P_l(cos \frac{4}{3} \theta_c)
-\frac{1}{8}P_l(cos 2 \theta_c)]$ \\
\tableline
\end{tabular}
\end{center}
\ssp
\caption[Summary of analytic expressions for diagonal window functions for
simple chopping strategies.]{Summary of analytic expressions for diagonal window functions for simple chopping strategies. All are multiplied by $\tilde B^2(l)$. The flat space approximations are the same except the Legendre polynomials $P_l(cos\theta_c)$ are replaced by Bessel functions $J_0(l\theta_c)$.}
\label{tbl:wl_chops}
\end{table*}

\subsection{Python V Window Functions}

\begin{figure}[t!]
\centerline{\epsfxsize=13.9cm \epsfbox{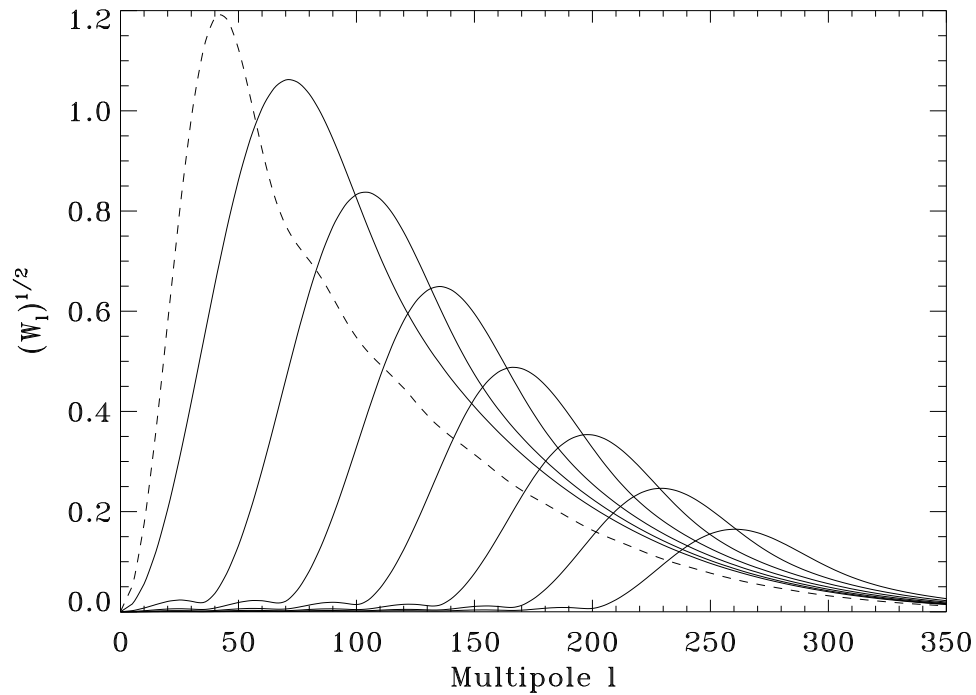}}
\ssp
\caption[Diagonal window functions for PyV.]{Diagonal window functions for PyV at a declination of $-49^{\circ}$. The unapodized cosine modulation is plotted with a dashed line and the apodized cosine modulations are plotted with solid lines.}
\label{fig:winplot}
\end{figure}

For the PyV observing strategy, equation \ref{eq:wkij5} becomes
\begin{equation}
W_{lij} = {(1/2\pi)} \int 
e^{-i {\vec k} \cdot ({\vec x_{i}} - {\vec x_{j}})}
\tilde B({\vec k}) \tilde B^{*}({\vec k})
d\phi
\label{wlij_pyv}
\end{equation}
where $\tilde B({\vec k})$ is the Fourier transform of the
beam map for the given modulation, ${\vec x_{i}}$
the position of field $i$, and ${\vec k} = l (\cos\phi,\sin\phi)$.
These functions are computed for all pairs of fields and channels.
Figure \ref{fig:winplot} shows the PyV diagonal window functions.
A different modulated beam map is used at each declination because the total
chopper throw, which is constant in azimuth degrees, subtends
fewer real degrees on the sky at higher declinations.

\spacing{2}		

\section{KL Decomposition}

Karhunen-Loeve (KL) decomposition
(see for example Tegmark et al. 1996), also
known as the signal-to-noise eigenmode method (Bond 1994),
transforms the data into a basis
in which modes of the data can be sorted by
their signal-to-noise (S/N).

This technique is useful for testing the self-consistency of
data sets, for examining modes with a
particular S/N range, and for data compression.

Calculation of the determinant in equation \ref{eq:like}
is an $N^3$ process
and calculation ${\bf C}^{-1}{\vec d}$ is an $N^2$ process
(see for example the backslash command in MATLAB),
where $N$ is the number of data points.
By eliminating the modes which do not contribute to parameter estimation,
(i.e. the ones with low S/N), KL decomposition can
compress data sets with a minimal loss of
information, speeding up the data analysis
process typically by 3 orders of magnitude.

The steps for KL decomposition are as follows:

\begin{enumerate}

\item Let ${\bf C^{N^{\prime}}}={\bf C^N}+{\bf C^C}$.
\item Calculate $({\bf C^{N^{\prime}}})^{-1/2}$, for example by Cholesky decomposition.
\item Calculate the matrix
\begin{equation}
{\bf M}=({\bf C^{N^{\prime}}})^{-1/2}{\bf C^T}({\bf C^{N^{\prime}}})^{-t/2}.
\label{eq:m_kl}
\end{equation}
\item Calculate ${\bf R}$, the matrix of eigenvectors of ${\bf M}$.
\item Sort ${\bf R}$ by eigenvalue. The eigenvalues
are a measure of signal/noise.
\item The compression matrix ${\bf B^{\prime}}$ is the
first $N_{\rm keep}$ rows of ${\bf B}={\bf R}({\bf C^{N^{\prime}}})^{-1/2}$,
where $N_{\rm keep}$ is the number of KL modes to keep.

\item Then the theory matrix in the KL basis is
\begin{equation}
{\bf C^T_{KL}}=({\bf B^{\prime}}){\bf C^T}({\bf B^{\prime}})^t,
\label{eq:ct_kl}
\end{equation}
the data in the KL basis is
\begin{equation}
{\vec d_{KL}}=({\bf B^{\prime}}){\vec d},
\label{eq:d_kl}
\end{equation}
and the noise matrix in the KL basis is the identity matrix $I$.

\end{enumerate}

\section{Constraint Matrices For Offsets}

The subtraction of offsets can be
accounted for as an additional term in the covariance
matrix (e.g., Bond et al. 1998).
For PyV, we must take into account the subtraction
of the chopper synchronous offset and the ground shield
synchronous offset.

In both cases, the subtracted data in field $i$ is given by:
\begin{equation}
\begin{array}{ll}
{\hat d^i} & = \frac{1}{N_{\rm files}}\sum_{n=1}^{N_{\rm files}} (d^i_n - \kappa f^i_n)\\
& \equiv {\bar d^i} - \kappa {\bar f^i}\\
\end{array}
\label{eq:ccgs_def}
\end{equation}
where
${\bar f^i}$ is the mean of the offset vector
which has been subtracted from field $i$.
Both constraint matrices thus have the form
\begin{equation}
C^{C}_{ij} = \kappa \langle {\bar f_i} {\bar f_j} \rangle.
\label{eq:cc_gen}
\end{equation}

\begin{figure}[h!]
\centerline{\epsfxsize=11.0cm \epsfbox{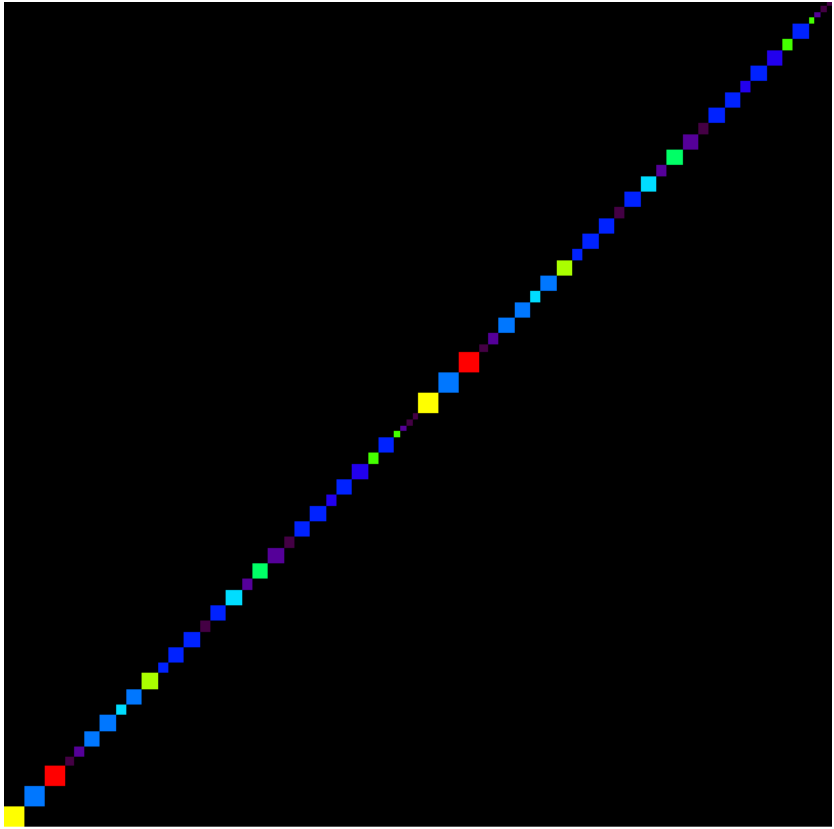}}
\ssp
\caption[Image of chopper offset constraint matrix.]{Image of chopper offset constraint matrix. $C^{CH}(1,1)$ is in the bottom left corner of the image. The matrix is 690 x 690 for the 345 fields and 2 channels in the reduced, co-added data. Points 1:345 correspond to channel 45 and points 346:690 correspond to channel 12.}
\label{fig:cchop_im}
\end{figure}

The chopper synchronous offset is constant over a file and
the same chopper offset has been subtracted from
fields $i$ and $j$, thus ${\bar f^i}={\bar f^j}$
and the constraint matrix takes the form:
\begin{equation}
{\bf C^{CH}}=(\kappa_1/N_{\rm files})
\label{eq:cchop}
\end{equation}
for fields taken with the same set of files and zero
for fields not taken with the same
set of files. 
$N_{\rm files}$ is the number of files which observed the fields.
Even if the offset subtracted is not perfectly estimated,
taking $\kappa_1$ to be large ensures that we have 
no sensitivity to modes of the data
that could have come from the chopper.
An image of the chopper constraint matrix
is shown in Figure \ref{fig:cchop_im}.

\begin{figure}[h!]
\centerline{\epsfxsize=11.0cm \epsfbox{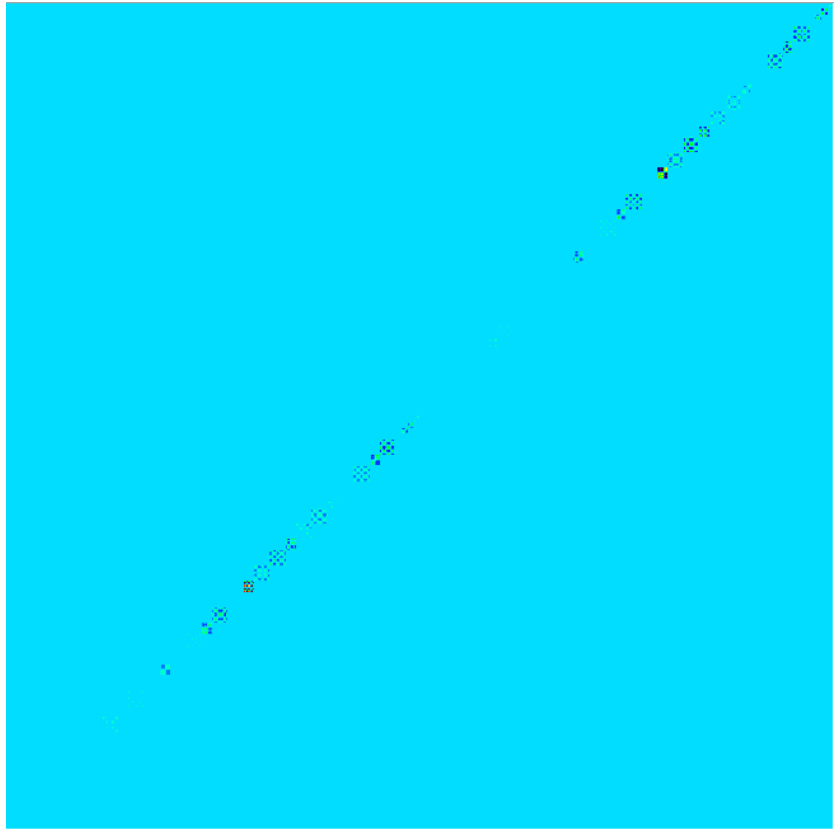}}
\ssp
\caption[Image of ground shield constraint matrix.]{Image of ground shield constraint matrix for modulation 2. $C^{GS}(1,1)$ is in the bottom left corner of the image. The matrix is 690 x 690 for the 345 fields and 2 channels in the reduced, co-added data. Points 1:345 correspond to channel 45 and points 346:690 correspond to channel 12.}
\label{fig:ccgs_im}
\end{figure}

In the case of the ground shield synchronous offset,
${\bar f^i} \neq {\bar f^j}$.
The ground shield constraint matrix is:
\begin{equation}
C^{GS}_{ij} = \kappa_2 \langle {\bar f_i} {\bar f_j} \rangle.
\label{eq:ccgs}
\end{equation}
for fields $i$ and $j$ taken with the same set of files,
and is zero for fields not taken with the same set of files.
For each field $i$, ${\bar f_i}$ is known from the data
reduction process (section 3.5).
There is a different ground
shield constraint matrix for each modulation.
Again, $\kappa_2$ is taken to be large.
An image of the ground shield
constraint matrix is shown in Figure \ref{fig:ccgs_im}.

We will refer to the total constraint matrix for PyV
as ${\bf C^C}={\bf C^{CH}}+{\bf C^{GS}}$.
\chapter{Single Modulation Analysis}

The angular power spectrum is first estimated from the PyV data set
by considering each modulation separately.
An analysis which includes cross-modulation correlations
and which estimates the angular power spectrum
simultaneously in a series of $l$-space bands is discussed in Chapter 6.
This single modulation analysis is simpler computationally
and conceptually than the cross-modulation analysis.
The reduced PyV data set
contains 690 pixels $\times$ 8 modulations = 5520 data points.
Computing the likelihood is considerably faster if 8 690$\times$690
matrices rather than a 5520$\times$5520 matrix is used.
Additionally,
the cross-modulation noise was difficult to model, as will be
discussed in Chapter 6.

\section{Power Spectrum Constraints}

\begin{figure}[ht!]
\centerline{\epsfxsize=11.0cm \epsfbox{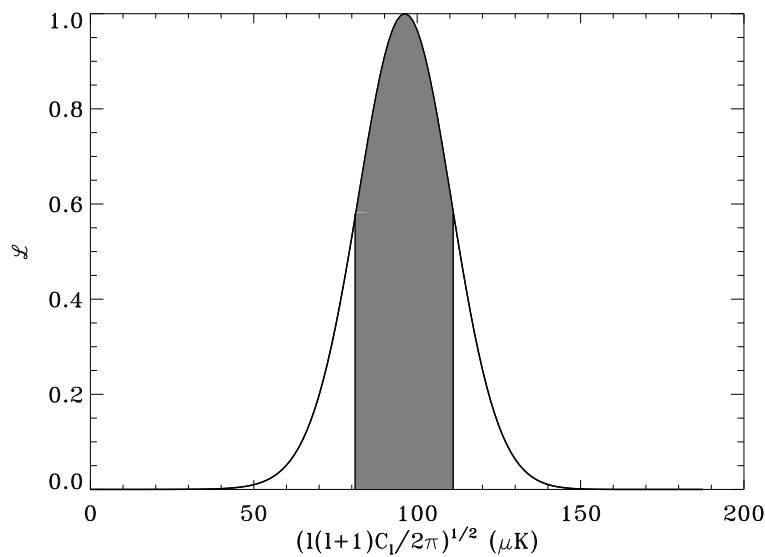}}
\ssp
\caption[Getting limits from a likelihood curve.]{Getting limits from a likelihood curve. The likelihood curve and 1$\sigma$ limits for modulation 6 are shown as a typical example. The shaded area is 68\% of the total area under the curve.}
\label{fig:likelims}
\end{figure}

A flat power spectrum is one for
which ${\cal C} \equiv (l(l+1)C_{l}/2\pi)$ is constant.
For each of the eight modulations, we compute 
the likelihood (${\cal L}$) as a function
of ${\cal C} \equiv (l(l+1)C_{l}/2\pi)$
following equation \ref{eq:like}.
Limits are obtained from the likelihood curves following the
prescription of Ganga et al. (1996), by integrating $\cal L$
over ${\cal C}^{1/2}$ starting with the most likely value of ${\cal C}^{1/2}$
and slicing at equal values of $\cal L$ until 68\% of the total area
under the curve is reached for 1$\sigma$ limits or 95\% of the total area
under the curve is reached for 2$\sigma$ limits. An example likelihood
curve and resulting limits are shown in Figure \ref{fig:likelims}.

\begin{figure}[h!]
\centerline{\epsfxsize=13.9cm \epsfbox{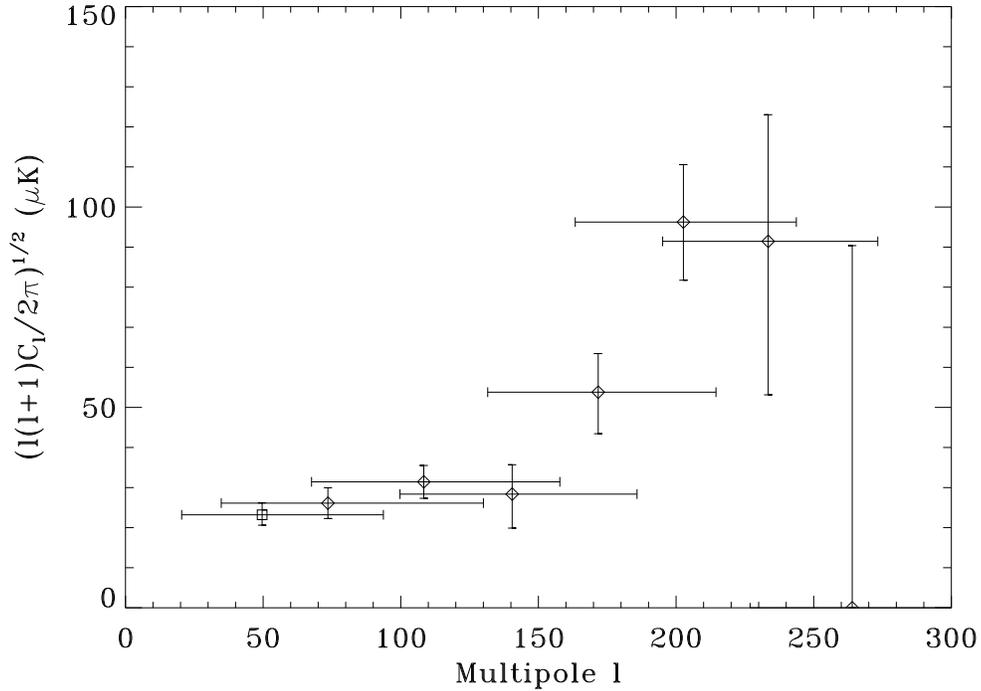}}
\ssp
\caption[PyV angular power spectrum constraints from the single modulation analysis.]
{PyV angular power spectrum constraints from the single modulation analysis.
The detections have $1 \sigma$ error bars and the upper limit has $2 \sigma$
error bars.
The unapodized cosine modulation is plotted with an open square
and the apodized cosine modulations are plotted
with diamonds. The error bars include
statistical uncertainties only and do not include uncertainties in the
calibration or beam size.
The $l$ range of each modulation is determined by the half-maximum points of
$(W_l)^{1/2}$. Low $l$ values correspond to large angular
scales and high $l$ values correspond to small angular scales. CMB power is
clearly rising from low to high $l$ up to the sensitivity cutoff of PyV.}
\label{fig:pyv_powerspec}
\end{figure}

The CMB power spectrum is shown in Figure \ref{fig:pyv_powerspec}
and is given in Table \ref{tbl:pyv_powerspec}. The band powers
shown in Figure \ref{fig:pyv_powerspec} are calculated for each modulation
separately. The error bars include
statistical uncertainties only and do not include uncertainties in the
calibration or beam size.
The calibration uncertainty 
allows all band powers to shift by the same amount (i.e. the calibration
errors are correlated). Given the beam size uncertainty
of approximately $0.015^\circ$, the band power for a given
modulation can move roughly by a factor of
$\exp ( \pm l(0.425)(0.015)(\pi/180) )$, only a $3\%$
effect at $l=200$. 
The effective $l$ of each modulation is given by
\begin{equation}
l_e = \frac{I(lW_l)}{I(W_l}
\label{eq:leff1}
\end{equation}
where
\begin{equation}
I(W_l) = \sum_l \frac{(l+\frac{1}{2})W_l}{l(l+1)}.
\label{eq:leff2}
\end{equation}
The $l$ range of each modulation is determined by the half-maximum points of
$(W_l)^{1/2}$ and is indicated as horizontal error bars.

\begin{table*}
\begin{center}
\begin{tabular}{|ccc|}
\tableline
mode & $l_{e}$    &  $(l(l+1)C_{l}/2\pi)^{1/2}$ \\
\tableline
1 &  $50^{+44}_{-29}$   &  $23^{+3}_{-3}$ \\
2 &  $74^{+56}_{-39}$   &  $26^{+4}_{-4}$ \\
3 &  $108^{+49}_{-41}$  &  $31^{+5}_{-4}$ \\ 
4 &  $140^{+45}_{-41}$  &  $28^{+8}_{-9}$\\
5 &  $172^{+43}_{-40}$  &  $54^{+10}_{-11}$ \\
6 &  $203^{+41}_{-39}$  &  $96^{+15}_{-15}$\\
7 &  $233^{+40}_{-38}$  &  $91^{+32}_{-38}$\\
8 &  $264^{+39}_{-37}$  &  $0^{+91}_{-0}$  \\
\tableline
\end{tabular}
\end{center}
\ssp
\caption[Angular power spectrum constraints from the single modulation likelihood analysis.]{Angular power spectrum constraints from the single modulation likelihood analysis. Band powers are in $\mu K$.}
\label{tbl:pyv_powerspec}
\end{table*}

\spacing{2}
	
\section{Noise Model}

Our noise model assumes the covariance between fields taken with
different sets of files is negligible because of the chopper offset
removal and because of the long time (at least 10 hours) between measurements.
An analysis comparing the noise covariance estimated from data which
had not yet been co-added over all cycles and the
noise covariance estimated from data which had been co-added over all cycles
indicates that PyV noise is dominated by detector noise. However, the
long term drifts due to the atmosphere, which add to the variance
as well as induce small correlations between fields taken with the
same set of files, could be important,
especially for power spectrum estimation.

The noise covariance between fields
taken with the same set of files is first
estimated by taking the usual covariance on the co-added data:
\begin{equation}
N_{\rm files} C^N_{ij}=\frac{1}{N_{\rm files}-1}
\sum_{n=1}^{N_{\rm files}}
(d^i_n-{\bar d^i})(d^j_n-{\bar d^j})
\label{eq:stat_cov1}
\end{equation}
where
\begin{equation}
{\bar d^i}=\frac{1}{N_{\rm files}} \sum_{n=1}^{N_{\rm files}}d^i_n,
\label{eq:stat_cov2}
\end{equation}
$d^i$ is the data from field $i$, $d^j$ is the data from field $j$
and $N_{\rm files}$ is the number of files in the set containing fields
$i$ and $j$.
However, since there were typically only 100 files
taken for each field, the sample
variance on the noise estimate is $\sim 1/(100)^{1/2}$, or 10\%, which will
severely bias estimates of band power. To obtain
a better estimate of the noise, we averaged the
variances for each set of files and then scaled the off-diagonal
elements of the covariance
to the average variance in a given set based
on a model derived from the entire PyV data set.

\begin{figure}[h!]
\centerline{\epsfxsize=11.0cm \epsfbox{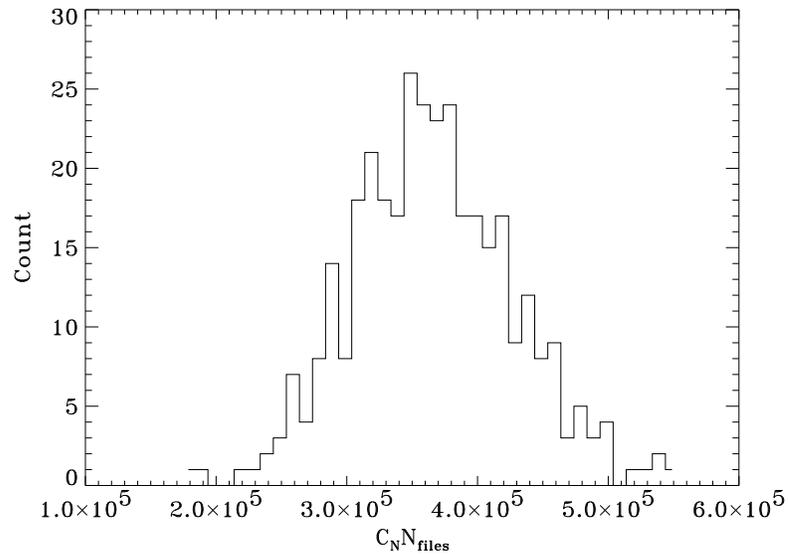}}
\ssp
\caption[Histogram of $C_{N}N_{\rm files}$ for a stare lag of zero.]{Histogram of $C_{N}N_{\rm files}$ for a stare lag of zero, modulation 5, channel 45.}
\label{fig:ch45_b_1_6}
\end{figure}

\begin{figure}[h!]
\centerline{\epsfxsize=11.0cm \epsfbox{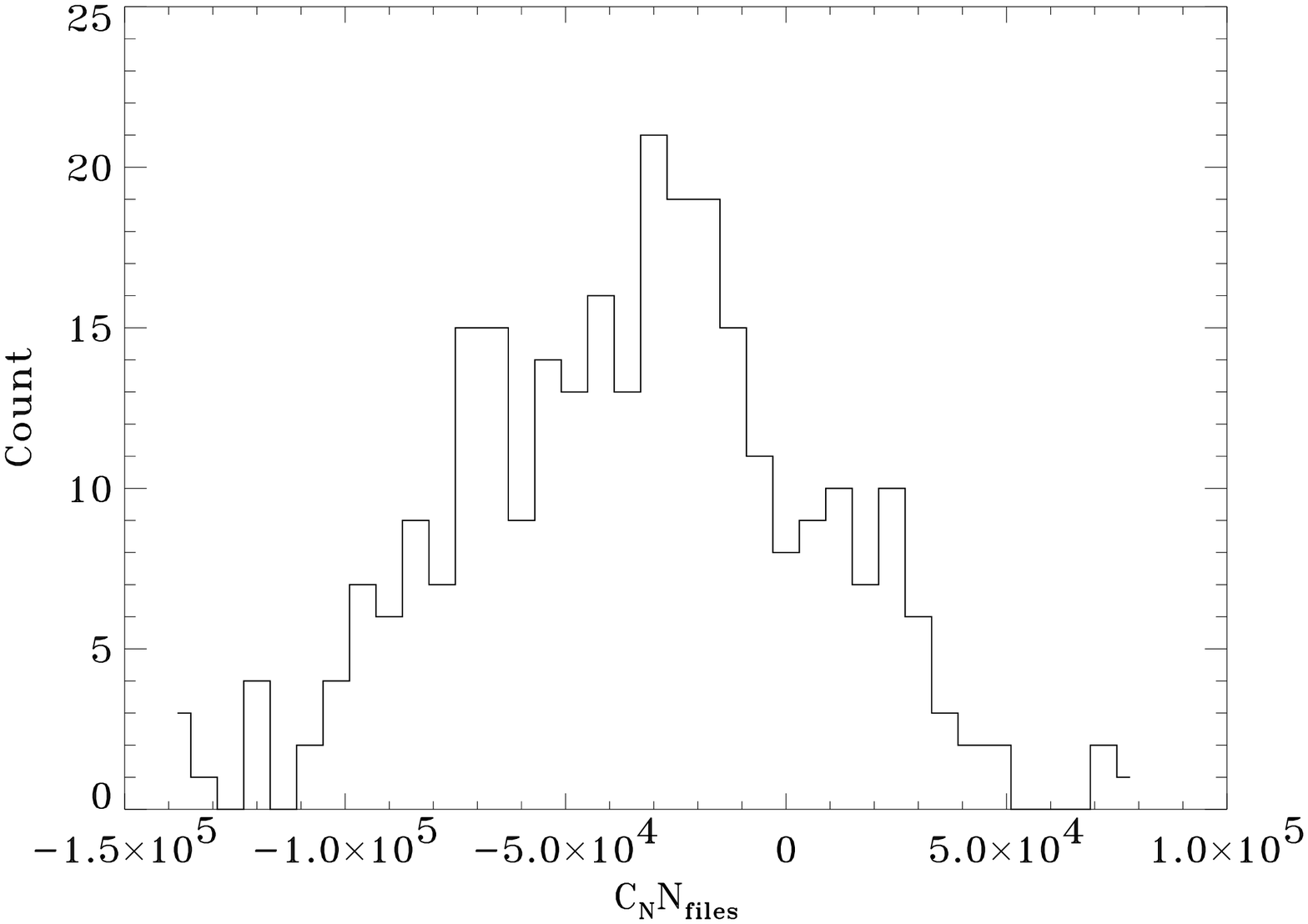}}
\ssp
\caption[Histogram of $C_{N}N_{\rm files}$ for a stare lag of two.]{Histogram of $C_{N}N_{\rm files}$ for a stare lag of two, modulation 5, channel 45.}
\label{fig:ch45_b_3_6}
\end{figure}

\begin{figure}[h!]
\centerline{\epsfxsize=11.0cm \epsfbox{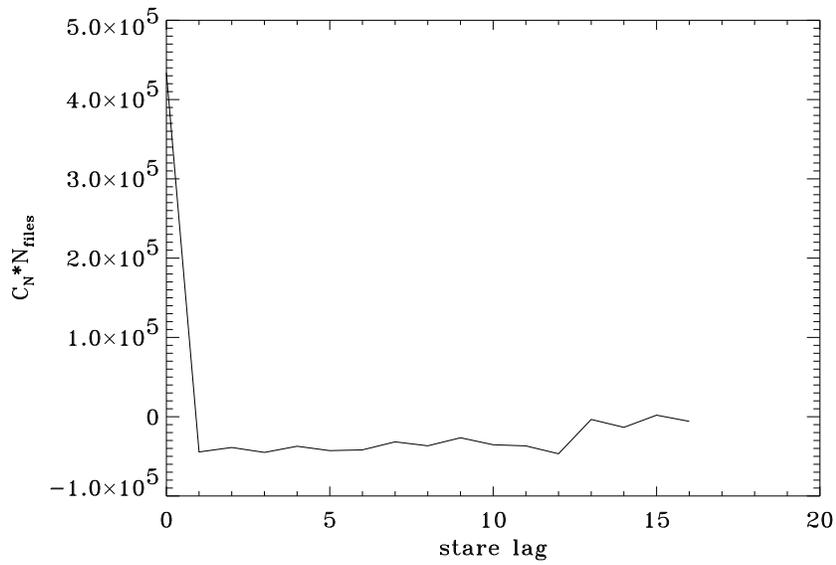}}
\ssp
\caption[$C_{N}N_{\rm files}$ vs. stare lag.]{$C_{N}N_{\rm files}$ vs. stare lag for modulation 5, channel 45 as an example. $C_{N}N_{\rm files}$ is not flat, but is tapered for non-zero lags.}
\label{fig:taper_han3_45_6}
\end{figure}

\begin{figure}[h!]
\centerline{\epsfxsize=11.0cm \epsfbox{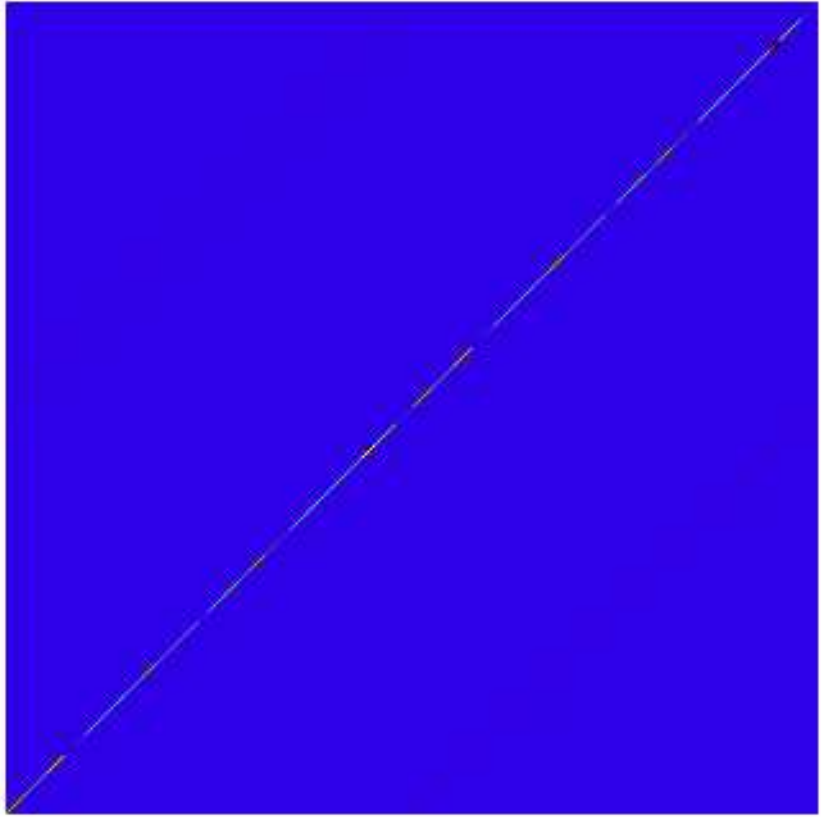}}
\ssp
\caption[Noise covariance matrix, {\bf $C^N$}, for mod 2.]{Noise covariance matrix, {\bf $C^N$}, for modulation 2. $C^N(1,1)$ is in the bottom left. The matrix is 690 x 690 for the 345 fields and 2 channels in the reduced, co-added data. Points 1:345 correspond to channel 45 and points 346:690 correspond to channel 12.}
\label{fig:cn_im}
\end{figure}

In order to derive a noise model for the single modulation analysis
we examined histograms of ${\bf C^N}N_{\rm files}$ for each modulation
as a function of stare lag from 0 to +16 for channels 12 and 45
and from -16 to +16 for the cross-covariance between channels 12 and 45.
Examples are shown in Figures \ref{fig:ch45_b_1_6} and \ref{fig:ch45_b_3_6}.
We also examined plots of the mean of ${\bf C^N}N_{\rm files}$
for each modulation as a function of stare lag.
If the off-diagonal correlations were due entirely to
the chopper offset subtraction and white noise, the
noise matrix would be of the form:
\begin{equation}
C^N_{ij}N_{\rm files}=(\delta_{ij}-\frac{1}{N_{\rm stares}})\sigma^2
\label{eq:cn_white}
\end{equation}
for fields $i$ and $j$ in the same observing set
and noise variance $\sigma^2$.
As illustrated in Figure \ref{fig:taper_han3_45_6}
${\bf C^N}N_{\rm files}$ is not flat as in equation \ref{eq:cn_white},
but has some taper,
indicating that long term drifts, perhaps due to atmosphere,
could be important.
The cross-channel covariances are small but non-zero.
A sample noise covariance matrix is shown in Figure \ref{fig:cn_im}.


\section{Internal Consistency Checks}

Several self-consistency checks were performed on
each modulation of the data set, using
our best estimate of the noise covariance.
We performed these same tests using the preliminary noise covariance,
which suffers from sample variance, and found that the
consistency tests failed.
The tests indicate that the data set used is self consistent
and that our best estimate of the noise covariance is indeed
a good model for the noise.

\subsection{$\chi^2$ Tests}

The $\chi^2={\vec d}^t{\bf C}^{-1}{\vec d}$, where ${\vec d}$ is
the data vector and
${\bf C}={\bf C^T}+{\bf C^N}+{\bf C^C}$ is the total
covariance, is consistent with
its expected value (the number of degrees of freedom) in
all modulations. Table \ref{tbl:chi2_indiv} lists $\chi^2$
for each modulation.

\begin{table*}
\begin{center}
\begin{tabular}{|cc|}
\tableline
mode & $\chi^2$\\
\tableline
1 & 615\\
2 & 660\\
3 & 603\\
4 & 600\\
5 & 667\\
6 & 710\\
7 & 640\\
8 & 609\\
\tableline
\end{tabular}
\end{center}
\caption{$\chi^2$ at best fit band power. Expected dof = 690 $-$ 31 $\times$ 2 = 628}
\label{tbl:chi2_indiv}
\end{table*}

\subsection{$\beta$ Tests}

Since the two channels do not observe exactly the same points
on the sky, a direct $\chi^2$ test between them
could not be done.
Instead, the probability enhancement factor, $\beta$,
(Knox et al. 1998) is computed between data from the two channels
for each modulation.

The value of $\beta$ is given by
\begin{equation}
\frac{1}{2}ln|{\bf C}|+\frac{1}{2}{\vec d}^t{\bf C}^{-1}{\vec d}
-\frac{1}{2}ln|{\bf C_{11}}|-\frac{1}{2}{\vec d_1}^t{\bf C_{11}}^{-1}{\vec d_1}
-\frac{1}{2}ln|{\bf C_{22}}|-\frac{1}{2}{\vec d_2}^t{\bf C_{22}}^{-1}{\vec d_2}
\label{eq:beta}
\end{equation}
where ${\bf C_{11}}$ is the total covariance for the first channel,
${\bf C_{22}}$ is the total covariance for the second channel,
and ${\bf C}$ is the total covariance for all of the pixels in
each modulation.
The expected value of $\beta$ is given by
\begin{equation}
<\beta>_0=\frac{1}{2}ln{\left ( {{|{\bf C_{11}}||{\bf C_{22}}|}\over{|{\bf C}|}} \right ) }
\label{eq:beta_0}
\end{equation}
and the variance of $\beta$ is
\begin{equation}
<(\beta-\beta_0)^2>_0=Tr({\bf w_{12}w_{21}})
\label{eq:beta_var}
\end{equation}
where
\begin{equation}
{\bf w_{21}} \equiv {\bf C^T_{21}}({\bf C^T_{11}}+{\bf C^N_{11}})^{-1}
\label{eq:beta_w21}
\end{equation}
and ${\bf C^T_{21}}$ is the theoretical cross-covariance between
the two channels.

The value of $\beta$ falls within the expected range for all modulations
when the best estimate noise matrix is used.
This is not our strongest test because the signal-to-noise is weaker when
each channel and modulation are considered separately.

\subsection{KL Decomposition}

\begin{figure}[h!]
\centerline{\epsfxsize=11.0cm \epsfbox{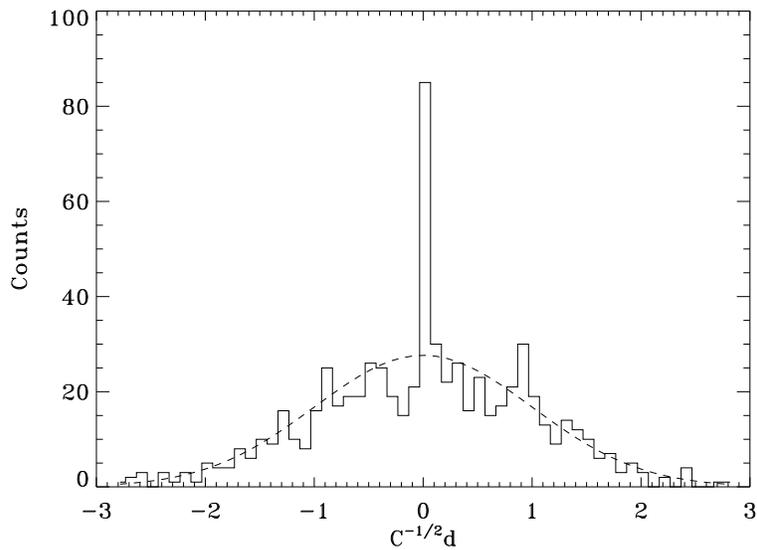}}
\ssp
\caption[Distribution of ${\bf C^{-1/2}}{\vec d}$ in the KL basis for modulation 7 using our best estimate noise matrix.]{Distribution of ${\bf C^{-1/2}}{\vec d}$ in the KL basis for modulation 7 using our best estimate noise matrix. The large spike at ${\bf C^{-1/2}}{\vec d}=0$ is due to modes which have been set to zero by the chopper constraint matrix. The histogram follows the expected Gaussian distribution.}
\label{fig:kl_sim24_fr8}
\end{figure}

\begin{figure}[h!]
\centerline{\epsfxsize=11.0cm \epsfbox{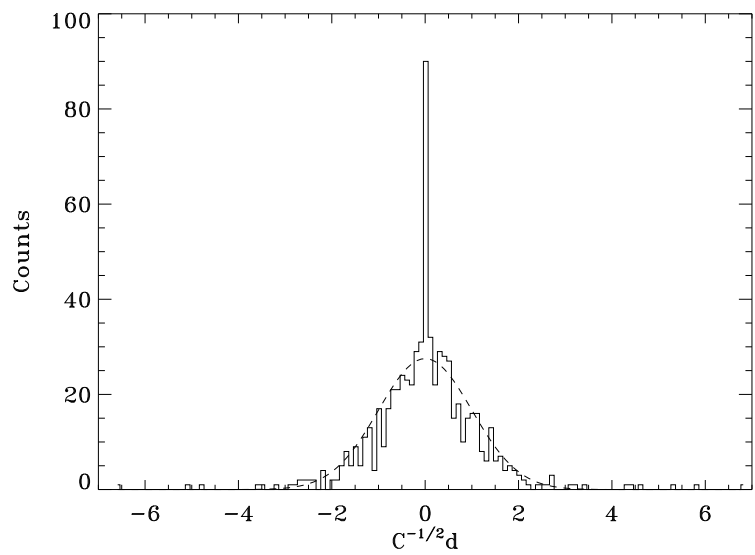}}
\ssp
\caption[Distribution of ${\bf C^{-1/2}}{\vec d}$ in the KL basis for modulation 7 using a preliminary noise matrix.]{Distribution of ${\bf C^{-1/2}}{\vec d}$ in the KL basis for modulation 7 using a preliminary noise matrix. The large spike at ${\bf C^{-1/2}}{\vec d}=0$ is due to modes which have been set to zero by the chopper constraint matrix. The number of counts in the central region are systematically lower than the number in the expected Gaussian distribution and there are many high $\sigma$ points. This preliminary noise matrix also failed the $\chi^2$ and $\beta$ tests.}
\label{fig:kl_klold_fr8}
\end{figure}

Finally, the data set was transformed into the KL
(signal-to-noise eigenmode) basis. In that basis,
${\bf C}^{-1/2}{\vec d}$ should be Gaussian distributed with
$\sigma=1$ and a total area equal to the number of degrees of freedom.
Histograms of ${\bf C}^{-1/2}{\vec d}$ are consistent with Gaussian
distributions for all of the modulations for the best estimate
noise matrices and are inconsistent with Gaussian
distributions for the preliminary noise matrices.
Figures \ref{fig:kl_sim24_fr8} and \ref{fig:kl_klold_fr8}
show histograms for one of the modulations using the
best estimate noise matrix and preliminary noise matrix respectively.

\chapter{Cross-modulation Analysis}

An analysis including the cross-modulation correlations
is presented in this chapter.
This analysis is not simply a trivial expansion
of the single modulation analysis presented in
Chapter 5, but requires more sophisticated
techniques for the estimation of the
noise model and of the likelihood.
Following the procedure of Wilson et al. (1999),
the likelihood is simultaneously estimated in a
series of 8 bands from a map which was constructed including
all of the cross-modulation correlations.
The results are consistent with those from
the single modulation analysis.
However, since this analysis includes the
correlations between modulations, the constraints
on the power spectrum can be used directly
for cosmological parameter estimation.

\section{Noise Model and Consistency Checks}

\begin{figure}[t!]
\centerline{\epsfxsize=11.0cm \epsfbox{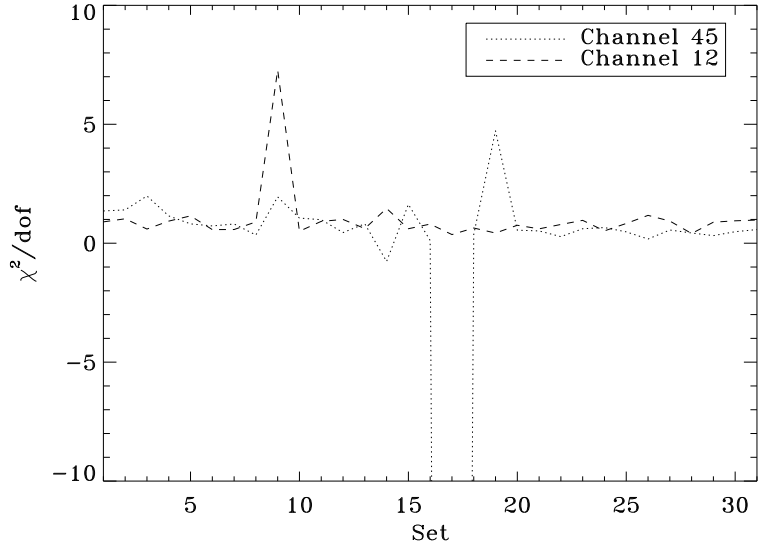}}
\ssp
\caption[$\chi^2$/dof for each set using an expansion of the noise model of section 5.2.]{$\chi^2$/dof for each set using an expansion of the noise model of section 5.2. $\chi^2={\vec d}({\bf C^N}+{\bf C^C})^{-1}{\vec d}^t$ and typical dof = 13 stares x 7 modulations - 7 constraints = 84. The $\chi^2$/dof are unacceptable, so a better model of the cross-modulation noise is necessary.}
\label{fig:big_chi2_dof_sim24}
\end{figure}

In order to derive a noise model for the cross-modulation analysis,
we expanded the procedure of section 5.2 for the cross-modulation
noise correlations. The value
of $\chi^2={\vec d}({\bf C^N}+{\bf C^C})^{-1}{\vec d}^t$,
where ${\bf C^N}$ includes noise correlations between all modulations,
was computed for each observing set and channel. The results
are shown in Figure \ref{fig:big_chi2_dof_sim24}
and indicate that a better
model of the cross-modulation noise is necessary.

In general, the elements of the
cross-modulation noise matrix can be expressed as
\begin{equation}
C^N_{ijmm^{\prime}}=C^M_{mm^{\prime}} C^F_{ij},
\label{eq:cnijmm}
\end{equation}
where the components of ${\bf C^M}$ are given by
\begin{equation}
C^M_{mm^{\prime}}={\vec M_m}^t {\bf C^S} {\vec M_{m^{\prime}}}
\label{eq:cnmm}
\end{equation}
and ${\vec M_m}$ is the timestream modulation
vector for modulation $m$, ${\bf C^S}$ describes
the noise in the timestream as a function of chopper sample $s$
and ${\bf C^F}$ is a model of the noise between fields $i$ and $j$.
Thus models for ${\bf C^M}$ and ${\bf C^F}$ are needed.
For clarity, ${\bf C^S}$ is a 128 $\times$ 128 matrix,
${\bf C^M}$ is a 8 $\times$ 8 matrix, and ${\bf C^F}$
is a 690 $\times$ 690 matrix  for PyV.

In the single modulation analysis we effectively
assumed ${\bf C^S}$ was the identity matrix.
For white noise with the chopper offset
removed, ${\bf C^N}$ would be of the form:
\begin{equation}
C^N_{ijmm^{\prime}} \propto
(\delta_{ij}-\frac{1}{N_{\rm stares}})\sigma^2
{\vec M_m}^t {\vec M_{m^{\prime}}}
\label{eq:cnijmm_white}
\end{equation}
However, a single $\sigma$ does not characterize all
modulations. Rather, $\sigma$ decreases
with increasing modulation, so a more
sophisticated model for ${\bf C^S}$ is necessary.

\begin{figure}[t!]
\centerline{\epsfxsize=11.0cm \epsfbox{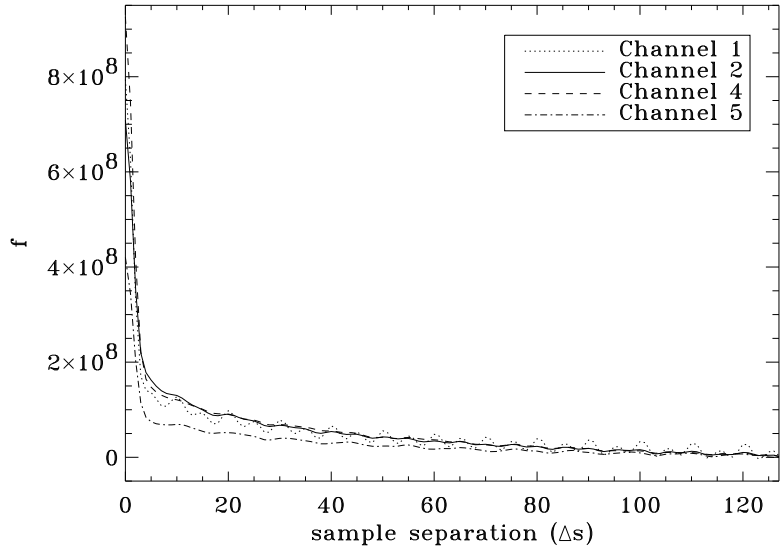}}
\ssp
\caption[$f(\Delta s)$ for set hb9.]{$f(\Delta s)$ for set hb9. $f(\Delta s)$ breaks around $\Delta s=6$, as would be expected from the beam size.}
\label{fig:f_delta_s}
\end{figure}

Assuming ${\bf C^S}$ only depends on
chopper sample separation $\Delta s=s-s^{\prime}$,
we can compute ${\bf C^S}$ from the following function:
\begin{equation}
f(\Delta s)=\sum_{s}d_{s}d_{s+\Delta s}.
\label{eq:fdeltas}
\end{equation}
For example, the $C^S_{12}$ component of ${\bf C^S}$ is given
by $f(\Delta s = 1)$.
In order to compute $f(\Delta s)$, a chopper synchronous offset is
first subtracted from the raw data. Then
$f(\Delta s)$ is calculated for each
channel, cycle and stare in a file. $f(\Delta s)$ is
then averaged over cycle, stare, and file.
Figure \ref{fig:f_delta_s} shows $f(\Delta s)$ for each channel
in one of the sets.

\begin{figure}[t!]
\centerline{\epsfxsize=7.0cm \epsfbox{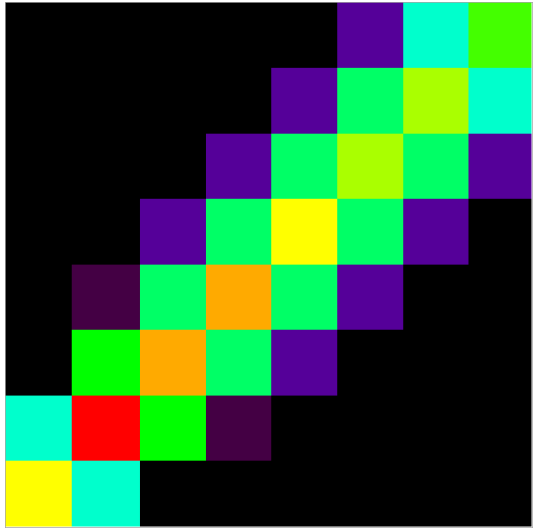}}
\ssp
\caption[${\bf C^M}$ for set hb9, channel 45.]{${\bf C^M}$ for set hb9, channel 45. Elements that are more than 2 modulations
apart are relatively uncorrelated.}
\label{fig:mfm}
\end{figure}

${\bf C^M}$ is calculated for each set and channel
following equation \ref{eq:cnmm}. ${\bf C^M}$ matrices for
channels 1 and 2 right and left are averaged, as are
${\bf C^M}$ matrices for
channels 4 and 5 right and left, yielding
${\bf C^M}$ matrices for channels 12 and 45 in each set.
As an example, ${\bf C^M}$ for one set and channel is shown
in Figure \ref{fig:mfm}.

\begin{figure}[t!]
\centerline{\epsfxsize=13.9cm \epsfbox{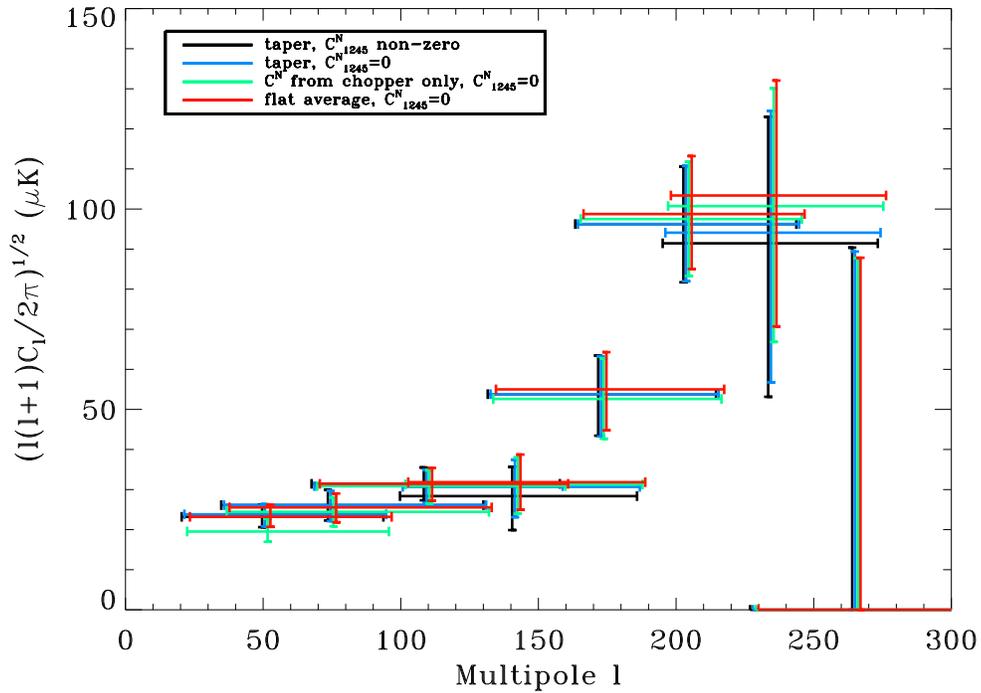}}
\ssp
\caption[Angular power spectra obtained using 4 different noise models.]{Angular power spectra obtained using  4 different noise models. Ignoring the correlations between channels 12 and 45 and the taper of $C_{N}N_{files}$ as a function of stare lag does not change the angular power spectrum significantly.}
\label{fig:powerspec_4sims}
\end{figure}

In order to get a simple form for ${\bf C^F}$, we investigated
the effect of ignoring the cross-channel correlations and
the taper as a function of stare lag on the single modulation
band powers. The results are shown in Figure \ref{fig:powerspec_4sims}.
Ignoring the correlations between channels 12 and 45 and
assuming the correlation between fields $i$ and $j$ come
only from the chopper offset subtraction does not change the angular
power spectrum significantly, so we assume ${\bf C^F}$
is of that form for the cross-modulation analysis.

\begin{figure}[t!]
\centerline{\epsfxsize=11.0cm \epsfbox{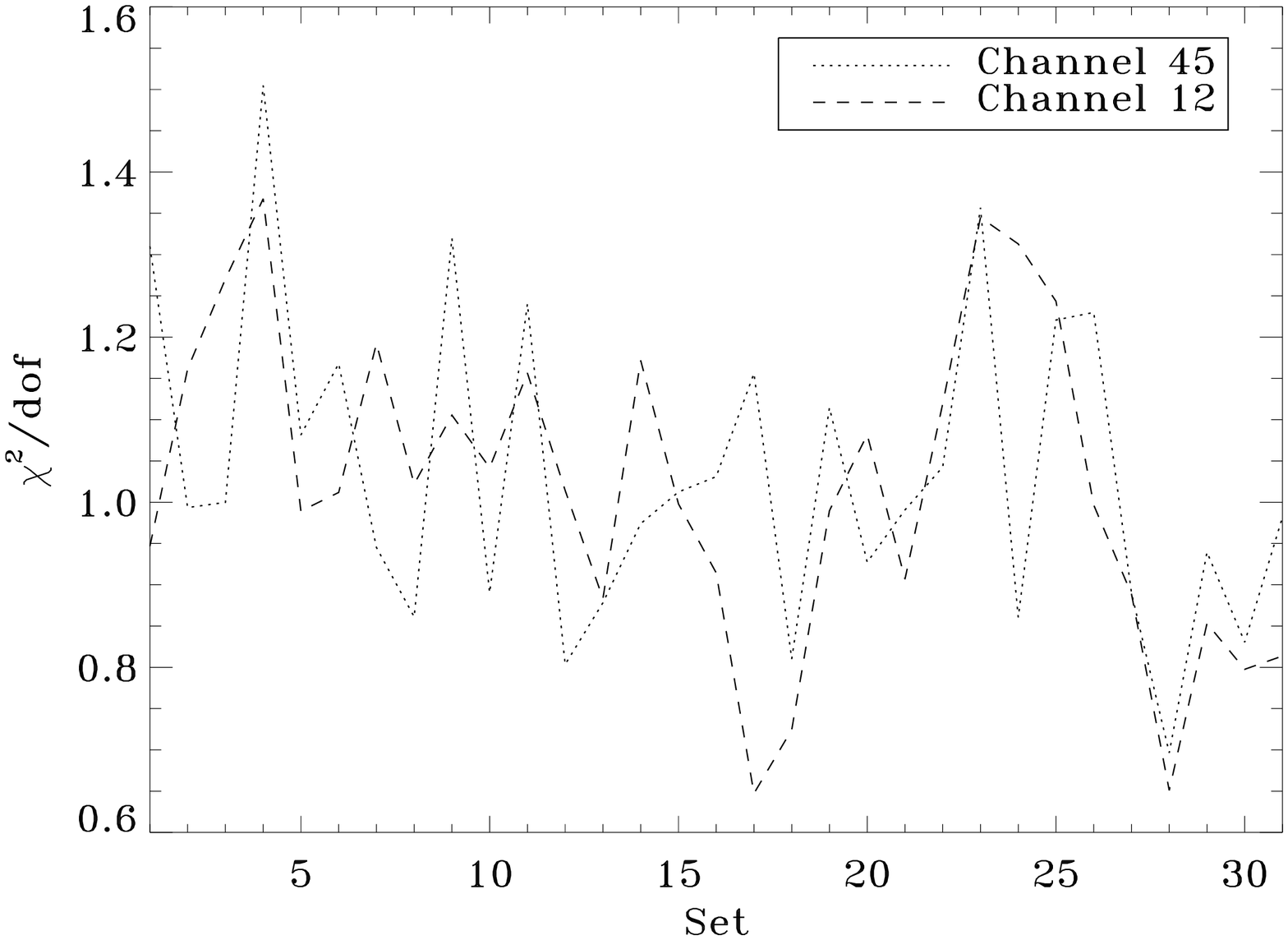}}
\ssp
\caption[$\chi^2$/dof for each set using the final noise matrix.]{$\chi^2$/dof for each set using the final noise matrix. $\chi^2={\vec d}({\bf C^N}+{\bf C^C})^{-1}{\vec d}^t$ and typical dof = 13 stares x 8 modulations - 8 constraints = 96. The vertical scale of the plot covers a smaller range than that of Figure \ref{fig:big_chi2_dof_sim24}.}
\label{fig:big_chi2_dof}
\end{figure}

\begin{figure}[t!]
\centerline{\epsfxsize=11.0cm \epsfbox{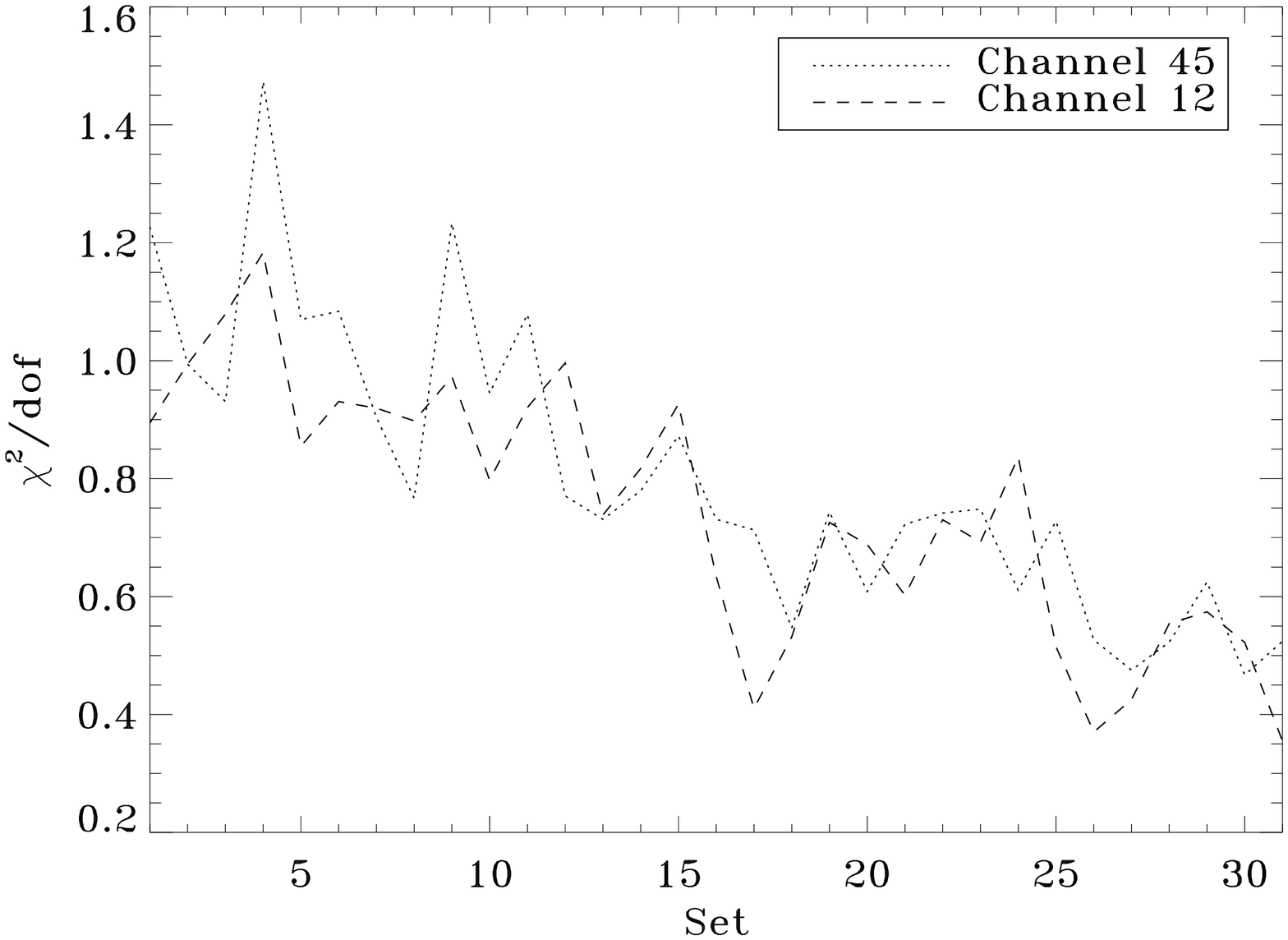}}
\ssp
\caption[$\chi^2$/dof for each set normalized to modulation 2.]{$\chi^2$/dof for each set normalized to modulation 2. $\chi^2={\vec d}({\bf C^N}+{\bf C^C})^{-1}{\vec d}^t$ and typical dof = 13 stares x 8 modulations - 8 constraints = 96. The declination of the sets roughly decreases with increasing set number. The vertical scale of the plot covers a larger range than that of Figure \ref{fig:big_chi2_dof}.}
\label{fig:big_chi2_dof_norm3}
\end{figure}

Finally, since this noise model is derived from sample-sample, it is larger
than the corresponding noise derived from the co-added data by a factor
of $\sim 10^4$, so $C^N_{ijmm^{\prime}}$ must be normalized to the
variance in the co-added data for each set. Since ${\bf C^M}$
accounts for the relative normalization of all of the modulations,
$C^N_{ijmm^{\prime}}$ must be normalized to the
variance in only one modulation of the co-added data for each set.
We normalize to modulation 8 because it is least potentially
affected by the ground shield.
Figure \ref{fig:big_chi2_dof} shows the $\chi^2$/dof for each set
using the final cross-modulation noise model.
If $C^N_{ijmm^{\prime}}$ is normalized to modulation 2 instead
of modulation 8, the $\chi^2$ depends on
elevation (Figure \ref{fig:big_chi2_dof_norm3}).

As another check on the noise matrix used in the cross-modulation
analysis, single modulation band powers were computed using
the $C^N_{ijmm}$ components of $C^N_{ijmm^{\prime}}$. They are
consistent with the band powers given in Chapter 5.

\section{Map Making}

The power spectrum is simultaneously estimated in
several $l$-space bands, this time including all
of the cross-modulation covariances, from a map
constructed from the modulated data.
The power spectrum could have been estimated using
covariance matrices of size 5520 $\times$ 5520
in each of the $l$-space bands, but transformation
into a map basis first reduces the computational load.
This technique was used in an analysis of the MSAM-I
experiment (Wilson et al. 1999), for which the power spectrum
estimated using the map is consistent with the
power spectrum estimated directly from the full
covariance matrices.

In order to construct the map, first recognize that the data
can be expressed as
\begin{equation}
{\vec d} =  {\bf B}{\vec T} + {\vec n}
\end{equation}
with noise covariance matrix $<nn> = {\bf N}$.
(The actual ${\bf N}$ used is ${\bf N}={\bf C^N}+{\bf C^C}$.)
To obtain the underlying temperature field ${\vec T}$,
the matrix ${\bf B}$ must be inverted. This inversion is carried
out by constructing the estimator $\hat T$ which minimizes
the $\chi^2$:
\begin{equation}
\chi^2 \equiv ({\vec d} - {\bf B} \hat T){\bf N^{-1}}
({\vec d} - {\bf B} \hat T).
\end{equation}
We find
\begin{equation}
\hat T = \tilde {\bf N} {\bf B} {\bf N^{-1}} {\vec d}.
\label{eq:lincom}
\end{equation}
This estimator will be distributed around the
true temperature due to noise,
where ${\bf \tilde N}$, the noise covariance
matrix for the map, is given by
\begin{equation}
{\bf \tilde N} \equiv <(\hat T - {\vec T})(\hat T - {\vec T}) > = {\bf \left(  B^T  N^{-1} B \right)^{-1}}.
\label{eq:cn_map}
\end{equation}

\subsection{Power Spectrum Constraints}
This map can be analyzed in the same manner as 
the modulated data, with the
advantage that the signal covariance
matrix is much simpler to compute
than using equations \ref{eq:cth} and \ref{eq:wkij5}.
In the map basis, the signal covariance matrix
simplifies to
\begin{equation}
<T_i T_j > = \sum_l {2l+1\over 4\pi} P_l(\cos(\theta_{ij}))C_l.
\label{eq:ct_map}
\end{equation}
Indeed, one way to think of a map is that it is the linear
combination of the data for which the signal (and therefore its
covariance) is independent of the experiment.
The noise covariance for the map (equation \ref{eq:cn_map}) accounts
for all of the experimental
processing.

\begin{figure}[t!]
\centerline{\epsfxsize=13.9cm \epsfbox{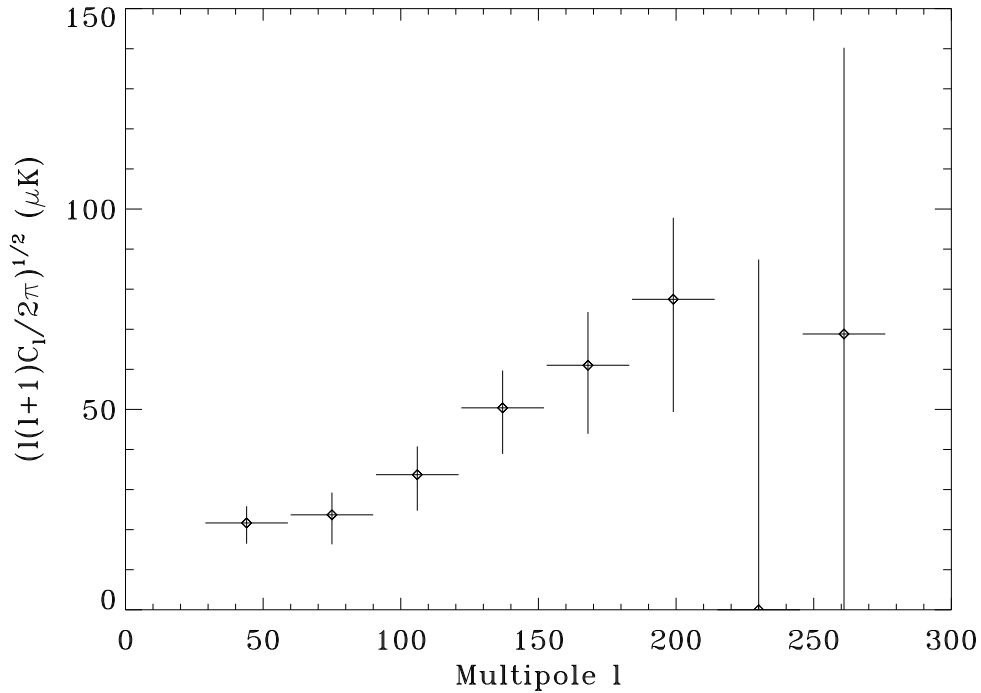}}
\ssp
\caption[PyV angular power spectrum constraints from the cross-modulation analysis.]{Angular power spectrum constraints from the cross-modulation analysis. Flat band power, $(l(l+1)C_{l}/2\pi)^{1/2}$ is simultaneously estimated in 8 $l$-space bands including cross-modulation theory and noise covariances. The error bars include statistical uncertainties only.}
\label{fig:crossmod_cl}
\end{figure}

The results of the likelihood analysis
are shown in Figure \ref{fig:crossmod_cl}
and are given in Table \ref{tbl:crossmod_cl}.

\begin{table*}
\begin{center}
\begin{tabular}{|cc|}
\tableline
$l$ & $(l(l+1)C_{l}/2\pi)^{1/2}$  \\
\tableline
$44^{+15}_{-15}$    &     $ 22_{-5}^{+4}$ \\
$75^{+15}_{-15}$    &     $ 24_{-7}^{+6}$ \\
$106^{+15}_{-15}$   &     $ 34_{-9}^{+7}$ \\
$137^{+15}_{-15}$   &     $ 50_{-12}^{+9}$ \\
$168^{+15}_{-15}$   &     $ 61_{-17}^{+13}$ \\
$199^{+15}_{-15}$   &     $ 77_{-28}^{+20}$ \\
$230^{+15}_{-15}$   &     $0.003_{-0.003}^{+87}$ \\
$261^{+15}_{-15}$   &     $ 69_{-69}^{+71}$  \\
\tableline
\end{tabular}
\end{center}
\ssp
\caption[Angular power spectrum constraints from the cross-modulation analysis.]{Angular power spectrum constraints from the cross-modulation analysis. Band powers are in $\mu K$.}
\label{tbl:crossmod_cl}
\end{table*}

\begin{figure}[t!]
\centerline{\epsfxsize=13.9cm \epsfbox{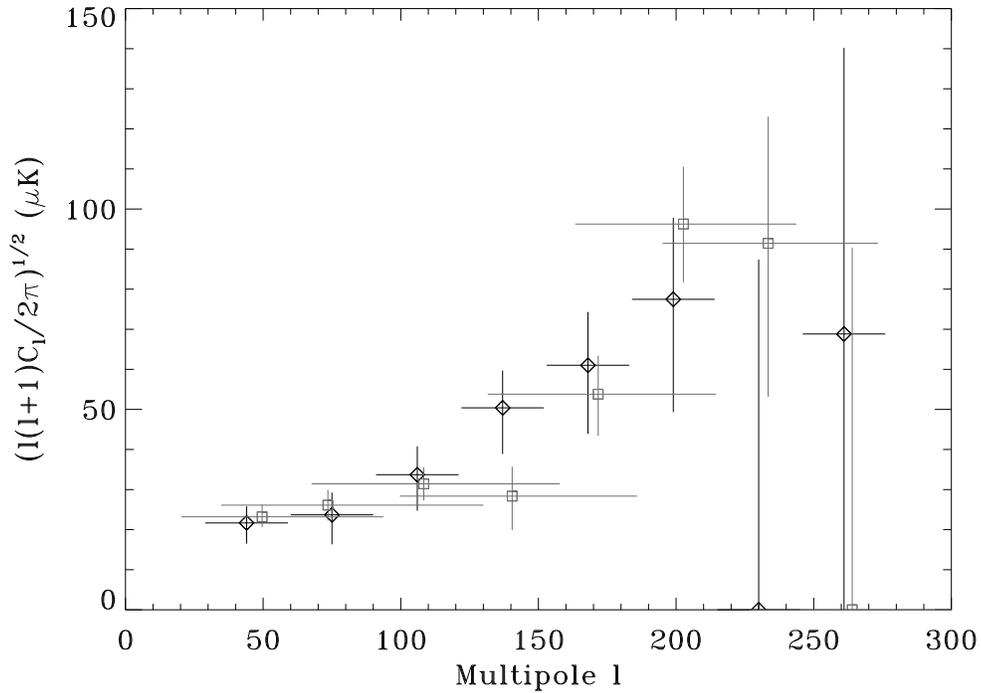}}
\ssp
\caption[Comparison of angular power spectra from single and cross-modulation analyses.]{Comparison of angular power spectra from single and cross-modulation analyses. Points obtained from the single modulation analysis are denoted by gray squares and points obtained from the cross-modulation are denoted by black diamonds.}
\label{fig:powerspec_sd_kc}
\end{figure}

The angular power spectra obtained using the cross-modulation and
single modulation analyses are consistent, as illustrated
in Figure \ref{fig:powerspec_sd_kc}.
At first glance, the band power in the fourth
modulation of the single modulation analysis
is higher than that of the fourth band of the
cross-modulation analysis.
However, the Knox minimum variance filter (Knox 1999)
shows that the fourth modulation of the single modulation
analysis has significant sensitivity
to lower $l$ scales. The Knox filter is a more complete
representation of angular sensitivity than just the diagonal window
functions, because it contains information from the full (off-diagonal)
covariance matrix.

\subsection{CMB Images}
\begin{figure}[t!]
\centerline{\epsfxsize=10.5cm \epsfbox{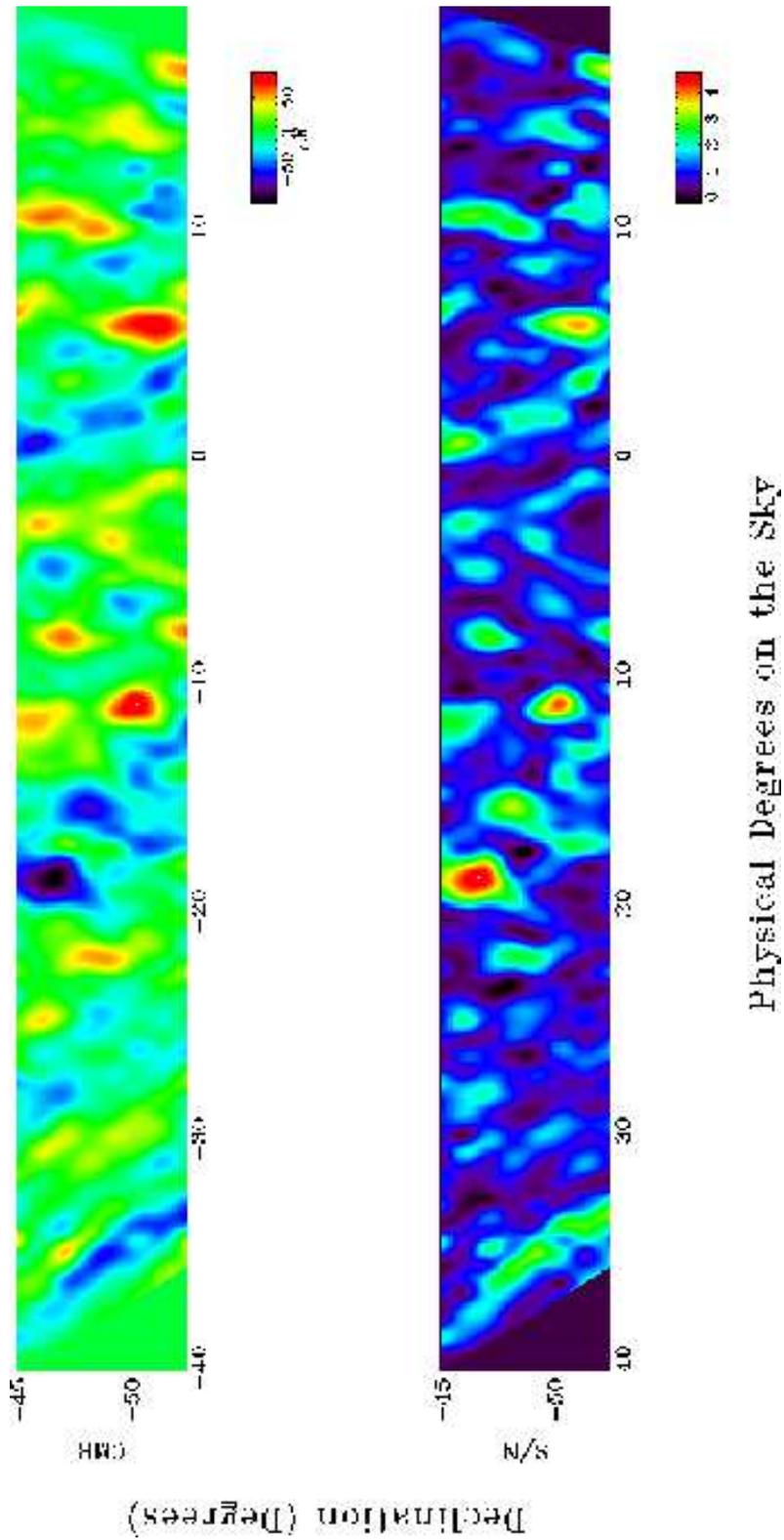}}
\ssp
\caption[Weiner-filtered CMB and S/N maps for the main PyV region.]{Weiner-filtered CMB and S/N maps for the main PyV region. The unfiltered map was used for power spectrum estimation.}
\label{fig:cmb_maps}
\end{figure}

The map can also be Weiner-filtered to produce a
realistic image of the sky. A Weiner-filtered map
of the main PyV region and the $S/N$ in the map
are shown in Figure \ref{fig:cmb_maps}.
The unfiltered map, not the Weiner-filtered map, was used for
power spectrum estimation.

\section{Comparisons With Other Experiments}

\subsection{Power Spectrum Comparison}
Flat band powers calculated from the subset of PyV data in the PyIII
region of sky are consistent with the PyIII flat band powers
to within the uncertainties.
Given the calibration and statistical uncertainties,
the lowest $l$ PyV modulation
agrees with the 
the smallest scale COBE measurements.

\subsection{Comparing PyV and PyIII Maps}
\begin{figure}[ht!]
\centerline{\epsfxsize=13.9cm \epsfbox{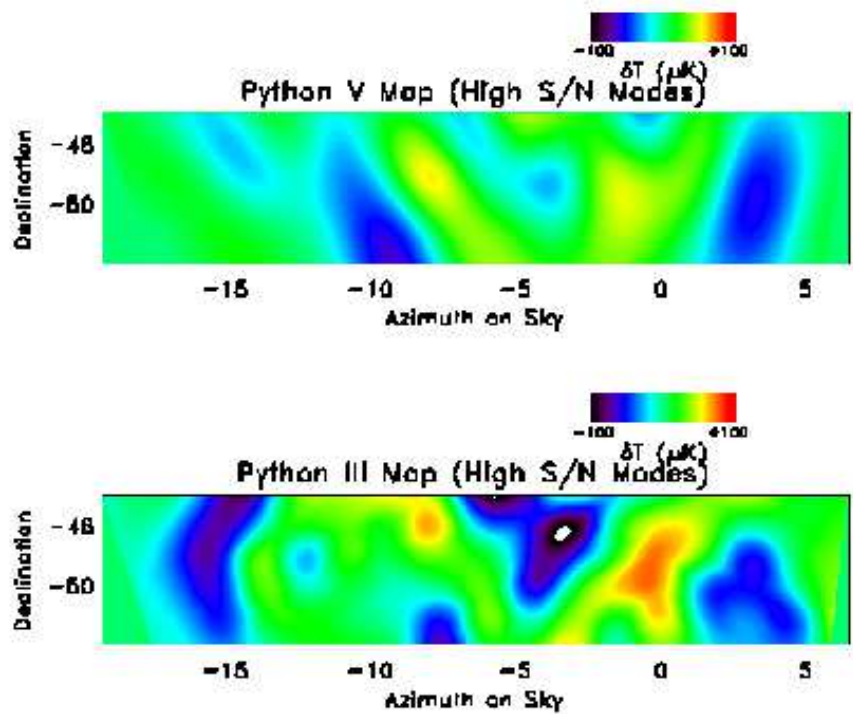}}
\ssp
\caption[A comparison of PyV and PyIII maps.]{A comparison of PyV and PyIII maps. Structures found in the PyV map
are also evident in the PyIII map,
implying that PyIII and PyV are consistent with each other.}
\label{fig:py53_maps}
\end{figure}

Using the technique of section 6.2.1, maps of the sky are
reconstructed from the PyIII data. The PyIII map is compared
to the PyV map in the same region of the sky in Figure
\ref{fig:py53_maps}.
Since PyIII has higher S/N than PyV, features which
appear in the PyV map should also appear in the PyIII map,
but features which appear in the PyIII map may not
necessarily appear in the PyV map.
In order to compare modes of the experiments which are
similar to each other, both maps are transformed
into the KL basis and only modes with $S/N>2$ are used.
It is not necessary to Weiner filter the maps (which removes
noise) if only the highest S/N modes are used.
Structures found in the PyV map of Figure \ref{fig:py53_maps}
are also evident in the PyIII map of Figure \ref{fig:py53_maps},
implying that PyIII and PyV are consistent with each other.

\chapter{Foregrounds}

The PyV data are cross-correlated with several foreground templates in
order to set limits on possible foreground contamination. The
templates used are the Schlegel et al. (1998) 100 micron dust map,
which is based on IRAS and DIRBE maps,
the Haslam et al. (1974) 408 MHz survey (synchrotron),
and the PMN survey (point sources).
Each foreground template map is smoothed to
PyV resolution, pixelized and
modulated according to the PyV observation scheme.

Two templates are created for the PMN survey.
We call one PMN, which is converted to
$\delta T_{\rm CMB}$ using the spectral indices given in the survey. The
other we call PMN0, which is
converted to a flux at 40 GHz assuming
a flat spectrum extrapolated from the flux measurement at 4.85 GHz.
The assumption of a flat spectrum is  
conservative in that it is likely to over-estimate the flux
at 40 GHz. Neither case is correct, since spectral indices have not
been measured for all of the sources, in which case a flat spectrum is
assumed, but we do know that some of them are not flat, so a flat
spectrum will be inappropriate. The two cases cover a
reasonable range of possibilities.

\begin{figure}[t!]
\centerline{\epsfxsize=13.9cm \epsfbox{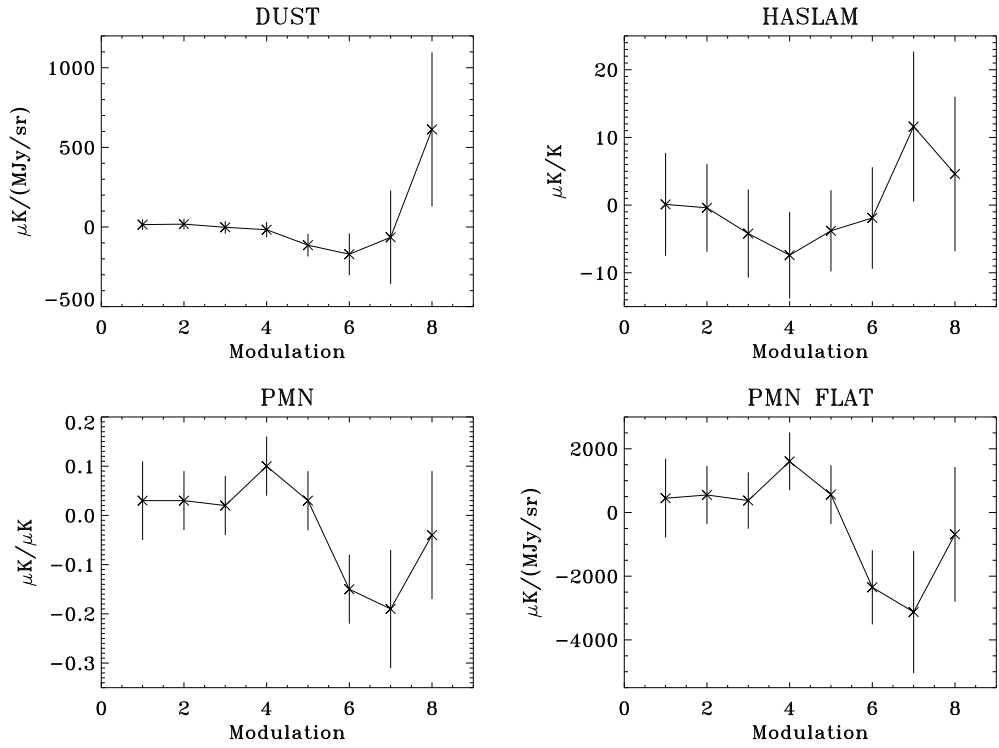}}
\ssp
\caption[Correlation coefficients and uncertainties for each foreground and modulation.]{Correlation coefficients and uncertainties for each foreground and modulation.}
\label{fig:foreg_correl}
\end{figure}

For each modulation and foreground template, a
correlation coefficient and uncertainty are calculated
following de~Oliveira-Costa et al. (1997):

\begin{equation}
\alpha = {{{\vec d}_{f}^{t} {\bf C}^{-1} {\vec d}_{p}} \over {{\vec d}_{f}^{t} {\bf C}^{-1} {\vec d}_{f}}}
;
\sigma^{2}_{\alpha} = {1 \over { {\vec d}_{f}^{t} {\bf C}^{-1} \ {\vec d}_{f}^{t}}}
\label{forg2}
\end{equation}
where ${\vec d}_{f}$ and ${\vec d}_{p}$ are the
modulated foreground and PyV data
respectively and ${\bf C}={\bf C^N}+{\bf C^C}+{\bf C^T}$
(Figure \ref{fig:foreg_correl}).

A weighted mean and uncertainty over all of the modulations
for each foreground are given in Table \ref{tbl:foreg_correl}.
In all cases there is no clear detection of foregrounds. 
The RMS of each modulation of each foreground was calculated
and then multiplied by the corresponding $1 \sigma$
error bar in Table \ref{tbl:foreg_correl} in
order to estimate an upper limit on the
foreground contribution to CMB band power.
The limits on contributions from
foregrounds are given in Table \ref{tbl:bp_foreg} and are
at least $\sim 10 \times$ smaller than the measured CMB bandpowers.

As a test on both the map-making procedure and the foreground
estimation procedure, a map of the dust was constructed from
the dust data which had been modulated and pixelized according to
PyV observing strategy. The original dust map was recovered.

If the diffuse morphology of the sky is not
constant as a function of wavelength, then these
templates do not reveal all of the foreground
contamination and more could be hidden in the PyV data.
A combined analysis of PyIII at 90 GHz and PyV at 40 GHz would
constrain the foreground contamination further.

\begin{table*}
\begin{center}
\begin{tabular}{|cccc|}
\tableline
Dust  & Haslam  & PMN  & PMN0 \\
$\mu K (MJy/sr)^{-1}$ & $\mu K / K$ & $\mu K / \mu K$ & $\mu K(MJy/sr)^{-1}$\\
\tableline
-3 $\pm$ 18  &  -2.0 $\pm$ 2.6  &  0.012 $\pm$ 0.024  &  195 $\pm$ 385\\
\tableline
\end{tabular}
\end{center}
\caption{Correlation coefficients and uncertainties for weighted means.} \label{tbl:foreg_correl}
\end{table*}

\begin{table*}
\begin{center}
\begin{tabular}{|ccccc|}
\tableline
Mode & Dust  &  Haslam & PMN  & PMN0\\
\tableline
1 &  1.0  & 0.5    & 0.3  & 0.3 \\
2 &  1.0  & 0.6    & 0.6  & 0.6 \\
3 &  1.1  & 0.8    & 1.0  & 1.0 \\
4 &  1.1  & 1.0    & 1.1  & 1.1 \\
5 &  1.9  & 2.7    & 2.7  & 2.7 \\
6 &  2.7  & 6.7    & 5.4  & 5.4 \\
7 &  2.0  & 6.9    & 5.9  & 5.9 \\
8 &  1.6  & 8.2    & 6.6  & 6.6 \\
\tableline
\end{tabular}
\end{center}
\caption{Upper limits on foreground contribution. All units are $\mu K$.} \label{tbl:bp_foreg}
\end{table*}

\begin{figure}[t!]
\centerline{\epsfxsize=13.9cm \epsfbox{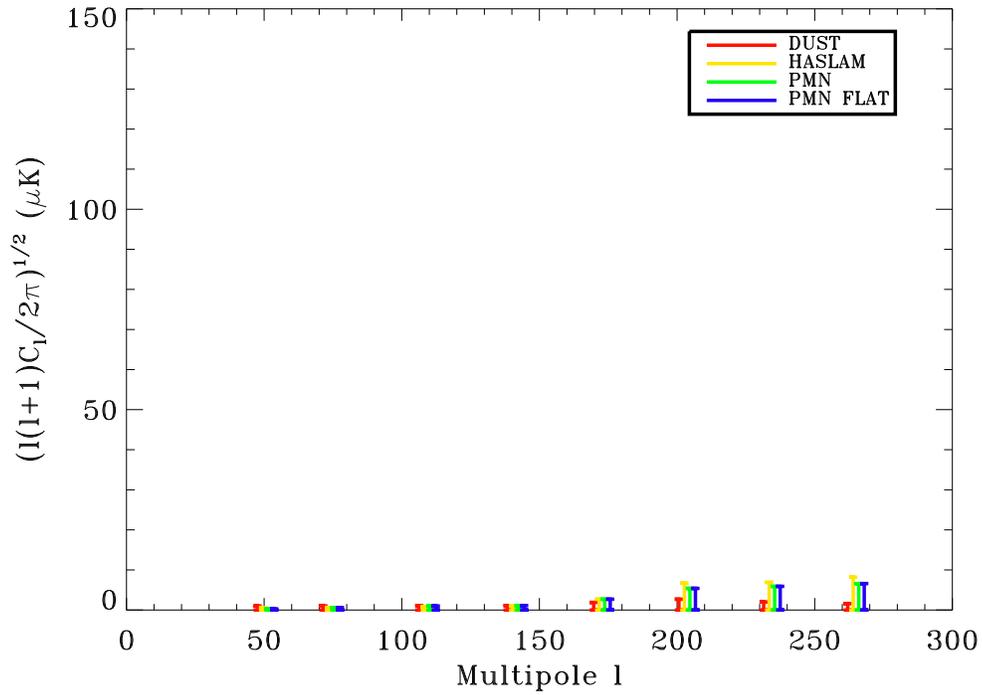}}
\ssp
\caption[Upper limits on foreground contribution. All units are $\mu K$.]{Upper limits on foreground contribution. All units are $\mu K$.}
\label{fig:bp_foreg}
\end{figure}

\chapter{Conclusions}

The goal of the Python V season was to
increase the sky coverage and
the range of observed angular scales of the Python experiment
thereby constraining the CMB angular power spectrum.

The PyV experiment fully samples 598 deg$^2$ of the microwave sky
and constrains the CMB angular power spectrum in the angular scale
range $40 \lesssim l \lesssim 260$.
The measurements pass internal consistency checks,
show little contamination from foreground radiation, and
are consistent with previous Python and COBE results.

\begin{figure}[t!]
\centerline{\epsfxsize=13.9cm \epsfbox{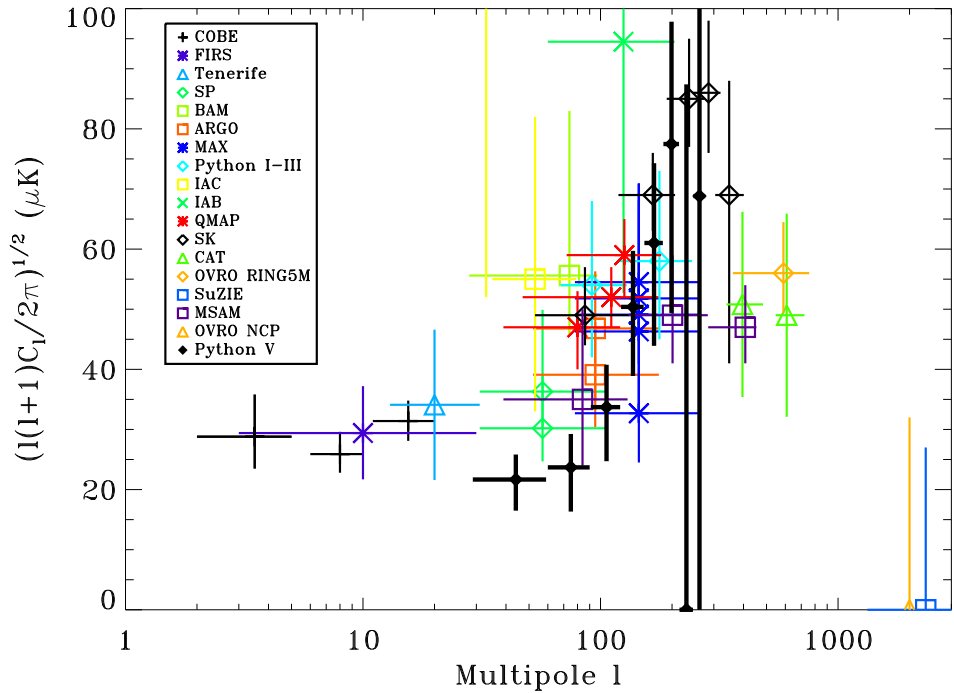}}
\ssp
\caption[CMB anisotropy measurements as of Feb 1999 with the PyV measurements overplotted.]{CMB anisotropy measurements as of Feb 1999. The points are the same as those in Figure \ref{fig:expt_summary}. The PyV measurements are overplotted.}
\label{fig:expt_summary_concl}
\end{figure}

The results of experiments previous to PyV collectively
imply a rise in the angular power spectrum, but the
PyV experiment definitively shows a rise in the spectrum
from from large ($l \sim 40$) to small ($l \sim 200$)
angular scales.
Figure \ref{fig:expt_summary_concl} summarizes
the constraints on the angular power spectrum
by experiments previous to PyV and
highlights the PyV results.

\begin{figure}[t!]
\centerline{\epsfxsize=13.9cm \epsfbox{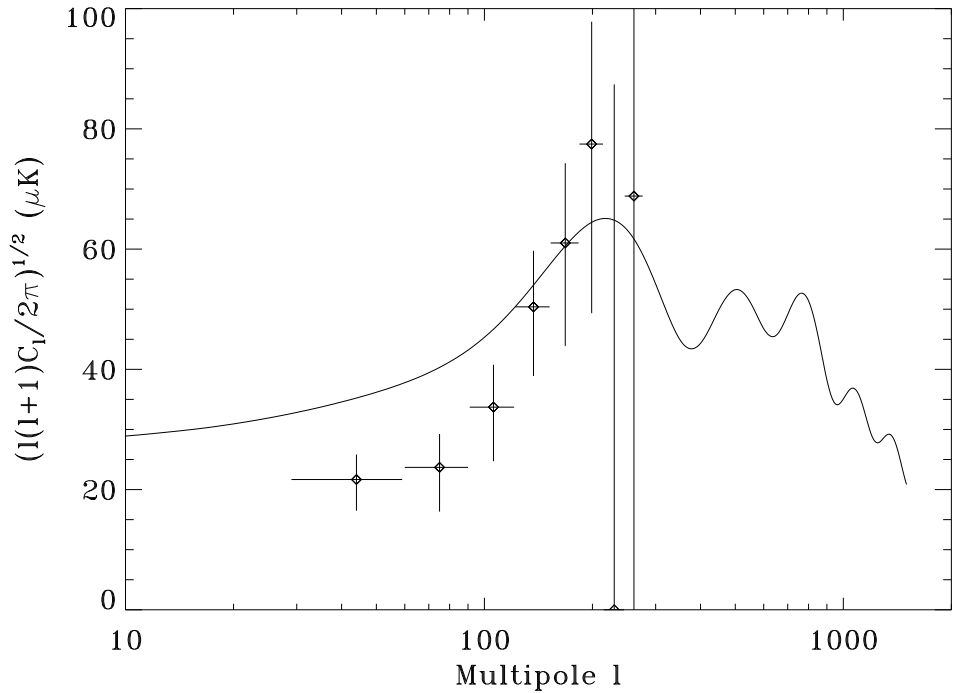}}
\ssp
\caption[PyV angular power spectrum and COBE-normalized standard CDM model.]{PyV angular power spectrum and COBE-normalized standard CDM model (density $\Omega=1.0$, Hubble constant $H_0=50$ km/s/Mpc, baryon density $\Omega_b=0.05$, and cosmological constant $\Lambda=0.0$). Only statistical uncertainties are included in the PyV points.}
\label{fig:powerspec_sd_scdm}
\end{figure}

The PyV data are inconsistent with a flat spectrum,
and instead show a rising spectrum, implying the acoustic scenario of
structure formation described in the introduction is correct.
The PyV data give confidence in results of estimation of
cosmological parameters, such as $\Omega$, from all previous CMB
experiments because the rise in the spectrum could no longer
be a calibration artifact. Since reionization of the
universe would damp power at smaller angular scales
(Hu and White 1995), the PyV data can rule out models
of homogeneous reionization with $z_{reion} \gtrsim 75$.
Earlier work on cosmic defect models
suggests there would be no peak in the CMB
angular power spectrum due to defects
(Allen et al. 1997, Pen et al. 1998),
in which case the PyV data would disfavor defect models.
However, there is still disagreement in the predictions,
with later work (Albrecht et al. 1999, Wu et al. 1998)
suggesting there could be a peak due to defects.
The PyV spectrum is shown in reference to a COBE-normalized
standard cold dark matter (sCDM) model
in Figure \ref{fig:powerspec_sd_scdm}.
The PyV spectrum has lower power at large
angular scales and a steeper slope than typical
sCDM models ($\chi^2/{\rm dof} \sim 2.5$ for the model shown).
Models with a higher baryon density and/or a cosmological
constant have a steeper slope than the model shown.
While future CMB experiments will estimate cosmological
parameters to high precision in the coming decade, the
robust rise in the Python V spectrum already tightly constricts
the range of viable models of structure formation.

%
%
%
%
%
%
%
\spacing{1}

\end{document}